\documentclass[prx,twocolumn,floatfix,superscriptaddress,amsmath,amssymb]{revtex4-1}

\usepackage{mathtools}
\usepackage{graphicx}
\usepackage{latexsym}
\usepackage{nicefrac,xfrac}
\usepackage{color}
\usepackage{bm}

\setcitestyle{numbers,square}

\begin{document}
\title{How spin-orbital entanglement depends on the spin-orbit coupling in a Mott insulator}

\author{      Dorota Gotfryd }
\affiliation{ \mbox{Institute of Theoretical Physics, Faculty of Physics,
              University of Warsaw, ul. Pasteura 5, PL-02093 Warsaw, Poland} }
\affiliation{ \mbox{Institute of Theoretical Physics, Jagiellonian University,
              Prof. S. {\L}ojasiewicza 11, PL-30348 Krak\'ow, Poland} }

\author{      Ekaterina M. P\"arschke       }
\affiliation{ Department of Physics, University of Alabama at Birmingham,
              Birmingham, Alabama 35294, USA }
\affiliation{Institute of Science and Technology Austria, Am Campus 1,
             A-3400 Klosterneuburg, Austria}

\author{     Ji\v{r}\'i Chaloupka }
\affiliation{Department of Condensed Matter Physics, Faculty of Science,
             Masaryk University, \\ Kotl\'a\v{r}sk\'a 2, CZ-61137 Brno,
             Czech Republic}
\affiliation{\mbox{Central European Institute of Technology, Masaryk University,
             Kamenice 753/5, CZ-62500 Brno, Czech Republic}}

\author{      Andrzej M. Ole\'s }
\affiliation{ \mbox{Institute of Theoretical Physics, Jagiellonian University,
              Prof. S. {\L}ojasiewicza 11, PL-30348 Krak\'ow, Poland} }
\affiliation{ Max Planck Institute for Solid State Research,
              Heisenbergstrasse 1, D-70569 Stuttgart, Germany }

\author{      Krzysztof Wohlfeld }
\affiliation{ \mbox{Institute of Theoretical Physics, Faculty of Physics,
              University of Warsaw, ul. Pasteura 5, PL-02093 Warsaw, Poland} }

\date{\today}

\begin{abstract}
The concept of the entanglement between spin and orbital degrees of
freedom plays a crucial role in understanding of various phases and
exotic ground states in a broad class of materials, including orbitally
ordered materials and spin liquids. We investigate how the spin-orbital
entanglement in a Mott insulator depends on the value of the
spin-orbit coupling of the relativistic origin. To this end, we
numerically diagonalize a one-dimensional spin-orbital model with the
`Kugel-Khomskii' exchange interactions between spins and orbitals on
different sites supplemented by the on-site spin-orbit coupling.
In the regime of small spin-orbit coupling w.r.t. the spin-orbital
exchange, the ground state to a large extent resembles the one obtained
in the limit of vanishing spin-orbit coupling. On the other hand, for
large spin-orbit coupling the ground state can, depending on the model
parameters, either still show negligible spin-orbital entanglement,
or can evolve to a highly spin-orbitally entangled phase with completely
distinct properties that are described by an effective XXZ model.
The presented results suggest that:
(i) the spin-orbital entanglement may be induced by large
on-site spin-orbit coupling, as found in the $5d$ transition metal
oxides, such as the iridates;
(ii)  for Mott insulators with weak spin-orbit coupling of
Ising-type, such as e.g. the alkali hyperoxides, the effects of the
spin-orbit coupling on the ground state can, in the first order of
perturbation theory, be neglected.
\end{abstract}

\maketitle

\section{Introduction}

\subsection{Interacting Quantum Many-body Systems:\\Crucial Role of Entanglement}

One of the main questions for a quantum interacting many-body system
concerns the nature of its ground state. Perhaps the most
fundamental question that can be formulated here is as follows:
Can the eigenstates of such a system be written in terms of a product
of the `local' (e.g. single-site) basis states? If this is the case,
then the ground state can be understood using the classical physics
intuition. Moreover, the low lying excited states can then be described
as weakly interacting quasiparticles, carrying the quantum numbers of
the constituents forming the system. Such physics is realized, for
instance, by the ions in conventional crystals, spins in ordered
magnets, or electrons in Fermi liquids~\cite{Khomskii2010}.

An interesting situation occurs, however, when the answer to the above
question is negative and we are left with a `fully quantum' interacting
many-body problem~\cite{Andrade2018,Zaanen2019}. In this case, the
classical intuition fails, numerical description of the
ground state may become exponentially difficult, and the low lying
eigenstates cannot be described as weakly interacting quasiparticles.
This can, for example, be found in the spin liquids stabilized in the
one-dimensional (1D) or highly frustrated magnets~\cite{Balents2010}
or incommensurate electronic systems with strong electron interactions
(the so-called non-Fermi liquids)~\cite{Varma2002}. In fact,
a large number of condensed matter studies are nowadays devoted to
the understanding of such `exponentially difficult' problems.

A way to characterize the fully quantum interacting many-body problem
is by introducing the concept of entanglement
\cite{Einstein1935,Zurek,Karol} and then by defining interesting
aspects of its quantum structure through the entanglement entropy
\cite{Bennett1996}. It is evident that studying entanglement always
requires first a definition of what is entangled with what, i.e.,
specifying the division of the system into the subsystems that
may become entangled. Perhaps the most widely performed division has
so far concerned splitting a lattice spin system into two subsystems
in real space \cite{Vidal2003}. Such studies enabled to identify the
relation between the entanglement of the spin system, its scaling with
the system size, and the product nature of the ground state, cf. Refs.
\cite{Vidal2003,Korepin2004,Wolf2006,Gioev2006,Lat04,Cin08,Thomale2010}.

\subsection{Spin-Orbital Entanglement in\\ Transition Metal Compounds}

Transition metal oxides involve often numerous competing degrees of
freedom. Examples are the three-dimensional (3D) ground states of
LaTiO$_3$, LaVO$_3$, YVO$_3$, Ba$_3$CuSb$_2$O$_9$
\cite{Khaliullin2000,Khaliullin2001,Kha05,Nakatsuji2012,Ishiguro2013,Man18},
where spin and orbital degrees of freedom are intertwined and entangled.
A similar situation occurs in MnP \cite{Pan19}, the first Mn-based
unconventional superconductor under pressure, and in some
two-dimensional (2D) model systems
\cite{Nor08,*Nor11,*Cha11,Corboz2012,Brz12}. In all these cases the
ground state can only be explained by invoking the joint spin-orbital
fluctuations. Consequently, the mean field decoupling separating
interactions into spin and orbital
degrees of freedom fails and cannot be used.

In this paper we study the {\it spin-orbital entanglement} which
manifests itself when a quantum many-body system with interacting spin
and orbital degrees of freedom is split into the subsystems with
separated degrees of freedom, i.e., one attempts to write interacting
spin and orbital wave functions separately \cite{Ole06,Ole12,Chen2007}.
The concept of entanglement has been first introduced in these systems
to understand the violation of the so-called Goodenough-Kanamori
rules~\cite{Goodenough1963,Kanamori1959}
in the ground states of several transition metal oxides with partially
filled $3d$ orbitals, strong intersite spin-orbital (super)exchange
interactions but typically negligible value of the on-site spin-orbit
coupling. It was also realized that entanglement is
important to understand the excited states where spin and orbital
variables are intertwined. Good examples are the temperature evolution
of the low energy excitations in LaVO$_3$ \cite{Miy02,Kha04} and the
renormalization of spin waves by orbital tuning in spin-orbital systems
due to the weak interactions with the lattice \cite{Sna19}.
Furthermore, the spin-orbital entanglement is also crucial to understand
the first unambiguous observations of the collective orbital excitation
(orbiton) in Sr$_2$CuO$_3$ and CaCu$_2$O$_3$ \cite{Schlappa2012,Bisogni2015}
and their interpretation in terms of the spin and orbital separation in
a 1D chain~\cite{Wohlfeld2011,Wohlfeld2013,Chen2015}.

Crucially, the spin-orbital entanglement is expected as well in the
oxides with strong on-site spin-orbit coupling---probably best
exemplified by the partially filled $4d$ and $5d$ orbitals as found in
the ruthenium and iridium oxides~\cite{WitczakKrempa2014,Ber19}, or in
the recently discovered $5d$ Ta chlorides \cite{Ish19}. For instance,
the concept of spin-orbital entanglement was recently invoked to
understand the inelastic x-ray spectrum of Sr$_3$NiIrO$_6$~\cite{Lef16},
the ground state of H$_3$LiIr$_2$O$_6$~\cite{Kit18} or, in a different
physical setting, the photoemission spectra of
Sr$_2$RuO$_4$~\cite{Vee14, Zha16}. Interestingly, the peculiarities of
the interplay between the strong on-site spin-orbit coupling and the
spin-orbital (super)exchange interactions allowed for the onset of
several relatively exotic phenomena in this class of compounds---such
as a condensed matter analogue of a Higgs boson in Ca$_2$RuO$_4$
\cite{Kha13,Jain2017} or the strongly directional, Kitaev-like,
interactions between the low energy degrees of freedom (pseudospins) in
some of the iridates or ruthenates on a quasi-2D honeycomb lattice
(Na$_2$IrO$_3$, Li$_2$IrO$_3$, $\alpha$-RuCl$_3$, H$_3$LiIr$_2$O$_6$)
\cite{Cha10,Winter2017,Kit18}.
The latter might be described to some extent by the exactly solvable
Kitaev model on the honeycomb lattice which, {\it inter alia}, supports
the onset of a novel spin-liquid ground state with fractionalized
`Majorana' excitations~\cite{Kitaev2006}.

\subsection{Main question(s) and organization of the paper}

It may come as a surprise that the concept of the spin-orbital
entanglement has so far been rigorously investigated only for the
systems where the spins and orbitals at neighboring sites interact,
as a result of the spin, orbital and spin-orbital
(super)exchange processes in Mott insulators
\cite{Ole06,Ole12,Chen2007,Lun12,You12,Lauchli2012,You2015,You2015b,Val19}.
This case is physically relevant to all Mott insulators with
{\it negligible} spin-orbit coupling of relativistic origin and
with active orbital degrees of freedom~\cite{Kugel1982}---e.g. to the
above mentioned case of transition metal oxides with partially filled
$3d$ orbitals.

On the other hand, to the best of our knowledge, such analysis has not
been done for the systems with {\it strong} on-site coupling between
spins and orbitals~\footnote{Note that in Ref.~\cite{Lun12} the impact
of relatively small spin-orbit coupling on the entanglement spectra was
discussed.}---as in the above-discussed case of the transition metal
oxides with partially filled $4d$ and $5d$ orbitals and strong spin-orbit
coupling.
\textcolor{black}{We stress that in this case the spin and orbital
degrees of freedom can get entangled as a result of both the nearest neighbor
exchange interactions as well as on-site spin-orbit coupling.}
In fact, one typically {\it implicitly} assumes that the
spin-orbital entanglement should be nonzero, since the spin ${\bf S}$
and orbital ${\bf L}$ operators couple at each site into a total angular
momentum ${\bf J} ={\bf S}+{\bf L} $ \cite{PaerschkeRay2018}.
The latter, `spin-orbital entangled' operators (also called pseudospins),
then interact as a result of the exchange processes in the
`relativistic' Mott insulators and are best described in terms of
various effective pseudospin models, such as for example the Kitaev-like
model discussed above~\cite{Jackeli2009}.
Finally, very few studies
discuss the problem of the evolution of a spin-orbital system between
the limit of weak and strong spin-orbit coupling~\cite{Ere11,Cho11,Svo17,Koga2018}.

Here we intend to bridge the gap between the understanding of the
spin-orbital physics in the above two limits. We ask the following
questions:
(i) what kind of evolution does the spin-orbital entanglement develop
with increasing spin-orbit coupling?
(ii) can one always assume that in the limit of strong spin-orbit
coupling the spin-orbital entanglement is indeed nonzero?
(iii) how does the spin-orbital entanglement arise in the limit of
the strong spin-orbit coupling?

To answer the above questions we formulate a minimal 1D model with
$S=\nicefrac{1}{2}$ spin and $T=\nicefrac{1}{2}$ orbital (pseudospin)
degrees of freedom. The model has the SU(2)$\otimes$SU(2) intersite
interactions between spins and orbitals which are supplemented by the
on-site spin-orbit coupling of the Ising character---its detailed
formulation as well as its relevance is discussed in Sec.~\ref{sec:model}.
Using exact diagonalization (ED), the method of choice described in
Sec. \ref{sec:methods}, we solve the 1D model and evaluate the various
correlation functions used to study the entanglement. Next, we present
the evolution of the ground state properties as a function of the model
parameters: in Sec.~\ref{sec:resultsA} for different values of the three
model parameters, and in Sec.~\ref{sec:resultsB} for a specific choice
of the relation between the two out of the three model parameters.
We then show two distinct paths of ground state evolution in Secs.
\ref{sec:caseA} and \ref{sec:caseB}. \textcolor{black}{The evolution of
the exact spectra of the periodic $L=4$ chain is analyzed in} Sec.
\ref{sec:resultsB} 3 \textcolor{black}{for increasing $\lambda$.} We discuss
obtained numerical results utilizing mapping of the model onto an
effective XXZ model in Sec.~\ref{sec:discussion}, which is valid in the
limit of the strong on-site spin-orbit coupling. We use the effective
model to understand:
(i) how the spin-orbital entanglement sets in the model system and
(ii) how it depends on the value of the on-site spin-orbit coupling
constant $\lambda$. The paper ends with the conclusions presented in
Sec.~\ref{sec:summa} and is supplemented by an Appendix which discusses
in detail the mapping onto the effective XXZ model in the limit of
large spin-orbit coupling $\lambda\to\infty$.

\subsection{Practical note on the organization of the paper}
\textcolor{black}{We note that, whereas the main results of the paper
are given in the extensive Sec.~\ref{sec:nume}, {\it some} of the main
results can be understood by using a mapping onto an effective
XXZ model in Sec.~\ref{sec:discussion}. Thus, we refer the interested
audience looking for the more physical and intuitive understanding of
(some of) the obtained numerical results to the latter section.
Finally, we stress that {\it all} the important results of the paper are
not only listed but also discussed in detail in Sec.~\ref{sec:summaa}.}

\section{Model}
\label{sec:model}

In this paper we study a spin-orbital model ${\cal H}$ defined in the
Hilbert space spanned by the eigenstates of the spin $S=\nicefrac{1}{2}$
and orbital (pseudospin) $T=\nicefrac{1}{2}$ operators at each lattice
site of a 1D chain with periodic boundary conditions.
The model Hamiltonian consists of two qualitatively distinct terms,
\begin{align}\label{eq:h}
{\cal H}={\cal H}_{\rm SE}+{\cal H}_{\rm SOC}.
\end{align}

The first term ${\cal H}_{\rm SE}$ describes the intersite
(super)exchange interactions between spins and orbitals.
The spin-orbital (`Kugel-Khomskii') exchange reads,
\begin{align}
{\cal H}_{\rm SE}&=
J\sum_i\left[\left(\textbf{S}_{i}\cdot\textbf{S}_{i+1}+\alpha\right)
       \left(\textbf{T}_{i}\cdot\textbf{T}_{i+1}+\beta\right)
       - \alpha\beta\right],
\label{ham_kk1}
\end{align}
where $J>0$ is the exchange parameter and the constants $\alpha$ and
$\beta$ are responsible for the relative strengths of the individual
spin and orbital exchange interactions. This 1D SU(2)$\otimes$SU(2)
symmetric spin-orbital Hamiltonian has been heavily studied in the 
literature---it is exactly solvable by Bethe Ansatz at the SU(4) point, 
i.e., when $\alpha=\beta=\nicefrac14$ \cite{Sutherland1975,Yamashita1998,Li1999},
has a doubly degenerate ground state at the so-called Kolezhuk-Mikeska
point $\alpha=\beta=\nicefrac34$ \cite{Kolezhuk1998,Kolezhuk2000},
and was studied using various analytical and numerical methods for
several other relevant values of $\{\alpha,\beta\}$ parameters
\cite{Arovas1995,Pati1998,Itoi2000,Zheng2001,Li2005,Chen2015}.
In particular, the entanglement between spin and orbital degrees of
freedom in such a class of Hamiltonians is extremely well-understood
\cite{Ole12,Chen2007,Lun12}. Last but not least, it was
suggested that this model may describe the low-energy physics found in
NaV$_{2}$O$_{5}$ and Na$_{2}$Ti$_{2}$Sb$_{2}$O~\cite{Pati1998},
CsCuCl$_3$ and BaCoO$_3$~\cite{Kugel2015}, as well as in the artificial
Mott insulators created in optical lattices~\cite{Gorshkov2010,Zhang2014}
and the so-called Coulomb impurity lattices~\cite{Dou2016}.

Altogether, this means that the spin-orbital exchange interaction has 
the simplest possible form \cite{Kugel1982} that can, nevertheless,
describe a realistic situation found in the transition metal oxides.
This, as already mentioned in
Introduction, constitutes the main reason behind the choice of this
form of spin-orbital intersite interaction. We note that the 
spin-orbital exchange would often has a more complex form. For instance, 
this would be the case, if e.g. three instead of two active orbitals 
were taken into account and the corrections from finite Hund's exchange 
were included (as relevant for the $5d$ iridates, whose spin-orbital 
exchange interactions are given by e.g.~Eq. (3.11) of Ref.~\cite{Kha05}).

The second term in the studied Hamiltonian (\ref{eq:h}) describes the on-site
interaction between the spin and orbital degrees of freedom and reads
\begin{align}
{\cal H}_{\rm SOC}&= 2\lambda\sum_{i}S^{z}_{i}T^{z}_{i}.
\label{ham_soc}
\end{align}
Here the parameter $\lambda$ measures the strength of the on-site
spin-orbit coupling term (of relativistic origin). The above Ising form
of the spin-orbital coupling was chosen as the simplest possible and yet
nontrivial one. Moreover, exactly such a form of the spin-orbit coupling
is {\it typically} realized in systems with {\it two} active orbitals.
This is the case of e.g. the active $t_{2g}$ doublets in
YVO$_3$~\cite{Horsch2003} and Sr$_2$VO$_4$~\cite{Jackeli2009b},
the molecular $\pi$ orbitals of KO$_2$~\cite{Solovyev2008}, or on
optical lattices. In fact, such a highly anisotropic form of spin-orbit
coupling is valid for any system with an active orbital doublet, either 
two $p$ ($p_x$ and $p_y$) or two $t_{2g}$ ($xz$ and $yz$) orbitals.

\section{Methods and Correlation functions}
\label{sec:methods}

As we are interested in quantum entanglement, and moreover, the exchange
Hamiltonian (\ref{ham_kk1}) itself bears a rather complex quantum
many-body term
\mbox{$\propto(\textbf{S}_i\textbf{S}_j)(\textbf{T}_i\textbf{T}_j)$},
we opt for the exact diagonalization (ED) method which preserves the
quantum fluctuations in the numerically found ground state.
More specifically, we choose the ED calculations, since:
(i) It allows us to investigate the system ground state in a numerically
exact manner and in a completely unbiased way which for the first study
of its kind is usually selected as the method of choice;
(ii) The analytically exact Bethe Ansatz approach can only be applied
to a few selected values of the model parameters;
(iii) The ED calculations can be relatively easily repeated for a number
of model parameters and can typically address the qualitative properties
of the ground state rather well;
(iv) We are interested here in rather local correlations which follow 
from the local spin-orbit and nearest neighbor exchange interactions. 

We calculate the properties of the ground state of model (\ref{eq:h})
on finite chains with periodic boundary conditions. We utilize chains 
of length $L=4n$ sites, where $n$ is an integer number, in order to 
avoid a degenerate ground state appearing in the case of a 
$(4n+2)$--site chain (see Table 1 of Ref.~\cite{Yamashita1998}). 
For chains $L=4$ a standard full ED procedure is performed, while for
$L=8$, 12, 16, and 20 sites we restrict
the ED calculations to the Lanczos method \cite{Koch}.
\begin{figure*}[t!]
\includegraphics[width=18cm]{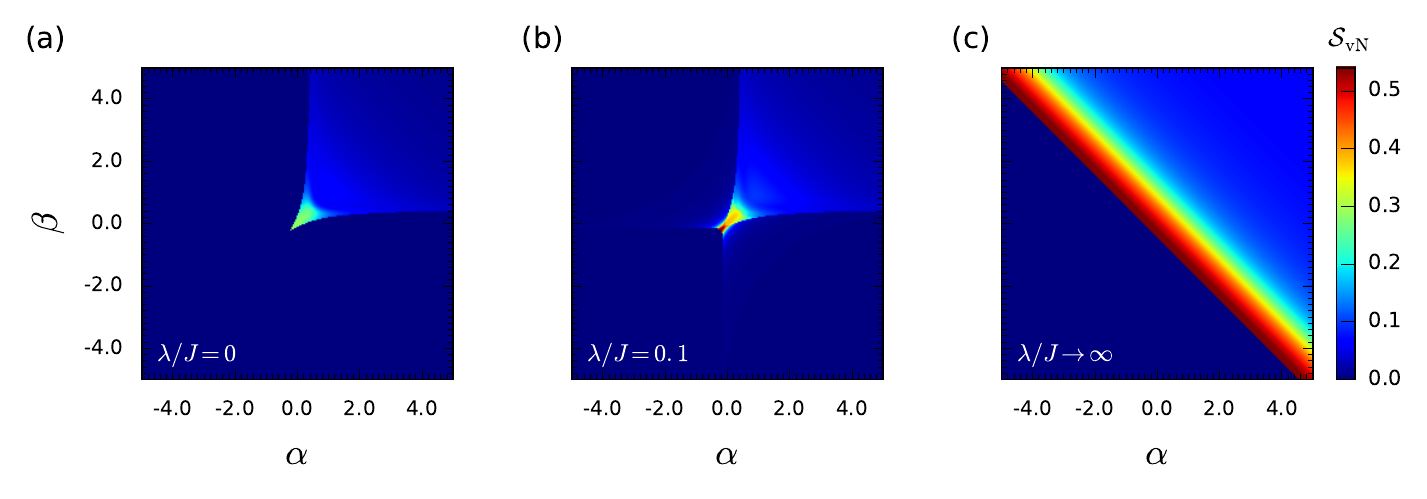}
\caption{The von Neumann spin-orbital entanglement entropy,
${\cal S}_{\rm vN}$ (\ref{vNS}), calculated using ED on
$L=12$--site
periodic chain for the spin-orbital model Eq.~(\ref{eq:h}) and for
the increasing value of the spin-orbit coupling $\lambda$:
(a) $\lambda/J=0$,
(b) $\lambda/J=0.1$, and
(c)~$\lambda/J\rightarrow\infty$.
}
\label{fig:en}
\end{figure*}

To capture the changes in the ground state of the spin-orbital model at
increasing spin-orbit coupling $\lambda$, we define and investigate the
following correlation functions which will be used, besides the von
Neumann spin-orbital entanglement entropy \cite{vNS}, to monitor the
evolution of the ground state with changing model parameters:

(i) The intersite spin-orbital correlation function ${\cal C}_{\rm SO}$:
\begin{align}
{\cal C}_{\rm SO}&=\frac{1}{L}\sum_{i=1}^{L}
\left\langle(\textbf{S}_{i}\cdot\textbf{S}_{i+1})
(\textbf{T}_{i}\cdot\textbf{T}_{i+1})\right\rangle \nonumber \\
&-\frac{1}{L}\sum_{i=1}^{L} \langle\textbf{S}_i\cdot\textbf{S}_{i+1}\rangle\,
  \langle \textbf{T}_{i} \cdot\textbf{T}_{i+1}\rangle ,
\label{eq:Ch}
\end{align}
which measures the intersite (nearest neighbor) correlation between the
spin and orbital degrees of freedom and has already been used in the
literature as a good {\it qualitative} estimate for the von Neumann
spin-orbital entanglement entropy~\cite{You2015,You2015b}. This
correlator can also be used to monitor the failure of the mean field
decoupling between the spins and orbital pseudospins once
${\cal C}_{\rm SO}\neq 0$~\cite{Ole06}.

(ii) The intersite spin $S$ or orbital $T$ correlation
function,
\begin{align}
\label{eq:S}
S=&\frac{1}{L}\sum_{i=1}^{L}
\left\langle{\bf S}_{i}\cdot {\bf S}_{i+1}\right\rangle,\\
\label{eq:T}
T=&\frac{1}{L}\sum_{i=1}^{L}
\left\langle{\bf T}_{i}\cdot {\bf T}_{i+1}\right\rangle,
\end{align}
which measures the intersite (nearest neighbor) correlation between the
spin (orbital) degrees of freedom and is therefore sensitive to the
changes in the ground state properties taking place solely in the spin
(orbital) subspace. We emphasize that these two functions are defined
on equal footing in the model with SU(2)$\otimes$SU(2) spin-orbital
superexchange.

(iii) The $\gamma$--component
$S^{\gamma\gamma}$ of spin scalar product:
\begin{align}
\label{Szz}
S^{\gamma\gamma}&=\frac{1}{L}\sum_{i=1}^{L}
\left\langle{S}^{\gamma}_{i}{S}^{\gamma}_{i+1}\right\rangle,
\end{align}
where $\gamma=x,y,z$. This function measures the component $\gamma$ of
the scalar product and thus allows one to investigate possible
anisotropy of the intersite (nearest neighbor) correlations between the
spin degrees of freedom. The orbital scalar product component
$T^{\gamma\gamma}$ is defined analogously to Eq. (\ref{Szz}).

(iv) Crucial for the systems with finite spin-orbit coupling is the
on-site spin-orbit correlation function ${\cal O}_{\rm SO}$:
\begin{align}
{\cal O}_{\rm SO}&=\frac{1}{L}\sum_{i=1}^{L}
\left\langle S_i^zT_i^z\right\rangle,
\label{eq:O}
\end{align}
which measures the correlations between the $z$ components of the spin
and orbital operators on the same site. The precise form of this
correlator is dictated by the Ising form (\ref{ham_soc}) of the
spin-orbit coupling present in Hamiltonian (\ref{eq:h}).
Conveniently, the function (\ref{eq:O}) is one of the generators of the
SU(4) group \cite{Li1999}, which proved to be quite useful for
examining the range of the SU(4)--symmetric ground state.

\section{Numerical results}
\label{sec:nume}
\subsection{von Neumann entropy in a general case}
\label{sec:resultsA}

The main goal of this paper is to determine how the spin-orbital
entanglement changes in the spin-orbital model~(\ref{eq:h}) upon
increasing the value of the spin-orbit coupling $\lambda$. To this end,
we first define the entanglement entropy calculated for a system that
is bipartitioned into two subsystems: $A$ and $B$. Typically such a
subdivision refers to two distinct parts of the real
\cite{Vidal2003,Korepin2004,Wolf2006,Gioev2006} or momentum
\cite{Thomale2010} space. Here, however, it concerns spin ($A$) and
orbital ($B$) degrees of freedom~\cite{Ole06,Ole12,Chen2007,You12}.

A standard measure of the entanglement entropy between subsystems $A$
and $B$ in the ground state $|\textrm{GS}\rangle $ of a system of size
$L$ is due to von Neumann \cite{vNS}. It is defined as
${\cal S}_{\rm vN}=-\textrm{Tr}_{A}\{\rho_{A}\ln\rho_{A}\}/L$,
and is obtained by integrating the density matrix,
$\rho_{A}=\textrm{Tr}_{B}|\textrm{GS}\rangle\langle GS|$ over subsystem
$B$. Consequently, in this paper we use the following definition of the
von Neumann spin-orbital entanglement entropy:
\begin{equation}
\label{vNS}
{\cal S}_{\rm vN}=-\frac{1}{L}\textrm{Tr}_S\{\rho_{S}\ln\rho_{S}\},
\end{equation}
where
\begin{equation}
\rho_{S}=\textrm{Tr}_{T}|\textrm{GS}\rangle\langle\textrm{GS}|
\end{equation}
is the reduced spin-only ($S$) density matrix with the orbital ($T$)
degrees of freedom integrated out.

The spin-orbital von Neumann entropy is calculated using ED on $L$--site
chain for model~(\ref{eq:h}) and is shown as function of the parameters
$\{\alpha,\beta\}$ for three representative values of the spin-orbit
coupling $\lambda$ in Fig.~\ref{fig:en}.
In perfect agreement with Refs. \cite{Chen2007,Lun12}, the von Neumann
entropy ${\cal S}_{\rm vN}$ is finite in a rather limited region of the
$\{\alpha,\beta\}$ parameters for $\lambda=0$, i.e., in the entangled
spin-orbital phase near the origin $\alpha=\beta=0$. The nonzero
entanglement in that case is well-understood and attributed to the
onset of the dominant antiferromagnetic (AF) and alternating orbital
(AO) fluctuations in the ground state without broken
symmetry~\cite{Chen2007,Lun12}, see discussion below in
Sec.~\ref{sec:summary}.

Interestingly, a finite but `small' spin-orbit coupling
$\lambda<\lambda_{\rm{CRIT}}$ ($\lambda_{\rm{CRIT}}$
is discussed in more detail in Sec.~\ref{sec:resultsB})
does not substantially increase the region in the
$\{\alpha,\beta\}$--parameter space for which the spin-orbital entropy 
is nonzero cf. Figs. \ref{fig:en}(a) and \ref{fig:en}(b).
A drastic change in the behavior of the spin-orbital von Neumann 
entropy only happens for the dominant spin-orbit coupling
$\lambda > \lambda_{\rm{CRIT}}$.
In this case the
region of nonzero spin-orbital entanglement is not only much larger but
also takes place for different values of the
$\{\alpha,\beta\}$--parameter space.
For instance, it is remarkable that the von Neumann entropy
in the case of
$\lambda > \lambda_{\rm{CRIT}}$
almost does not depend on $\alpha$ along the lines of constant
$\alpha+\beta$.
Moreover, finite entanglement is activated when
$\alpha+\beta>-1/2$---however,
the value of the von Neumann entropy
strongly decreases for $\alpha$ and $\beta$
located `above' the stripe given by the inequalities
$-1/2  \le \alpha + \beta \le 2$
and showing the highest value of entropy. In fact, it will be shown 
later (see Sec.~\ref{sec:discussion}) that the von Neumann entropy is 
expected to vanish in the limit of $\alpha+\beta\rightarrow\infty$.

Altogether, we observe that:
(i) in the limit of small
$\lambda < \lambda_{\rm{CRIT}}$
the spin-orbital
entanglement entropy does not change substantially w.r.t. the case with
vanishing spin-orbit coupling;
(ii) in the limit of large
$\lambda > \lambda_{\rm{CRIT}}$ the spin-orbital
entanglement can become finite even if it vanishes for $\lambda=0$;
though it can also happen that
(iii)~in~the limit of large
$\lambda > \lambda_{\rm{CRIT}}$ the spin-orbital
entanglement vanishes when $\alpha+\beta<-1/2$.

\begin{figure*}[t!]
\includegraphics[width=18cm]{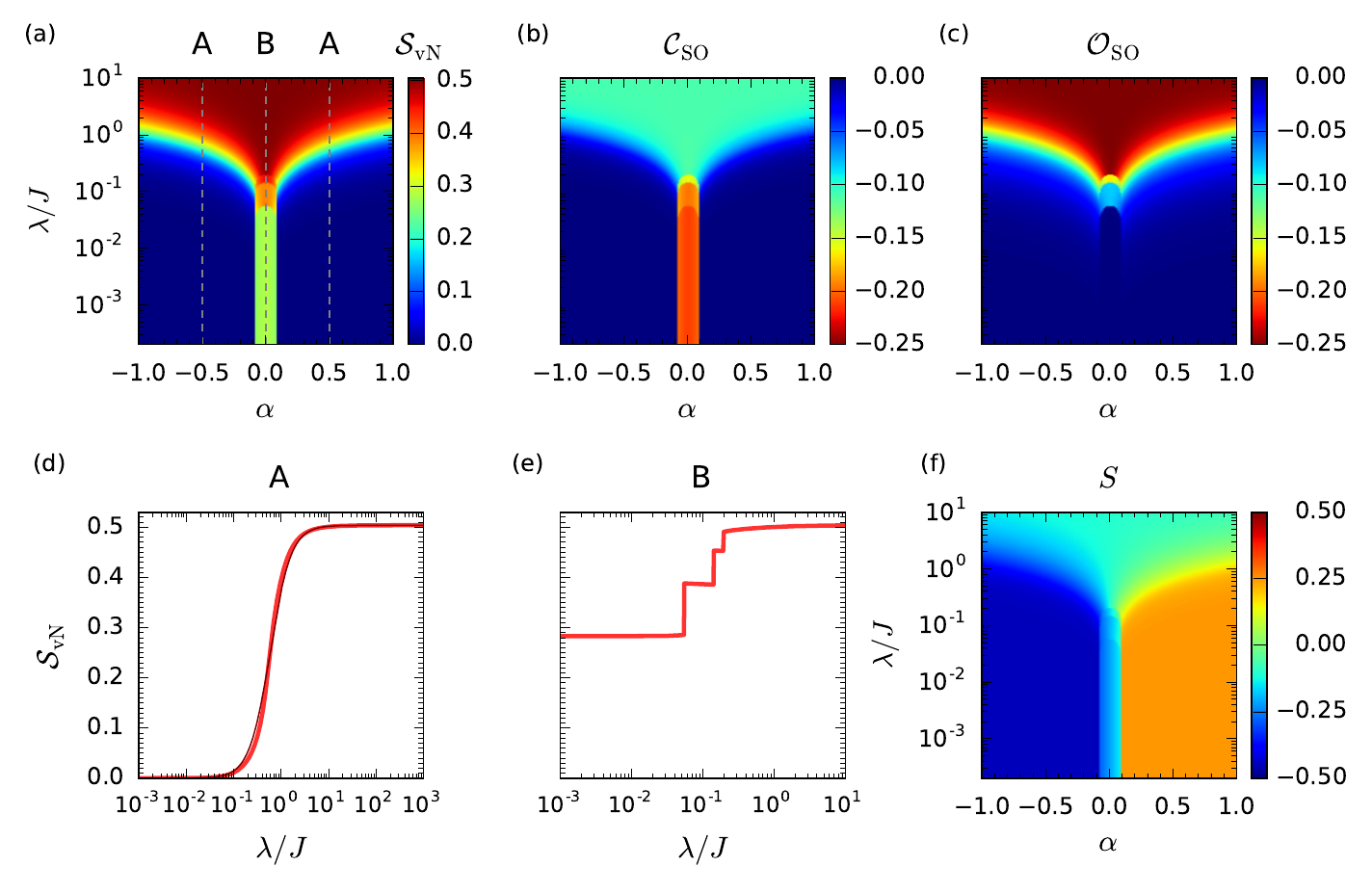}
\caption{
Evolution of the von Neumann spin-orbital entanglement entropy and the
three spin-orbital correlation functions in the ground state of 
Hamiltonian (\ref{eq:h}) with $\alpha=-\beta$, calculated using ED with 
the periodic $L=12$--site chain for logarithmically 
increasing spin-orbit coupling $\lambda$:
(a) the von Neumann spin-orbital entanglement entropy
    ${\cal S}_{\rm vN}$ (\ref{vNS}) for $\alpha\in[-1.0,1.0]$;
(b)~the intersite spin-orbital correlation function
    ${\cal C}_{\rm SO}$ (\ref{eq:Ch}) for $\alpha\in[-1.0,1.0]$;
(c) the on-site spin-orbit correlation function ${\cal O}_{\rm SO}$
    (\ref{eq:O}) for $\alpha\in[-1.0,1.0]$;
(d) the von Neumann spin-orbital entanglement entropy
    ${\cal S}_{\rm vN}$ (\ref{vNS}) obtained with $|\alpha|=0.5$ [cut A
    in panel (a)] and fitted with a logistic function (black thin line);
(e) the von Neumann spin-orbital entanglement entropy
    ${\cal S}_{\rm vN}$ (\ref{vNS}) obtained with $\alpha=0$ [cut B in
    panel (a)];
(f) the spin correlation function $S$ (\ref{eq:S}) for $\alpha\in[-1.0,1.0]$.
}
\label{fig:ent}
\end{figure*}

\subsection{von Neumann entropy for $\beta=-\alpha$:\\
Two distinct evolutions for increasing $\lambda$}
\label{sec:resultsB}

In order to better understand the physics behind the observations (i)
and (ii) discussed in the end of the previous subsection, here we study
in great detail the onset of the spin-orbital entanglement once
$\beta=-\alpha$. As shown in Fig.~\ref{fig:en}, for these values of the
model parameters the region of the nonzero spin-orbital entanglement
increases dramatically with the increasing value of the spin-orbit
coupling $\lambda$.

\begin{figure}[t!]
\includegraphics[width=9cm]{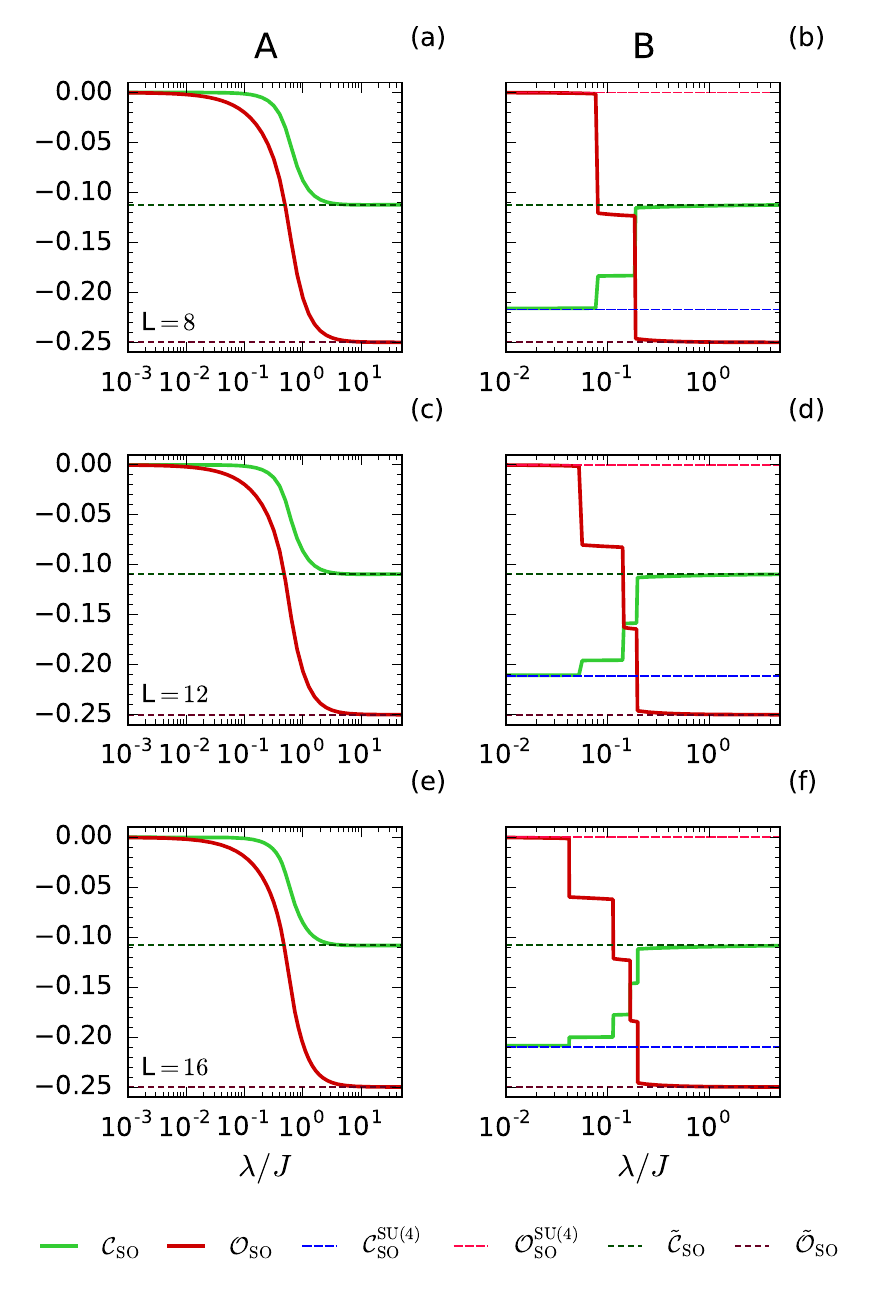}
\caption{
The intersite spin-orbital correlation function ${\cal C}_{\rm SO}$
(green lines) and the on-site spin-orbit correlation function
${\cal O}_{\rm SO}$ (red lines) as functions of $\lambda$ calculated
for: case A (left), i.e., $\alpha=0.5$ [(a), (c), (e)] and
case B (right), i.e., $\alpha=0$ [(b), (d), (f)].
The ED results are shown for periodic chains of length
$L=8$ (top), $L=12$ (middle), and $L=16$ (bottom).
The dashed lines represent the
asymptotic values of the above correlation functions:
(i) the exact SU(4)--point limit $\lambda=0$, $\alpha=\beta=0.25$
is denoted by ${\cal C}^{\rm SU(4)}_{\rm SO}$ and
${\cal O}^{\rm SU(4)}_{\rm SO}$ (blue and light--red dashed lines);
(ii) the $\lambda=\infty$, $\beta=-\alpha$ XY limit---by
${\cal\tilde{C}}_{\rm SO}$ and ${\cal\tilde{O}}_{\rm SO}$ (dark--green
and dark--red dashed lines). For further details see discussion in
Secs.~\ref{sec:caseA}-\ref{sec:caseB} and Sec.~\ref{sec:benchmarking}).
}
\label{fig:finite_C_O}
\end{figure}

We present in Fig.~\ref{fig:ent} the von Neumann spin-orbital
entanglement entropy ${\cal S}_{\rm vN}$
(\ref{vNS}) and the three spin-orbital correlation functions in the
ground state of Hamiltonian~(\ref{eq:h}) with \mbox{$\alpha=-\beta$},
calculated using ED on an
$L=12$--site chain.
We begin the analysis by comparing the values of the three
spin-orbital correlation functions (\ref{eq:Ch}), (\ref{eq:O}), and
(\ref{eq:S}) against the von Neumann spin-orbital entanglement entropy,
see Figs.~\ref{fig:ent}(b), \ref{fig:ent}(c), and \ref{fig:ent}(f). We
observe that only the intersite spin-orbital correlation function
$C_{\rm SO}$ can be used as a qualitative  measure for the von Neumann
entropy, consistent with previous studies~\cite{You2015,You2015b}.
In particular, the on-site spin-orbit correlation function
${\cal O}_{\rm SO}$ cannot be used to `monitor' the entanglement
entropy, for it measures the correlations between spins and orbitals
locally and on the Ising level only. Nevertheless, both
${\cal O}_{\rm SO}$ as well as ${\cal S}_{\rm vN}$ can be used to
identify various quantum phases obtained in the $\{\alpha,\lambda\}$
parameter space of the Hamiltonian, as suggested before for system
with negligible spin-orbit coupling \cite{You2015b}.

Next, we study the
evolution of the von Neumann spin-orbital entanglement entropy with
increasing spin-orbit coupling $\lambda$ for various values of the
parameter $\alpha$, see Fig.~\ref{fig:ent}(a). We observe that in the
representative \mbox{$\alpha\in[-1,1]$} interval there exist three
distinct regimes of the value of the von Neumann entropy:
(i) two compact areas in the $\{\alpha,\lambda\}$ parameter space for
which the von Neumann entropy is vanishingly small, which exist
in the large parameter range of $|\alpha|\gtrsim 0.1$ and
$\lambda/J\lesssim 10^{-1}-10^{0}$  [the bottom left and bottom right
parts of panel Fig.~\ref{fig:ent}(a)];
(ii)~one compact area in the $\{\alpha,\lambda\}$ parameter space for
which the von Neumann entropy takes maximal possible values,
which exists in the large parameter range
$\lambda/J\gtrsim 10^{-1}-10^{1}$ for all values of $\alpha$
[the top part of panel Fig.~\ref{fig:ent}(a)];
(iii)~the compact area
in the $\{\alpha,\lambda\}$ parameter space
for which the von Neumann entropy is neither negligible nor takes
maximal value, which exists in the relatively
small parameter range between
cases (i) and (ii). In order to understand the onset of these three
distinct regimes, we study below two qualitatively different cases
of the von Neumann entropy evolution with the increasing spin-orbit
coupling: case `A' with $|\alpha|=0.5$ and case `B' with $\alpha=0$
[shown with dashed lines in Fig.~\ref{fig:ent}(a)].

\begin{figure}[t!]
\includegraphics[width=9cm]{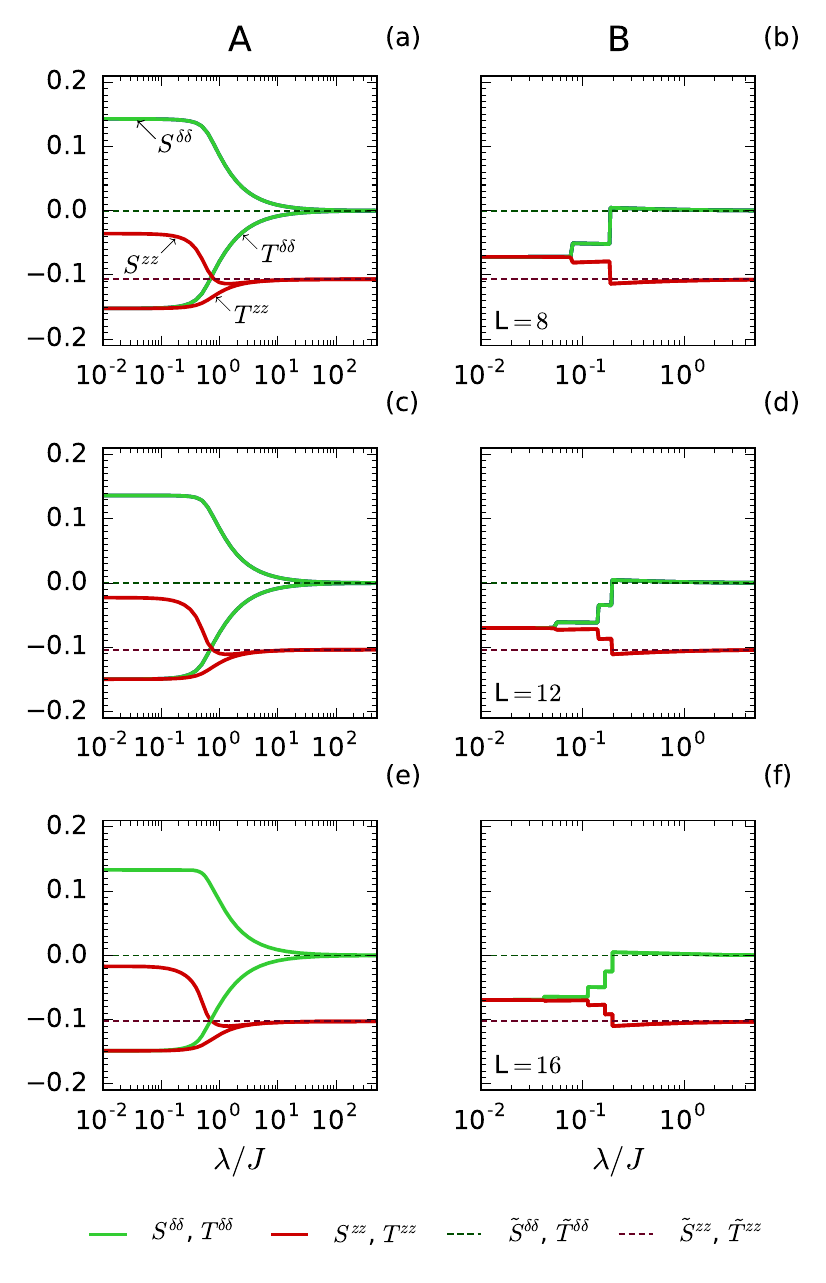}
\caption{
The anisotropic spin correlation function $S^{\gamma\gamma}$
(\ref{Szz}) and orbital correlation function
$T^{\gamma\gamma}$  for increasing $\lambda$.
The $S^{\delta\delta}$ and  $T^{\delta\delta}$ ($\delta=x, y$)
components are marked by green color while the $S^{zz}$ and $T^{zz}$
components are marked by red color. The correlation functions are
calculated for:
case A (left),  i.e., $\alpha=0.5$ [(a), (c), (e)] and
case B (right), i.e., $\alpha=0$   [(b), (d), (f)].
The ED results are shown for periodic chains of length
$L=8$ (top), $L=12$ (middle), and $L=16$ (bottom).
The asymptotic values of the correlation functions in
the limit $\lambda=\infty$ are shown for both $\alpha=0$ and $\alpha=0.5$
case and denoted as $\tilde{S}^{\delta\delta}$,$\tilde{S}^{zz}$  and
$\tilde{T}^{\delta\delta}$, $\tilde{T}^{zz}$,
see discussion in Sec.~\ref{sec:benchmarking} for further details.
}
\label{fig:anisotropic_S}
\end{figure}

\subsubsection{From a product state to highly entangled state}
\label{sec:caseA}

When $|\alpha|=0.5$ (case A) the evolution of the von Neumann
spin-orbital entanglement entropy ${\cal S}_{\rm vN}$ (\ref{vNS}) with
increasing spin-orbit coupling can be well-approximated by a logistic
function, see Fig.~\ref{fig:ent}(d).
The von Neumann entropy has infinitesimally small values for
$\lambda/J\lesssim 10^{-1}$, experiences a rapid growth for
$\lambda/J\in(10^{-1},10^1)$, and saturates at ca. $0.5$ for
$\lambda/J\gtrsim 10^1$.
Comparable behavior is observed for the intersite spin-orbital 
correlation (${\cal C}_{\rm SO}$), which, as already discussed, 
is a good and computationally not expensive qualitative measure 
for the von Neumann entropy, see Figs.~\ref{fig:finite_C_O}(a), 
\ref{fig:finite_C_O}(c), and \ref{fig:finite_C_O}(e).
Crucially, the latter calculations are obtained for the spin-orbital
chains of different length and [as well-visible in Figs.
\ref{fig:finite_C_O}(a), \ref{fig:finite_C_O}(c), \ref{fig:finite_C_O}(e)]
show relatively small finite-size effects.
This means that indeed the von
Neumann entropy ${\cal S}_{\rm vN}$ depends
here mainly on short-range processes and can remain
negligibly small for a finite value of the spin-orbit coupling even
in the thermodynamic limit. Finally, Figs.~\ref{fig:ent} and
\ref{fig:finite_C_O} allow us to define the critical value
$\lambda_{\rm{CRIT}}$ for case A as being located in an interval of
rapid growth of the spin-orbital entanglement: 
$\lambda_{\rm{CRIT}}/J\in (10^{-1}, 10^1)$.

While the nature of the quantum phase for large spin-orbit coupling
$\lambda > \lambda_{\rm{CRIT}}$ is discussed in detail in
Sec.~\ref{sec:benchmarking}, here we merely mention that in this case
the value of the spin-orbital entanglement entropy saturates at
about $0.5$ ($0.504$ for $L=12$ site chain)
per site.
Hence, we call this quantum phase a
{\it highly entangled state}. Besides, in this case also the absolute
value of the on-site spin-orbit correlation function ${\cal O}_{\rm SO}$
takes its maximal value, while the spin (and orbital) correlation
function $S$ ($T$) is weakly AF (AO).

Next, we focus on the properties of the ground state obtained for small
$\lambda<\lambda_{\rm{CRIT}}$. To this end we
investigate the evolution of the two other correlation functions, the
on-site spin-orbit correlation function ${\cal O}_{\rm SO}$ and the
spin correlation function $S$, for $\lambda < \lambda_{\rm{CRIT}}$ and
$|\alpha|=0.5$, see Figs.~\ref{fig:ent}(c), \ref{fig:ent}(f). We observe 
that whereas the on-site spin-orbit correlation function shows 
vanishingly small values in this limit, the spin correlation function
$S\simeq 0.25$ \mbox{($S\simeq -0.45$)} for $\alpha=0.5$ ($\alpha=-0.5$),
thus behaving similarly to the 1D FM (AF) chain, respectively.
We note that the (unshown) analogous nearest neighbor orbital
correlation function $T$ calculated for $\alpha = \pm 0.5$ takes
complementary values to the spin correlation function for
$\alpha=\mp 0.5$, i.e., $T=-S$.
Such behavior is again observed for chains of various lengths, with
$S(T)$ better approximating the expected AF Bethe Ansatz value for
larger chains, see Fig.~\ref{fig:anisotropic_S}. Altogether, this
shows that the quantum phase that is observed for
$\lambda < \lambda_{\rm{CRIT}}$ qualitatively resembles the phases
obtained in the limit of $\lambda =0$: the FM$\otimes$AO
(AF$\otimes$FO) for $\alpha=0.5$ ($\alpha = -0.5$), respectively.

The above discussion contains just one caveat. Let us look at the
evolution of the anisotropic spin (and orbital) correlation function
$S^{\gamma\gamma}$ (and $T^{\gamma\gamma}$) with the increasing spin-orbit
coupling $\lambda$, see Figs. \ref{fig:anisotropic_S}(a, c, e).
We notice that whenever $\lambda/J>0$ for $\alpha=0.5$ there exist
an anisotropy between  the $zz$ (solid red lines) and the planar
($xx$, $yy$, solid green lines) correlation functions---which
is absent for $\lambda=0$. However, for
$\lambda/J \lesssim 3\cdot 10^{-1}$ the anisotropy is only partial,
being absent in the strongly AF $T^{\gamma\gamma}$ correlations, in
contrast to the $S^{\gamma\gamma}$ correlations. In fact,
$S^{\delta\delta}$ (where $\delta=x, y$), stay positive as in
$\lambda=0$ case while $S^{zz}$ becomes negative. In this way the
energy coming from the finite spin-orbit coupling is `minimized' in
the ground state without {\it qualitatively} changing the nature of
the FM$\otimes$AO and AF$\otimes$FO ground states, allowing however
for a very small value of the spin--orbital entanglement.
This is the reason why, in what follows, this quantum phase is called
a {\it perturbed} FM$\otimes$AO  product state.

\subsubsection{From SU(4) singlet to a highly entangled state}
\label{sec:caseB}

We now investigate how the von Neumann spin-orbital entanglement entropy
${\cal S}_{\rm vN}$ evolves with the spin-orbit coupling once $\alpha=0$
(case B): i.e., from its finite value for the SU(4)--singlet ground state
at $\lambda=0$~\cite{Chen2007,Lun12} to an even higher value obtained in 
the limit of large $\lambda/J$ in the highly entangled state (i.e., 
the state already encountered in case A). To this end, we first note that 
the von Neumann entropy ${\cal S}_{\rm vN}$ at $\alpha=0$ changes
with the spin-orbit coupling in a qualitatively different
manner than in the case of $|\alpha| = 0.5$,
see Fig.~\ref{fig:ent}(e).
While we again encounter a monotonically growing function
in $\lambda$, which saturates at about $0.5$ for
$\lambda/J \gtrsim 0.2$, this function seems to be discontinuous at
three particular values of $\lambda$ and three `kinks'
(for $L=12$ sites) that can be easily identified in Fig.~\ref{fig:ent}(e). A similar 
behavior is encountered in the qualitative measure for the von Neumann 
entropy---the spin-orbital correlation function ${\cal C}_{\rm SO}$, 
see Fig.~\ref{fig:ent}(b) and Fig.~\ref{fig:finite_C_O}(b, d, f).
\begin{figure}[t!]
\includegraphics[width=9cm]{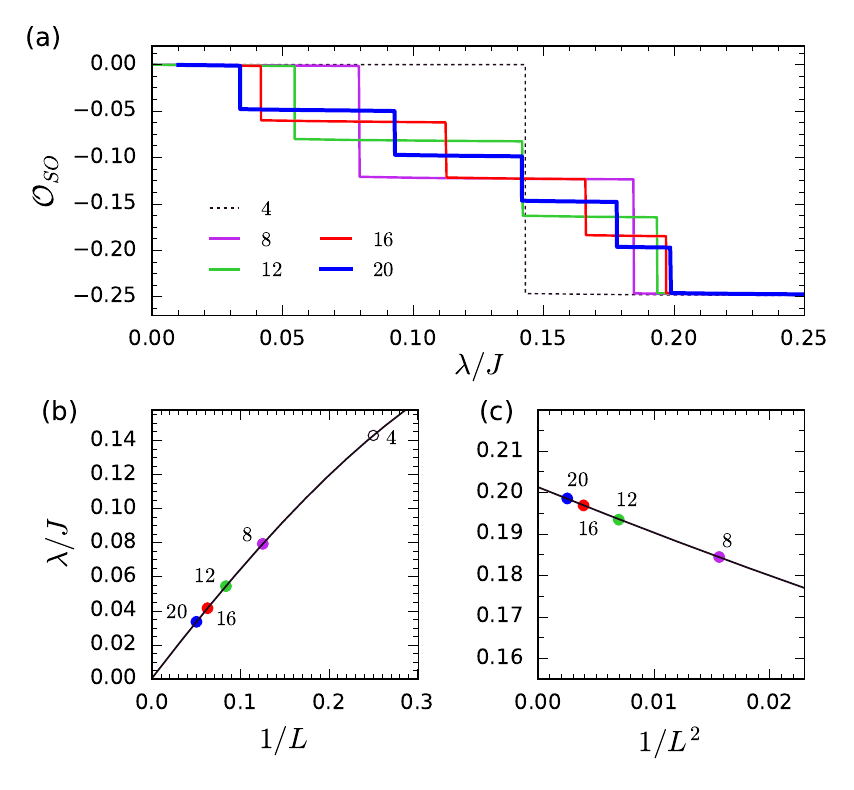}
\caption{
Finite-size scaling of the boundaries of the intermediate entangled 
state for the case B in Fig. 3:
(a) $L/4$ kinks for $L=4,8,12,16,20$ shown on an example of
${\cal O}_{\rm SO}$, 
(b) the decreasing position of the `first' kink as a function of $1/L$,
(c)~the increasing position of the `final' kink as a function of $1/L^2$.
These fits use the ED numerical results obtained for $L=4,8,12,16,20$ 
periodic chains presented as colorful dots in (b)\&(c); the lines are 
the fits (\ref{eq:fit}) to the numerical data.}
\label{fig:1_over_L}
\end{figure}

As a side note let us mention that once $\lambda=0$ and $\alpha=\beta=0$
the model~(\ref{eq:h}) has an SU(4)-symmetric ground state, as confirmed
by the remarkable convergence of the functions ${\cal C}_{\rm SO}$ and
${\cal O}_{\rm SO}$ to their asymptotic values
${\cal C}^{\rm SU(4)}_{\rm SO}$ and ${\cal O}^{\rm SU(4)}_{\rm SO}$
calculated at the exact SU(4) point \mbox{$\alpha=\beta=1/4$,} cf.
Fig.~\ref{fig:finite_C_O}(b, d, f). As the operator in the
${\cal O}_{\rm SO}$ function is one of the generators of the SU(4) group,
its zero expectation value in the ground state is not only related to
the absence of the spin-orbit coupling but also is a signature of the
SU(4)-symmetric {\it singlet} \cite{Li98}.

It is clearly visible in Fig.~\ref{fig:finite_C_O}(b, d, f) that the
${\cal C}_{\rm SO}$ and ${\cal O}_{\rm SO}$ correlations split from
their SU(4)-singlet asymptotes in the subsequent kinks, which occur with
the increasing value of the spin-orbit coupling. Interestingly, the
number of kinks grows and their position changes with the system size,
see Fig.~\ref{fig:finite_C_O}(b, d, f). In fact, $L/4$ kinks are
observed for a chain of length
$L=4,8,12,16,20$, see panel (a) of Fig. \ref{fig:1_over_L}.
This naturally suggests that in the infinite system the number of kinks 
will be infinite.

But what about the position of the `first' and the `last' kink in the
thermodynamic limit? To answer this intriguing question
with the available ED data we deduced qualitative values of $\lambda/J$
which define the regime where
correlations take intermediate values
and the entangled state is not yet dominated by the large spin-orbit
coupling $\lambda>\lambda_{\rm CRIT}$. 

The finite size scaling performed here,  
shown in panels (b) and (c) of Fig. \ref{fig:1_over_L}, 
uses a polynomial fit similar as for instance for the gap in the 1D 
half-filled Hubbard model \cite{Ole85}. Here the positions of the 
`first' and the `last' kink scale differently with the increasing length 
of $L=4n$ chain. Namely, the position of the `first' kink $k_1$ is 
almost linear in $1/L$ while the `final' kink's position $k_f$ scales 
almost linearly with $1/L^2$. By performing the fits, we have found that
\begin{eqnarray}
\label{eq:fit}
k_1 &=& 0.00004 + 0.69712\,x\,\, - 0.50147\,x^2\,, \nonumber \\
k_f &=& 0.20132 - 1.13007\,x^2 + 3.14415\,x^4\,,
\end{eqnarray}
where $x\equiv 1/L$. As a result, the position of the `first' kink 
$k_1$ converges to $\lambda/J=0$ when $L\to\infty$, and it is indeed 
reasonable to expect that infinitesimal $\lambda$ modifies weakly 
spin-orbital correlations in the thermodynamic limit. 
In contrast, the `final' kink $k_f$ would then shift to
$\lambda/J\simeq 0.201$.
Therefore, for the case B we define $\lambda_{\rm{CRIT}}$ as a single 
number: $\lambda_{\rm{CRIT}}/J\simeq 0.2$. Altogether, this means that 
the quantum phase encountered for $0<\lambda<\lambda_{\rm{CRIT}}$ does 
not disappear in the thermodynamic limit and that its spin-orbital
entanglement grows with the increasing spin-orbit coupling in a
continuous way. To contrast this intermediate phase with the one showing 
the maximal value of entanglement at $\lambda>\lambda_{\rm CRIT}$,
we call it an {\it intermediate entangled state}.

\begin{figure}[t!]
\includegraphics[width=9cm]{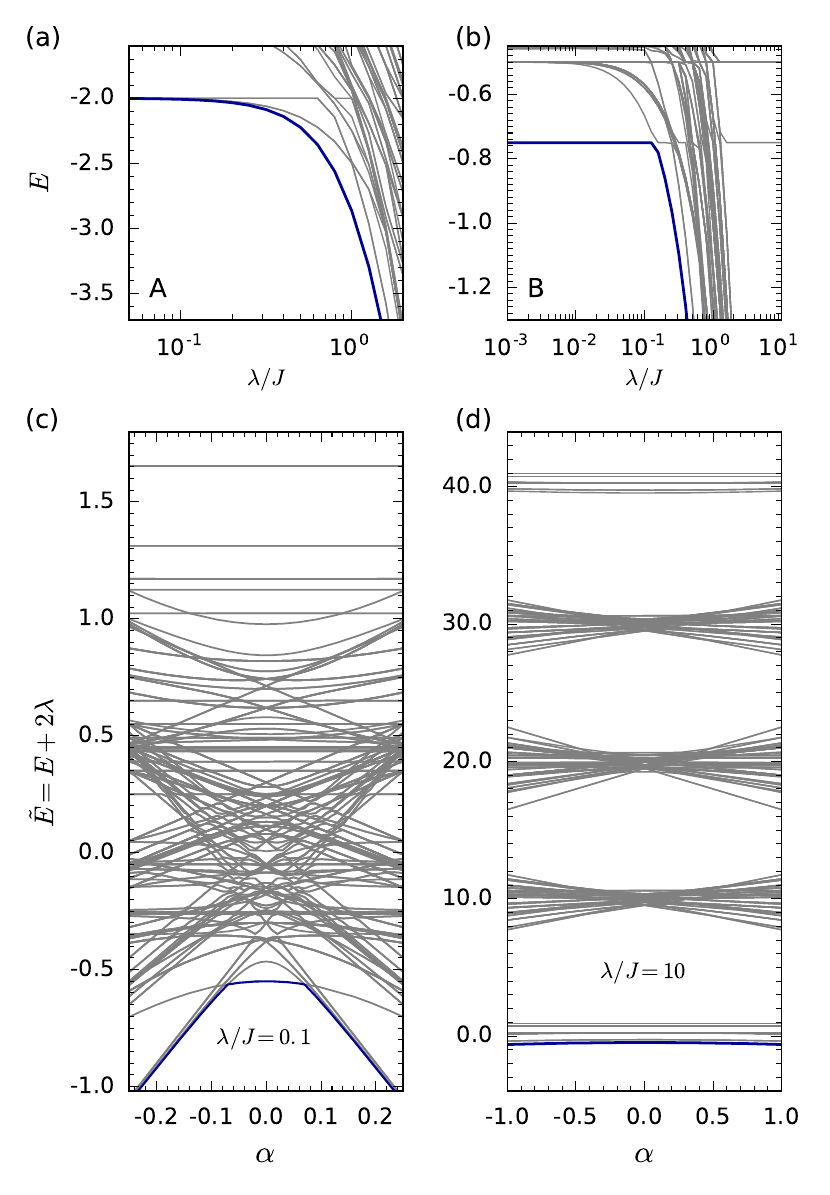}
\caption{
Top panels---the energy $E$ of the ground (blue) and low lying
excited states (gray) obtained for model (\ref{eq:h}) using ED for
periodic $L=4$-site chain as a function of increasing $\lambda/J$
[(a) case A for $\alpha=0.5$, (b) case B for $\alpha=0$]. \\
Bottom panels---complete energy spectra for small and large $\lambda$
[(c) $\lambda/J=0.1$ and (d) $\lambda/J=10$]. Note that we display here
$\tilde{E}$ (\ref{et}) to compare the spectra in similar energy range,
independently of the actual value of $\lambda$.
}
\label{fig:energy}
\end{figure}

To better understand the properties of this phase, we also consider the
spin correlation function $S$, the anisotropic spin $S^{\gamma\gamma}$,
and the orbital $T^{\gamma\gamma}$ correlation functions, see
Figs.~\ref{fig:ent}(c, f), \ref{fig:finite_C_O}(b, d, f), and
\ref{fig:anisotropic_S}(b, d, f). Similarly to the von Neumann entropy,
also ${\cal O}_{\rm SO}$ or ${\cal C}_{\rm SO}$ correlation
functions show kinks due to finite-size effects which are expected to
disappear in the thermodynamic limit. Noticeably, the behavior
of $S$, $S^{\gamma \gamma}$, and $T^{\gamma \gamma}$ is quite distinct
w.r.t. the one observed both for the highly entangled phase
and seemingly for the perturbed FM$\otimes$AO or AF$\otimes$FO phases.
This shows that the intermediate entangled phase observed at $\alpha=0$
and for $0<\lambda<\lambda_{\rm{CRIT}}$ is indeed qualitatively
different and constitutes a `genuine' quantum phase.

\subsubsection{Exact spectra for $L=4$ at increasing $\lambda$}
\label{sec:ex}

\textcolor{black}{
\begin{table}[b!]
\caption{
The energies of the ground state and eight first excited states
$\tilde{E}$ (\ref{et}), with their degeneracies $d$, obtained for the
periodic $L=4$ chain at $\alpha=\beta=0$ describing the spin-orbital
model Eq. (\ref{eq:h}) at $\lambda=0$ and for two representative values
of $\lambda$ ($\lambda/J=0.1$ and 10), standing for weak and strong
spin-orbit coupling.
}
\begin{ruledtabular}
  \begin{tabular}{lclclc}
\multicolumn{2}{c}{$\lambda=0$} & \multicolumn{2}{c}{$\lambda/J=0.1$} & \multicolumn{2}{c}{$\lambda/J=10$} \\
$\;\;\;\;\;\tilde{E}$&  $d$ &$\;\;\;\;\;\tilde{E}$& $d$ &$\;\;\;\;\;\tilde{E}$& $d$ \\ \hline
   $-0.75$      &   1  &   $-0.55$      &   1  &   $-0.45736$   &  1  \\
   $-0.50$      &  28  &   $-0.46570$   &   1  &   $-0.25$      &  2  \\
   $-0.45711$   &   9  &   $-0.37965$   &   4  &$\;\;\;0.24367$ &  1  \\
   $-0.43301$   &  12  &   $-0.36944$   &   8  &$\;\;\;0.24684$ &  2  \\
   $-0.40139$   &   1  &   $-0.30$      &  15  &$\;\;\;0.24688$ &  4  \\
   $-0.25$      &  48  &   $-0.26229$   &   2  &$\;\;\;0.24691$ &  1  \\
$\;\;\;0.0$     &  76  &   $-0.25345$   &   2  &$\;\;\;0.25$    &  2  \\
$\;\;\;0.25$    &  34  &   $-0.25$      &   2  &$\;\;\;0.74369$ &  2  \\
$\;\;\;0.43301$ &  12  &   $-0.24561$   &   2  &$\;\;\;0.94779$ &  1  \\
\end{tabular}
\end{ruledtabular}
\end{table}
}

We also note that the phase transition to the highly entangled phase
with increasing $\lambda$ is detected by level crossing in
Fig.~\ref{fig:energy}(b) and by the discontinuity in the derivative
$\left(\frac{\partial E}{\partial\lambda}\right)$, which appears as the
only kink for \mbox{$L=4$ -- site} chain, 
cf.~Fig.~\ref{fig:1_over_L}(a).
Other phase transitions occur
by varying $\alpha$---here at $\lambda/J=0.1$ a phase transition is
found from the FM (FO) phase with $S_{\rm tot}=2$ and $T_{\rm tot}=0$
($S_{\rm tot}=0$ and $T_{\rm tot}=2$) to an entangled SU(4) phase
(with all 15 generators being equal to 0) at $|\alpha|\simeq 0.08$,
see Fig. \ref{fig:energy}(c).
At this latter phase transition one finds also a discontinuous change of
the von Neumann entropy \cite{You2015b}. For convenience, we introduce
here the energy $\tilde{E}$ which does not decrease with increasing
$\lambda$ as in Fig. \ref{fig:energy}(b). For a chain of length $L=4$
it is defined as follows,
\begin{equation}
\tilde{E}\equiv E+2\lambda.
\label{et}
\end{equation}

To get more insight into the evolution of the spectra with increasing
spin-orbit coupling $\propto\lambda$, we consider in more detail the
exact spectra of the $L=4$ periodic chain, see Table I. At $\lambda=0$
the ground state is the SU(4) singlet with energy $E=-0.75$. The
degeneracies of the excited states follow from the $S$ and $T$ quantum
numbers. Indeed, several states with higher values of $S$ and $T$
exhibit huge degeneracies. Weak spin-orbit coupling $\lambda/J=0.1$
perturbation of the superexchange introduces the splittings of
degenerate excited states and in fact the spectrum is quite dense,
see Fig.~\ref{fig:energy}(c) and the data in Table I. However, the SU(4)
singlet ground state is still robust as
shown by the correlation functions $C_{\rm SO}$ and $O_{\rm SO}$
which do not change from their $\lambda=0$ values,
see the dotted line in Fig. \ref{fig:1_over_L} (a).
It indicates that the spin-orbit term does not
align here spin $S^z$ and orbital $T^z$ components. The energy of the
ground state (\ref{et}) is just moved by $2\lambda$ from $-0.75$ to
$\tilde{E}=-0.55$ (Table I), and the spin-orbit coupling does not
modify the ground state.

At $\lambda_{\rm CRIT}=0.14219J$ the energies of the
ground state and of the lowest energy excited state cross and a
completely different situation arises---then the von Neumann entropy
 changes in a discontinuous way to the value
corresponding to the strongly entangled state (for $L=4$), and the
spin-orbit correlation $O_{\rm SO}$ drops to $-0.25$, see
Fig. \ref{fig:1_over_L} (a). 
Since $\lambda_{\rm CRIT}$ is defined based on the changes
{\it in the ground state},  for
$\lambda>\lambda_{\rm CRIT}$ the
full energy spectrum
does change further
and consists of several bands of states, separated by gaps of the
order of $\lambda$, see Fig. \ref{fig:energy}(d). The energies and their
degeneracies within the lowest band of states are shown in the last two
columns of Table I at large $\lambda/J=10$ for the chain of length
$L=4$.

We have verified that the 16 low energy states displayed in Table I for
$\lambda/J=10$ collapse to the spectrum of the XY model in the limit of
$\lambda\to\infty$, with the degeneracies 1, 2, 10, 2, 1, as expected
and discussed in more detail in Sec. \ref{sec:xxz}. Thus, in general,
the spectrum consists of energy bands with the energies increasing in
steps of $\simeq\lambda$, depending on the number of sites at which the
spin-orbit coupling aligns the expectation values of spin and orbital
operators, $\left\langle S_i^zT_i^z\right\rangle$, at each site $i$.
In this regime the spectra are dominated by the spin-orbit coupling.
Note that the highly entangled phase can indeed be regarded as a
qualitatively unique phase, irrespectively of the value of
$\alpha$---provided that $\beta=-\alpha$ and that
\mbox{$\lambda>\lambda_{\rm{CRIT}}$.} 

\begin{figure}[t!]
\includegraphics[width=\columnwidth]{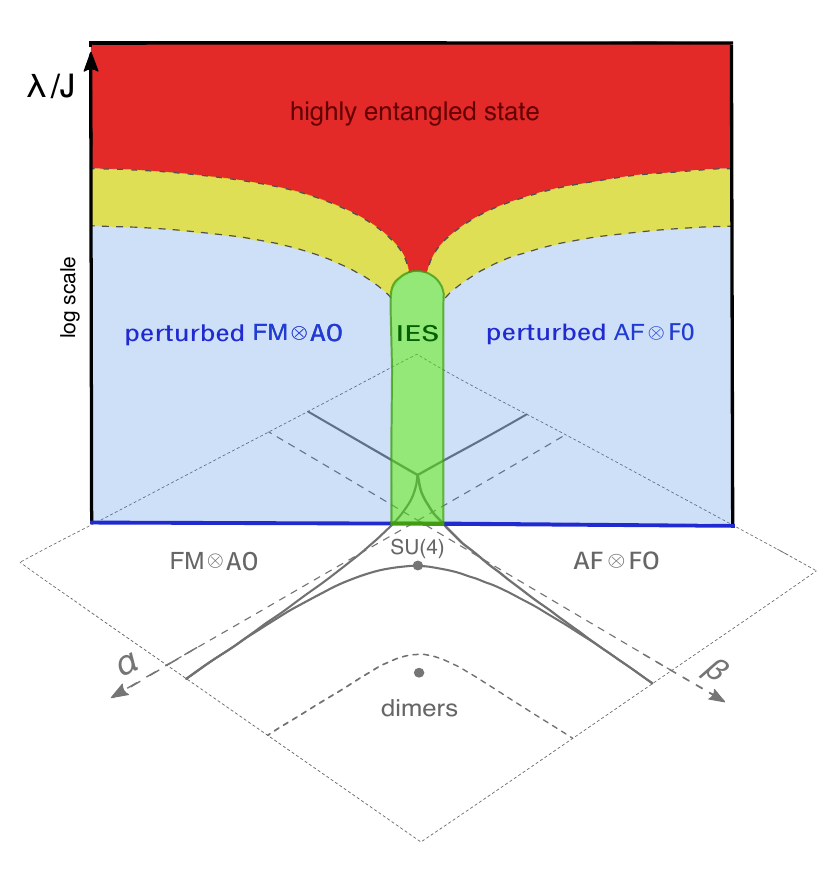}
\caption{Schematic quantum phase diagram of Hamiltonian (\ref{eq:h})
in the here--discussed regime of the parameters.
The limit of $\alpha=-\beta$ is depicted by the colorful vertical plane
and is based on the results from Sec.~\ref{sec:resultsB}: whereas the
four distinct phases are depicted with their names and separated by
solid lines (IES stands for the intermediate entangled state), the two
crossover regimes are denoted by yellow color and separated by the
dashed lines. The limit $\lambda=0$ is depicted by the horizontal plane
and is adopted from Fig. 1 of Ref.~\cite{Lun12}---see text for further
details. We note that the shape of the phase boundaries depends on the
logarithmic scale of $\lambda$, chosen here for convenience.
\textcolor{black}{The schematic phase diagram is based on the ED results
on small clusters, see text for the validity of these results in the thermodynamic limit.}
}
\label{fig:art}
\end{figure}

\subsubsection{Summary}
\label{sec:summary}

We have discussed in detail the evolution of the spin-orbital
entanglement, and its impact on the quantum phases, with the increasing
value of the spin-orbit coupling $\lambda$ for two representative
values of the parameter $\alpha$. We can now extend the above reasoning
to the other values of $\alpha$, keeping $\beta = - \alpha$. However,
in order to obtain a quantum phase diagram of the model we still need
to investigate whether the transitions between the obtained ground
states could be regarded as phase transitions or are rather just of the
crossover type. Dependence of the ground states energy on the model
parameters (see Fig.~\ref{fig:energy}) as well as the analytic
characteristics of the von Neumann entropy [see
Fig.~\ref{fig:ent}(d-e); cf.~Refs~\cite{Osterloh2002, Xun2008}] suggest
that the transitions along cuts A and B [Fig.~\ref{fig:ent}(a)] are of
distinct character. Whereas in case A the energy (as well as the von
Neumann entropy) shows an analytic behavior across the transition,
\textcolor{black}{[Fig. \ref{fig:energy}(a)]},
in case B such behavior (both in energy as well as in von Neumann
entropy) is clearly non-analytic
\textcolor{black}{[Fig. \ref{fig:energy}(b)]}.
This points to a crossover (phase) transition in case A~(B),
\mbox{respectively.}

Altogether, this allows us to draw, on a qualitative level, a quantum
phase diagram in the $\{\alpha,\lambda\}$ parameter space (with
$\beta=-\alpha$), see Fig.~\ref{fig:art} (colorful vertical plane).
As already discussed in Sec.~\ref{sec:nume}, there are four distinct
ground states (first two shown in Fig.~\ref{fig:art} in blue,
and the other two in green and red, respectively):
(i) the perturbed FM$\otimes$AO state for $\alpha\gtrsim 0.08$ and
$\lambda<\lambda_{\rm{CRIT}}$,
(ii) the perturbed AF$\otimes$FO state for $\alpha\lesssim -0.08$ and
$\lambda<\lambda_{\rm{CRIT}}$,
(iii) the intermediate entangled state for $|\alpha|\lesssim 0.08$ and
$0<\lambda<\lambda_{\rm{CRIT}}$, and
(iv) the highly entangled state for $\lambda>\lambda_{\rm{CRIT}}$ and
for all values of $\alpha$.
The latter state is discussed in more detailed in
Sec.~\ref{sec:discussion}. The four clearly distinct states are
supplemented by two crossover regimes (shown in yellow in Fig.
\ref{fig:art}), which separate phases (i-ii) from phase (iv)---see also
discussion above.

It is instructive to place the above phase diagram in the context of the
one already known from the literature and obtained for Hamiltonian
(\ref{eq:h}) in the limit of the vanishing spin-orbit coupling $\lambda$
but varying values of both $\alpha$ and $\beta$
\cite{Sutherland1975,Arovas1995,Yamashita1998,Pati1998,Kolezhuk1998,
Li1999,Itoi2000,Kolezhuk2000,Zheng2001,Li2005,Chen2007,Lun12,Chen2015}.
As can be seen on the horizontal plane of Fig.~\ref{fig:art}, the
$\lambda=0$ phase diagram consists of three simple product phases 
(AF$\otimes$FO, FM$\otimes$AO and FM$\otimes$FO) as well as two 
spin-orbital entangled phases (cf. Fig.~\ref{fig:en}): a phase with 
previously mentioned `global'  SU(4)-symmetric singlet ground state and 
gapless excitations \cite{Li1999} and a phase with the ground state 
breaking the $Z_{2}$ symmetry and opening a finite gap by forming the 
two nonequivalent patterns of the spin and orbital dimers 
\cite{Chen2007,Lun12}. We would like to emphasize at this point that
the finite size effects for the spin-orbital model (at $\lambda=0$) 
calculated on chains
of length $L=16$ (the maximal size studied in~Ref. 
\cite{Lun12}) and $L=20$ (the maximal size studied here) 
are already relatively small~\cite{Lun12}.
This may suggest that the schematic phase diagram of Fig.~\ref{fig:art}
is {\it qualitatively} correct also in the thermodynamic limit.

\section{Discussion: the limit of large $\lambda$}
\label{sec:discussion}

\subsection{Effective XXZ model}
\label{sec:xxz}

To better understand the numerical results obtained in
Sec.~\ref{sec:nume} for the Hamiltonian (\ref{eq:h}) in the limit of the
large spin-orbit coupling, $\lambda>\lambda_{\rm{CRIT}}$, we derive an
effective low-energy description of the system. In fact, as already
discussed in Introduction, such an approach has become extremely popular
in describing the physics of the iridium oxides \cite{Jackeli2009}, for
it has lead to the description of the latter in terms of effective
Heisenberg or Kitaev-like models. To obtain such an effective
description for the case of large spin-orbit coupling,
$\lambda>\lambda_{\rm{CRIT}}$, we first obtain the eigenstates of the
spin-orbit coupling Hamiltonian (\ref{ham_soc}): these are two doublets,
separated by the gap $\Delta E=\lambda$. Next, we restrict the Hilbert
space to the lowest doublet
$\{|{\uparrow}-\rangle,|{\downarrow}+\rangle\}$, where
$|{\downarrow}\rangle$ ($|-\rangle$) denotes the state with
\mbox{$S^z=-\nicefrac12$} ($T^z=-\nicefrac12$) quantum number. Lastly,
we project the intersite Hamiltonian (\ref{ham_kk1}) onto the lowest
doublet (see Appendix for details) and obtain the following effective
model:
\begin{align}
{\cal H}_{\rm eff}=\frac{J}{2} \sum_{i}\!\Big(
\tilde{J}^{x}_{i}\tilde{J}^x_{i+1}\!+\tilde{J}^{y}_{i}\tilde{J}^y_{i+1}\!
+2(\alpha\!+\!\beta)\tilde{J}^{z}_{i}\tilde{J}^{z}_{i+1} \Big),
\label{Heff_gen}
\end{align}
where $\tilde{J}^z_i= - \frac12
\left(n_{i,|\uparrow-\rangle}+n_{i,|\downarrow+\rangle}\right)$ is an
effective $\tilde{J^z}=\nicefrac12$ pseudospin operator.

Interestingly, it turns out that this effective Hamiltonian describes
exactly a spin $\nicefrac12$ XXZ chain. Moreover, in the limit of
$\alpha=-\beta$ the Ising interaction in Eq. (\ref{Heff_gen})
disappears and we obtain an AF XY model. Thus, resembling the iridate
case~\cite{Jackeli2009}, the effective model in the limit of large
spin-orbit coupling has a surprisingly simple form.

\begin{figure*}[t!]
\includegraphics[width=18cm]{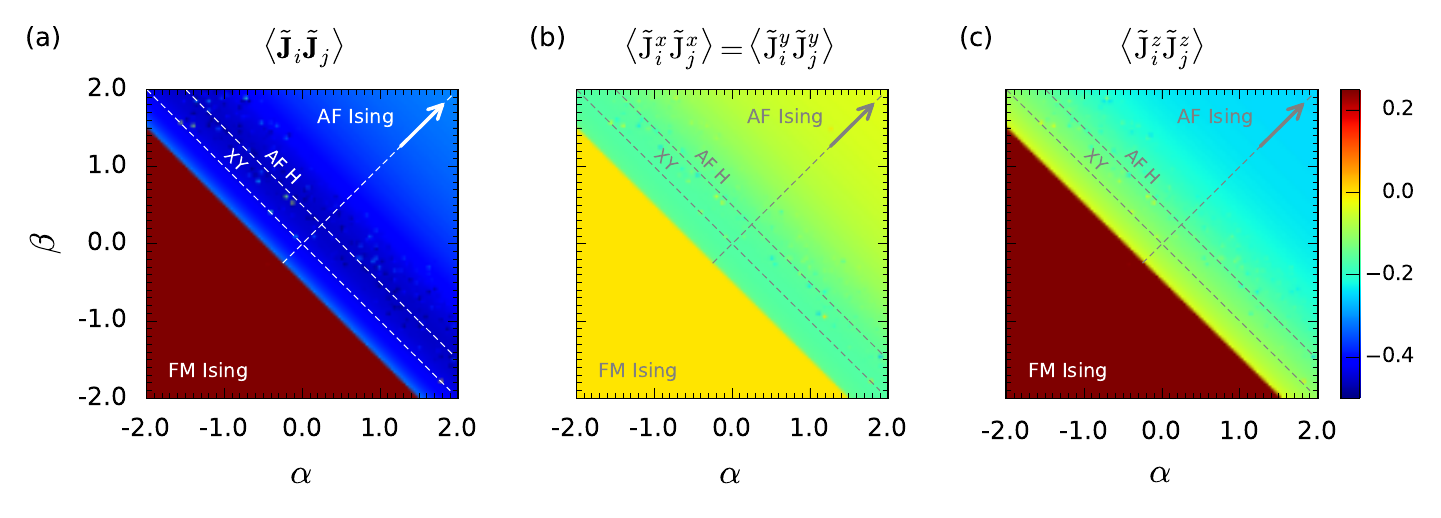}
\caption{The zero-temperature phase diagram of the effective XXZ model
(\ref{Heff_gen}) as a function of the model parameters $\alpha$
and $\beta$ obtained using ED on a $L=10$ site chain.
The panels present the correlations:
(a) $\langle \tilde{\bf J}_{i}\tilde{\bf J}_{j}\rangle$;
(b) $\langle \tilde{\rm J}^{\delta}_{i}\tilde{\rm J}^{\delta}_{j}\rangle$
with $\delta=x$, $y$;
(c)~$\langle \tilde{\rm J}^{z}_{i}\tilde{\rm J}^{z}_{j}\rangle$.
The labels depict various {\it ground states} of the 1D XXZ model:
AF H---the Heisenberg antiferromagnet,
AF Ising---the Ising antiferromagnet,
XY---the XY
antiferromagnet,
FM Ising---the Ising ferromagnet.
}
\label{fig:9}
\end{figure*}

\subsection{Validity of the effective XXZ model:\\ benchmarking $\alpha=-\beta$ case}
\label{sec:benchmarking}

First, let us show that the effective XXZ model indeed gives the
correct description of the ground state of the full spin-orbital model
(\ref{eq:h}) in the limit of
$\lambda > \lambda_{\rm{CRIT}}$. To this end, we
compare the spin-orbital correlation functions calculated using the
effective and the full models.

We first express the spin-orbital correlation function
${\cal C}_{\rm SO}$, the on-site spin-orbit correlation function
${\cal O}_{\rm SO}$, and the anisotropic spin (orbital) correlation
functions $S^{\gamma \gamma}$ ($T^{\gamma \gamma}$) in the basis
spanned by the two lowest doublets per site
$\{|{\uparrow}-\rangle,|{\downarrow}+\rangle\}$---see the Appendix for
the explicit formula. Next, we compare the values of the correlation
functions in the two special $\alpha=-\beta$ cases, already discussed
above:
(i) case A with $|\alpha|=0.5$, and
(ii) case B with $\alpha=0$.
As can be seen in Fig.~\ref{fig:finite_C_O} and Fig.
\ref{fig:anisotropic_S}, the correlation functions calculated using the
two distinct models agree extremely well once
$\lambda/J\gtrsim 1$--$100$ ($\lambda/J \gtrsim 0.2$) in case A (B),
respectively. We note that calculations performed for other values of
the $\{\alpha,\beta\}$ parameters (unshown) also show that the effective
model describes the ground state properties in the limit of
$\lambda > \lambda_{\rm{CRIT}}$ well. Moreover, once
$\lambda/J\simeq 10^{6}$, the ground and lowest lying excited states
are {\it quantitatively} the same in the full and the effective models.

\subsection{Why the spin-orbital entanglement can vanish}

Having derived the effective model---and having shown its validity---we
now discuss how it can help us with understanding one of the crucial
results of the paper: How can the spin-orbital entanglement vanish
in the limit of large spin-orbit coupling
$\lambda>\lambda_{\rm{CRIT}}$?

We start by expressing the measure for the spin-orbital entanglement
for nearest neighbors, the spin-orbital correlation ${\cal C}_{\rm SO}$,
in the basis of the effective model (see Appendix for details):
\begin{align}\label{eq:ChXXZ}
\tilde{\cal{C}}_{\rm SO} = \frac{1}{2L}\,\sum_{i=1}^{L}\left[
\langle\tilde{J}^{x}_{i}\tilde{J}^{x}_{i+1}\!
+\tilde{J}^{y}_{i}\tilde{J}^{y}_{i+1}\rangle\!
-2\langle \tilde{J}^{z}_{i}\tilde{J}^{z}_{i+1} \rangle^2\!
+\frac{1}{8}\right]\!,
\end{align}
where the averages are calculated in the ground state.

To evaluate Eq.~(\ref{eq:ChXXZ}), we calculate expectation values of the
effective pseudospin operators using ED, which we show in
Fig.~\ref{fig:9}. (We note in passing that the presented ED results for
an XXZ $L=10$ site chain agree well with those which were published
earlier, cf. Ref.~\cite{Hijii2005}.) The obtained ground state of the
effective Hamiltonian (\ref{Heff_gen}) for $\alpha+\beta<-1/2$ is
described by a ferromagnetic Ising state, where
$\langle\tilde{J}^z_i\tilde{J}^z_{i+1}\rangle=\nicefrac14$ and all other
correlations vanish. Substituting these into Eq.~(\ref{eq:ChXXZ})
explains why ${\cal{C}}_{\rm SO} =0$ in the ground state of model
(\ref{eq:h}) in the limit of large $\lambda > \lambda_{\rm{CRIT}}$ and
when restricted to $\alpha+\beta<-1/2$. In conclusion, the spin-orbital
entanglement for $\alpha+\beta<-1/2$ vanishes because not only the
on-site interaction between spins and orbitals but also the intersite
interactions in the ground state are of purely Ising type, and
effectively the ground state is just a product state with no
spin-orbital entanglement.

\subsection{Why the spin-orbital entanglement can be finite}
\label{Why_SOE_finite}
The effective model (\ref{Heff_gen})
can also be used to explain the presence of finite spin-orbital
entanglement in the limit of large spin-orbit coupling
$\lambda > \lambda_{\rm{CRIT}}$ while it vanishes in the $\lambda=0$ limit.
Let us first look at the already discussed in detail
$\beta=-\alpha$ case:

In this case and in the $\lambda=0$ limit, the term in (\ref{ham_kk1})
which is {\it explicitly} responsible for the spin-orbital entanglement,
$\propto(\textbf{S}_i\textbf{S}_j)(\textbf{T}_i\textbf{T}_j)$, can become
relatively small for large $\alpha$ or $\beta$ due to the presence of
the $\alpha\textbf{T}_i\textbf{T}_j$ and $\beta\textbf{S}_i\textbf{S}_j$
terms. Consequently, the region of significant spin-orbital entanglement
is quite small without spin-orbit coupling along the $\beta=-\alpha$
line, see Fig. \ref{fig:en}(a).
This situation, however, drastically changes in the limit of large
$\lambda > \lambda_{\rm{CRIT}}$, as discussed below.

Specifically, downfolding the exchange Hamiltonian (\ref{ham_kk1}) term
by term onto the effective Hamiltonian (\ref{Heff_gen}) should reveal
the origin of the spin-orbital entanglement in the large spin-orbit
coupling limit.
First, the $\alpha\textbf{T}_i\textbf{T}_j$ and
$\beta\textbf{S}_i\textbf{S}_j$ terms of Eq. (\ref{ham_kk1}) upon
projecting onto spin-orbit coupled basis produce
$\alpha\tilde{J}^z_i\tilde{J}^z_j$ and $\beta\tilde{J}^z_i\tilde{J}^z_j$,
resulting in the Ising terms in the effective model (\ref{Heff_gen}).
Note that in the case that $\beta=-\alpha$, these Ising terms disappear.
Second, the term responsible for the spin-orbital entanglement, i.e.,
$(\textbf{S}_i\textbf{S}_j)(\textbf{T}_i\textbf{T}_j)$ (cf. above),
reduces exactly to the XY terms in the effective model. These terms do
not vanish once $\beta=-\alpha$. In fact, in this special limit the
whole effective Hamiltonian is obtained from the term that is fully
responsible for the spin-orbital entanglement in the original Hamiltonian.
Finally, as the ground state of the XY Hamiltonian carries
`spatial entanglement' in pseudospins $\tilde{J}$, we expect the
spin-orbital entanglement to be finite in the limit of large spin-orbit
coupling \mbox{$\lambda>\lambda_{\rm{CRIT}}$} and once $\beta=-\alpha$.

The above reasoning is confirmed by calculating the two contributions to
the intersite spin-orbital correlation function $\tilde{\cal{C}}_{SO}$
in the effective model once \mbox{$\beta=-\alpha$.} This can be done
analytically for the XY model:
\mbox{$\langle\tilde{J}^{z}_{i}\tilde{J}^{z}_{i+1}\rangle=-1/\pi^2$} and
\mbox{$\langle\tilde{J}^{x}_{i}\tilde{J}^{x}_{i+1}\rangle=
 \langle\tilde{J}^{y}_{i}\tilde{J}^{y}_{i+1}\rangle=-1/(2\pi)$}.
(These results agree with the correlations calculated using ED and
presented in Fig.~\ref{fig:9}.)

The above discussion can now be extended to the case that
$\beta\neq -\alpha$ and $\alpha+\beta > -1/2 $, for which finite, though
increasingly small for large and positive $\alpha+\beta$, spin-orbital
entanglement can be observed, see Fig. \ref{fig:en}(c). Such result can
be understood by using the effective model and by noting that the
intersite spin-orbital correlation $\tilde{\cal{C}}_{SO}$ is always
finite provided that  $\alpha+\beta$ is finite and $\alpha+\beta>-1/2$.
This is because in this limit:
(i) the correlations
$\langle\tilde{J}^{x}_{i}\tilde{J}^{x}_{i+1}\rangle=
\langle\tilde{J}^{y}_{i}\tilde{J}^{y}_{i+1}\rangle$
are nonzero, see Fig.~\ref{fig:9}(b);
(ii) $\langle\tilde{J}^{z}_{i}\tilde{J}^{z}_{i+1}\rangle \neq 1/4$, see
Fig.~\ref{fig:9}(c).
It is then only in the limit $\alpha+\beta\rightarrow\infty$ that the
spin-orbital entanglement can vanish, for the ground state of the XXZ
model is `pure' Ising antiferromagnet. (A completely different situation
occurs once $\alpha+\beta<-1/2 $, i.e., for the FM ground state of the
effective XXZ model, as already discussed in the previous subsection---that
explains why the spin-orbital entanglement can `sometimes' vanish even
in the limit of large spin-orbit coupling, $\lambda>\lambda_{\rm{CRIT}}$.)

\section{Conclusions}
\label{sec:summa}

\subsection{Entanglement induced by spin-orbit coupling}
\label{sec:summaa}
In conclusion, in this paper we studied the spin-orbital entanglement
in a Mott insulator with spin and orbital degrees of freedom. We
investigated how the spin-orbital entanglement gradually changes
with the increasing value of the on-site spin-orbit coupling.
The results, obtained by exactly diagonalizing a 1D model with the
intersite SU(2)$\otimes$SU(2) spin-orbital superexchange $\propto J$ and the
on-site Ising-type spin-orbit coupling $\propto\lambda$, reveal that:
\begin{enumerate}
\item For small
 $\lambda < \lambda_{\rm{CRIT}}$ \footnote{$ \lambda_{\rm{CRIT}}$ depends on
 the particular values of the model parameters,  see Sec. \ref{sec:resultsB}.}:
\begin{enumerate}
\item In general, the spin-orbital entanglement in the ground state
is not much more robust than
in the $\lambda=0$ case;
\item
If the ground state had finite spin-orbital entanglement
for $\lambda=0$, it is driven into a novel spin-orbital
strongly entangled phase upon increasing $\lambda$;
\item
If the ground state did {\it not} show spin-orbital entanglement for
$\lambda=0$, it still shows none or negligible spin-orbital entanglement
upon increasing $\lambda$.
\end{enumerate}
\item In the limit of large
$\lambda > \lambda_{\rm{CRIT}}$:
\begin{enumerate}
\item In general, the spin-orbital entanglement in the ground state
is far more robust than in the $\lambda=0$ case;
\item
The ground state may be driven into a novel spin-orbitally entangled
phase {\it even} if it does not show spin-orbital entanglement for
$\lambda=0$;
\item
The ground state may still show vanishing spin-orbital entanglement,
but only if the quantum fluctuations vanish in the ground state of
an effective model (as is the case of an Ising ferromagnet).
\end{enumerate}
\end{enumerate}

The statements mentioned under point 2. above,
concerning  $\lambda > \lambda_{\rm{CRIT}}$,
 constitute, from the
purely theoretical perspective, the main results of this paper.
In particular, they mean that:
(i) the spin-orbital entanglement between spins and orbitals on
different sites can be triggered by a joint action of the on-site
spin-orbit coupling (of relativistic origin)
and the spin-orbital exchange (of the `Kugel-Khomskii'--type);
(ii) and yet, the onset of the spin-orbital entanglement in such a model
does not have to be taken `for granted', for it can vanish even in the
large spin-orbit coupling limit.

Crucially, we have verified that the spin-orbital entanglement can be
induced by the spin-orbit coupling, for the latter interaction {\it may}
enhance the role played by the spin-orbitally entangled
$(\textbf{S}_{i}\textbf{S}_{j})(\textbf{T}_{i}\textbf{T}_{j})$
term by `quenching' the bare spin
$(\textbf{S}_{i}\textbf{S}_{j})$ and orbital
$(\textbf{T}_{i}\textbf{T}_{j})$ exchange terms in an effective
low-energy Hamiltonian valid in this limit.
Interestingly, such mechanism can be valid even if the spin-orbit
coupling has a purely `classical' Ising form (as for example in the case
discussed in this paper). For a more intuitive explanation of these
results, in Sec.~\ref{sec:discussion} we presented  a detailed analysis
of the effective low-energy pseudospin XXZ model.

\subsection{Consequences for correlated materials}

The results presented here may play an important role in the
understanding of the correlated systems with non-negligible spin-orbit
coupling---such as e.g. the $5d$ iridates, $4d$ ruthenates, $3d$
vanadates, the $2p$ alkali hyperoxides, and other
to-be-synthesized materials. To this end, we argue that, even though
obtained for a specific 1D model, some of the results presented here
are to a large extent valid also for these 2D or 3D systems:

First, this is partially the case for the results obtained in the limit
of large $\lambda > \lambda_{\rm{CRIT}}$.
In particular, the mapping to the effective XXZ model is also valid in
2D and 3D cases. Moreover, one can easily verify that the spin-orbital
correlation function [$\tilde{\cal{C}}_{\rm SO}$, Eq.~(\ref{eq:ChXXZ})],
which measures spin-orbital entanglement never vanishes also in the 2D
and 3D cases, unless the quantum fluctuations completely disappear
(as is the case of the 2D or 3D Ising ferromagnet or antiferromagnet).
Therefore, the main conclusions from Secs. \ref{sec:nume}C and \ref{sec:nume}D
are also valid in 2D and 3D cases and consequently also point 2 of the
concluding Section~\ref{sec:summaa} holds.
This means that, for example, the results obtained here would apply to
any Mott insulator with two active $t_{2g}$ orbitals with small Hund's
coupling and with $\lambda > \lambda_{\rm{CRIT}}$
(such as e.g. Sr$_2$VO$_4$~\cite{Jackeli2009b}).

Naturally, the question
remains to what extent one could use the reasoning discussed here to the
understanding of the spin-orbital ground state of the probably most
famous Mott insulators with active orbital degrees of freedom and large
spin-orbit coupling---the $5d$ iridates (such as e.g. Sr$_2$IrO$_4$
\cite{Ber19}, Na$_2$IrO$_3$, Li$_2$IrO$_3$, etc.~\cite{Winter2017}).
Here we suggest that, while the situation in the iridates might be quite
different in detail and requires solving a distinct spin-orbital model with
three active $t_{2g}$ orbitals and an SU(2)-symmetric spin-orbit coupling
(which is beyond the scope of this work),
we {\it expect} point 2(b) of the concluding Section~\ref{sec:summaa}
to hold also in this case: 
in fact, the quantum nature of the Heisenberg spin-orbit coupling 
of the iridates (in contrast to the classical Ising spin-orbit coupling studied in this paper), 
should only facilitate the onset of the spin-orbital entanglement. Thus, we suggest that
in principle also for the iridates the ground state may be driven into a novel
spin-orbitally entangled phase even if it does not show spin-orbital
entanglement for $\lambda =0$.

Second, we suggest that also the fact that the spin-orbit coupling does
not induce additional spin-orbital entanglement in the limit of small
$\lambda<\lambda_{\rm{CRIT}}$ will carry on to higher dimensions and to
spin-orbital models of lower symmetry---for {\it a priori} there is no
reason why the tendency observed in a 1D (and highly symmetric) model,
towards a`more classical' behavior should fail in dimensions higher
than one (and for more anisotropic models). Thus, in general the
spin-orbital entanglement of the systems with weak spin-orbit coupling
$\lambda<\lambda_{\rm{CRIT}}$
and Ising-like spin-orbit coupling \cite{Horsch2003}, such as e.g. the 
alkali hyperoxides with two active `molecular' $2p$ orbitals
(e.g. KO$_2$~\cite{Solovyev2008}), should not qualitatively depend on
the value of spin-orbit coupling. This means that, to simplify the
studies one may, in the first order of approximation, neglect the
spin-orbit coupling in the effective models for these materials.

\acknowledgments

We thank Clio Agrapidis, Wojciech Brzezicki, Cheng-Chien Chen,
George Jackeli, Juraj Rusna\v{c}ko, and Takami Tohyama for
insightful discussions.
The calculations were performed partly at the Interdisciplinary
Centre for Mathematical and Computational Modeling (ICM),
University of Warsaw, under grant No.~\mbox{G72-9}.
This research was supported in part by PLGrid Infrastructure
(Academic Computer Center Cyfronet AGH Krak\'ow).
We kindly acknowledge support by the Naro\-do\-we Centrum Nauki (NCN,
Poland) under Projects Nos. 2016/22/E/ST3/00560 and 2016/23/B/ST3/00839.
E.~M.~P. acknowledges funding from the European Union's Horizon 2020
research and innovation programme under the Maria Sk\l{}odowska-Curie
grant agreement No. 754411.
J. Ch. acknowledges support by M\v{S}MT \v{C}R under NPU II project
CEITEC 2020 (LQ1601). Computational resources were supplied by the
project ``e-Infrastruktura CZ'' (e-INFRA LM2018140) provided within
the program Projects of Large Research, Development and Innovations
Infrastructures. A.~M.~Ole\'s is grateful for an Alexander von
Humboldt Foundation Fellowship \mbox{(Humboldt-Forschungspreis)}.

\appendix

\section*{Appendix: Effective XXZ model}

Let us consider the Hamiltonian (\ref{eq:h}) of the main text:
\begin{align}\label{eq:h:App}
{\cal H}={\cal H}_{\rm SE}+{\cal H}_{\rm SOC},
\end{align}
where the intersite interaction ${\cal H}_{\rm SE}$ and on-site
spin-orbit coupling are described by
\begin{align}
\label{ham_kk1:App}
{\cal H}_{\rm SE}&=J\sum_i\left[
    \left(\textbf{S}_i\!\cdot\!\textbf{S}_{i+1}\!+\alpha\right)\!
    \left(\textbf{T}_i\!\cdot\!\textbf{T}_{i+1}\!+\beta\right)
    -\alpha\beta\right],\\
\label{ham_kk2:App}
{\cal H}_{\rm SOC}&= 2\lambda\sum_{i}S^{z}_{i}T^{z}_{i}.
\end{align}
The characteristic scales for ${\cal H}_{\rm SE}$ and
${\cal H}_{\rm SOC}$ are intersite exchange parameter $J$ and on-site
SOC $\lambda$, respectively. In the strong spin-orbit coupling limit,
$\lambda > \lambda_{\rm{CRIT}}$, ${\cal H}_{\rm SE}$
can be considered as a perturbation to ${\cal H}_{\rm SOC}$. The
eigenstates of the full Hamiltonian (\ref{eq:h:App}) in zeroth-order
are then obtained by the diagonalization of the on-site spin-orbit part
${\cal H}_{\rm SOC}$. In our simple case ${\cal H}_{\rm SOC}$ is already
diagonal with two doubly--degenerate energies $\pm\lambda/2$.
The corresponding eigenstates defined by total momentum
$\tilde{J}$  form two doublets. The lower energy doublet consists of
\begin{align}
\tilde{J}_{\downarrow}&=|{+{\downarrow}}\rangle,
 \nonumber \\
\tilde{J}_{\uparrow}&=|{-{\uparrow}}\rangle,
\nonumber
\end{align}
while the higher  doublet is given by:
\begin{align}
\tilde{J}^{'}_{\uparrow}&=|{+{\uparrow}}\rangle,
 \nonumber\\
\tilde{J}^{'}_{\downarrow}&=|{-{\downarrow}}\rangle.
 \nonumber
\end{align}

Here, $|{\uparrow}\rangle$ ($|+\rangle$) denotes the state with
\mbox{$S^z=\nicefrac12$} \mbox{($T^z=\nicefrac12$)} quantum number.
The on-site \mbox{basis} transformation between the spin and orbital
\mbox{$\{|T^{z},S^{z}\rangle\}=\{|{+{\uparrow}}\rangle,
|{+{\downarrow}}\rangle,|{-{\uparrow}}\rangle,|{-{\downarrow}}\rangle\}$}
basis and spin-orbit coupled {$\{\tilde{J}_{\downarrow},\tilde{J}_{\uparrow},
\tilde{J}^{'}_{\uparrow},\tilde{J}^{'}_{\downarrow}\}$}
basis consisting of two doublets is described by a unitary matrix
\begin{align}
U=\left(
\begin{array}{rrrr}
 0 & 0 & 1 & 0 \\
 1& 0 & 0 & 0 \\
 0 & 1 & 0 & 0 \\
 0 & 0 & 0 & 1 \\
\end{array}
\right).
\end{align}

We then project the Hamiltonian (\ref{ham_kk1:App}) onto spin-orbit
coupled basis $\{\tilde{J}$,$\tilde{J}^{'}\}$:
${\cal H}_{\rm SE}^{\rm SOC}=U^\dag {\cal H}_{\rm SE} U$. As we are
interested in the low-energy physics, we truncate Hilbert
space to the lowest doublet $\tilde{J}$ and obtain effective
Hamiltonian (\ref{Heff_gen}) from the main text:
 \begin{align}
{\cal H}_{\rm eff}=\frac{J}{2}\sum_{i}\!\Big(
\tilde{J}^x_{i}\tilde{J}^x_{i+1}\!+\tilde{J}^y_{i}\tilde{J}^y_{i+1}\!
+2(\alpha\!+\!\beta)\tilde{J}^{z}_{i}\tilde{J}^{z}_{i+1} \Big).
\label{Heff_app}
\end{align}

To analyze the effective model (\ref{Heff_app}) and obtain important
correlation functions, we first need to establish a link between operators
describing correlation functions in original $\{ |T^{z}, S^{z}\rangle\}$
basis and spin-orbit coupled $\{\tilde{J}$,$\tilde{J}^{'}\}$ basis.
To this end, we project each of the spin/orbital operators,
${\cal O}_r=\{S^{\gamma}_r,T^{\gamma}_r\}$, $\gamma=\{x,y,z\}$,
$r=\{i,i+1\}$ entering the original correlation functions  (\ref{eq:Ch})
-- (\ref{eq:O}) onto spin-orbit coupled basis:
${\cal O}_r^{\rm SOC}=U^\dag {\cal O}_r U$. As most of the correlation
functions include intersite terms, the result shall be written as a
$16\times16$ matrix, spanned by
$\{\tilde{J}$,$\tilde{J}^{'}\}_i\times\{\tilde{J}$,$\tilde{J}^{'}\}_j$
basis.

We then once again drop out the high-energy doublet on each site and
obtain correlation functions as $4\times4$ matrices defined in Hilbert
space of $\{\tilde{J}\}_i\times\{\tilde{J}\}_j$:
\begin{equation}
\tilde{S}=\tilde{T}=\left\langle \begin{pmatrix}
\frac{1}{4} & 0 & 0 & 0\\ 0 & -\frac{1}{4} & 0 & 0 \\ 0 & 0 &
-\frac{1}{4} & 0 \\ 0 & 0 & 0 & \frac{1}{4} \end{pmatrix}
\right\rangle = \langle\tilde{J}^{z}_{i}\tilde{J}^{z}_{j}\rangle,
\end{equation}
\begin{equation}
\tilde{S}^{\delta\delta}=\tilde{T}^{\delta\delta}=\left\langle
\begin{pmatrix} 0 & 0 & 0 & 0\\ 0 & 0 & 0 & 0 \\ 0 & 0 & 0 & 0
\\ 0 & 0 & 0 & 0 \end{pmatrix}\right \rangle = 0,
\end{equation}
where $\delta=\{x,y\}$,
\begin{equation*}
\tilde{S}^{zz}=\tilde{T}^{zz}=\left\langle \begin{pmatrix} \frac{1}{4}
& 0 & 0 & 0\\ 0 & -\frac{1}{4} & 0 & 0 \\ 0 & 0 & -\frac{1}{4} & 0 \\
0 & 0 & 0 & \frac{1}{4} \end{pmatrix}\right \rangle =\left\langle
\tilde{J}^{z}_{i}\tilde{J}^{z}_{j}\right\rangle,
\end{equation*}
\begin{eqnarray*}
\tilde{\cal C}_{\rm SO}\!=\left\langle \begin{pmatrix} \frac{1}{16}
& 0 & 0 & 0\\ 0 & \frac{1}{16} & \frac{1}{4} & 0 \\ 0 &  \frac{1}{4}
& \frac{1}{16} & 0 \\ 0 & 0 & 0 & \frac{1}{16} \end{pmatrix}
\right \rangle -\!\left[\left \langle \begin{pmatrix} \frac{1}{4}
& 0 & 0 & 0\\ 0 & -\frac{1}{4} & 0 & 0 \\ 0 & 0 &
-\frac{1}{4} & 0 \\ 0 & 0 & 0 & \frac{1}{4} \end{pmatrix}
\right \rangle\right]^{2} \\ \\
 =\frac{1}{2}\langle \tilde{J}^{x}_{i}\tilde{J}^{x}_{j}
 + \tilde{J}^{y}_{i}\tilde{J}^{y}_{j}\rangle + \frac{1}{16}
 - \langle \tilde{J}^{z}_{i}\tilde{J}^{z}_{j}\rangle ^{2}.
\end{eqnarray*}
\newline
To express the on-site spin-orbit correlation function ${\cal O}_{SO}$,
which does not include intersite terms,
in the same basis, we multiply it by a $2\times2$ identity matrix
representing the neighboring site:
\begin{equation*}
\tilde{\cal O}_{SO,i}\otimes \text{id}_{j}=\left\langle \begin{pmatrix}
-\frac{1}{4} & 0 & 0  & 0\\ 0 & -\frac{1}{4} & 0 & 0  \\ 0 & 0 &
-\frac{1}{4}& 0 \\ 0 &0  & 0 & -\frac{1}{4} \end{pmatrix}\right \rangle
=-\frac{1}{4}.
\end{equation*}


\begin{thebibliography}{92}%
\makeatletter
\providecommand \@ifxundefined [1]{%
 \@ifx{#1\undefined}
}%
\providecommand \@ifnum [1]{%
 \ifnum #1\expandafter \@firstoftwo
 \else \expandafter \@secondoftwo
 \fi
}%
\providecommand \@ifx [1]{%
 \ifx #1\expandafter \@firstoftwo
 \else \expandafter \@secondoftwo
 \fi
}%
\providecommand \natexlab [1]{#1}%
\providecommand \enquote  [1]{``#1''}%
\providecommand \bibnamefont  [1]{#1}%
\providecommand \bibfnamefont [1]{#1}%
\providecommand \citenamefont [1]{#1}%
\providecommand \href@noop [0]{\@secondoftwo}%
\providecommand \href [0]{\begingroup \@sanitize@url \@href}%
\providecommand \@href[1]{\@@startlink{#1}\@@href}%
\providecommand \@@href[1]{\endgroup#1\@@endlink}%
\providecommand \@sanitize@url [0]{\catcode `\\12\catcode `\$12\catcode
  `\&12\catcode `\#12\catcode `\^12\catcode `\_12\catcode `\%12\relax}%
\providecommand \@@startlink[1]{}%
\providecommand \@@endlink[0]{}%
\providecommand \url  [0]{\begingroup\@sanitize@url \@url }%
\providecommand \@url [1]{\endgroup\@href {#1}{\urlprefix }}%
\providecommand \urlprefix  [0]{URL }%
\providecommand \Eprint [0]{\href }%
\providecommand \doibase [0]{http://dx.doi.org/}%
\providecommand \selectlanguage [0]{\@gobble}%
\providecommand \bibinfo  [0]{\@secondoftwo}%
\providecommand \bibfield  [0]{\@secondoftwo}%
\providecommand \translation [1]{[#1]}%
\providecommand \BibitemOpen [0]{}%
\providecommand \bibitemStop [0]{}%
\providecommand \bibitemNoStop [0]{.\EOS\space}%
\providecommand \EOS [0]{\spacefactor3000\relax}%
\providecommand \BibitemShut  [1]{\csname bibitem#1\endcsname}%
\let\auto@bib@innerbib\@empty
\bibitem [{\citenamefont {Khomskii}(2010)}]{Khomskii2010}%
  \BibitemOpen
  \bibfield  {author} {\bibinfo {author} {\bibfnamefont {D.~I.}\ \bibnamefont
  {Khomskii}},\ }\href@noop {} {\emph {\bibinfo {title} {Basic Aspects of the
  Quantum Theory of Solids: Order and Elementary Excitations}}}\ (\bibinfo
  {publisher} {Cambridge University Press},\ \bibinfo {address} {Cambridge},\
  \bibinfo {year} {2010})\BibitemShut {NoStop}%
\bibitem [{\citenamefont {Andrade}\ \emph {et~al.}(2018)\citenamefont
  {Andrade}, \citenamefont {Krikun}, \citenamefont {Schalm},\ and\
  \citenamefont {Zaanen}}]{Andrade2018}%
  \BibitemOpen
  \bibfield  {author} {\bibinfo {author} {\bibfnamefont {T.}~\bibnamefont
  {Andrade}}, \bibinfo {author} {\bibfnamefont {A.}~\bibnamefont {Krikun}},
  \bibinfo {author} {\bibfnamefont {K.}~\bibnamefont {Schalm}}, \ and\ \bibinfo
  {author} {\bibfnamefont {J.}~\bibnamefont {Zaanen}},\ }\href {\doibase
  10.1038/s41567-018-0217-6} {\bibfield  {journal} {\bibinfo  {journal} {Nature
  Physics}\ }\textbf {\bibinfo {volume} {14}},\ \bibinfo {pages} {1049}
  (\bibinfo {year} {2018})}\BibitemShut {NoStop}%
\bibitem [{\citenamefont {Zaanen}(2019)}]{Zaanen2019}%
  \BibitemOpen
  \bibfield  {author} {\bibinfo {author} {\bibfnamefont {J.}~\bibnamefont
  {Zaanen}},\ }\href {\doibase 10.21468/SciPostPhys.6.5.061} {\bibfield
  {journal} {\bibinfo  {journal} {SciPost Phys.}\ }\textbf {\bibinfo {volume}
  {6}},\ \bibinfo {pages} {61} (\bibinfo {year} {2019})}\BibitemShut {NoStop}%
\bibitem [{\citenamefont {Balents}(2010)}]{Balents2010}%
  \BibitemOpen
  \bibfield  {author} {\bibinfo {author} {\bibfnamefont {L.}~\bibnamefont
  {Balents}},\ }\href {https://doi.org/10.1038/nature08917} {\bibfield
  {journal} {\bibinfo  {journal} {Nature}\ }\textbf {\bibinfo {volume} {464}},\
  \bibinfo {pages} {199} (\bibinfo {year} {2010})}\BibitemShut {NoStop}%
\bibitem [{\citenamefont {Varma}\ \emph {et~al.}(2002)\citenamefont {Varma},
  \citenamefont {Nussinov},\ and\ \citenamefont {van Saarloos}}]{Varma2002}%
  \BibitemOpen
  \bibfield  {author} {\bibinfo {author} {\bibfnamefont {C.}~\bibnamefont
  {Varma}}, \bibinfo {author} {\bibfnamefont {Z.}~\bibnamefont {Nussinov}}, \
  and\ \bibinfo {author} {\bibfnamefont {W.}~\bibnamefont {van Saarloos}},\
  }\href {\doibase https://doi.org/10.1016/S0370-1573(01)00060-6} {\bibfield
  {journal} {\bibinfo  {journal} {Physics Reports}\ }\textbf {\bibinfo {volume}
  {361}},\ \bibinfo {pages} {267 } (\bibinfo {year} {2002})}\BibitemShut
  {NoStop}%
\bibitem [{\citenamefont {Einstein}\ \emph {et~al.}(1935)\citenamefont
  {Einstein}, \citenamefont {Podolsky},\ and\ \citenamefont
  {Rosen}}]{Einstein1935}%
  \BibitemOpen
  \bibfield  {author} {\bibinfo {author} {\bibfnamefont {A.}~\bibnamefont
  {Einstein}}, \bibinfo {author} {\bibfnamefont {B.}~\bibnamefont {Podolsky}},
  \ and\ \bibinfo {author} {\bibfnamefont {N.}~\bibnamefont {Rosen}},\ }\href
  {\doibase 10.1103/PhysRev.47.777} {\bibfield  {journal} {\bibinfo  {journal}
  {Phys. Rev.}\ }\textbf {\bibinfo {volume} {47}},\ \bibinfo {pages} {777}
  (\bibinfo {year} {1935})}\BibitemShut {NoStop}%
\bibitem [{\citenamefont {Ollivier}\ and\ \citenamefont {Zurek}(2001)}]{Zurek}%
  \BibitemOpen
  \bibfield  {author} {\bibinfo {author} {\bibfnamefont {H.}~\bibnamefont
  {Ollivier}}\ and\ \bibinfo {author} {\bibfnamefont {W.~H.}\ \bibnamefont
  {Zurek}},\ }\href {\doibase 10.1103/PhysRevLett.88.017901} {\bibfield
  {journal} {\bibinfo  {journal} {Phys. Rev. Lett.}\ }\textbf {\bibinfo
  {volume} {88}},\ \bibinfo {pages} {017901} (\bibinfo {year}
  {2001})}\BibitemShut {NoStop}%
\bibitem [{\citenamefont {Bentsson}\ and\ \citenamefont
  {Zyczkowski}(2006)}]{Karol}%
  \BibitemOpen
  \bibfield  {author} {\bibinfo {author} {\bibfnamefont {I.}~\bibnamefont
  {Bentsson}}\ and\ \bibinfo {author} {\bibfnamefont {K.}~\bibnamefont
  {Zyczkowski}},\ }\href@noop {} {\emph {\bibinfo {title} {Geometry of Quantum
  States: An Introduction to Quantum Entanglement}}}\ (\bibinfo  {publisher}
  {Cambridge University Press},\ \bibinfo {address} {Cambridge},\ \bibinfo
  {year} {2006})\BibitemShut {NoStop}%
\bibitem [{\citenamefont {Bennett}\ \emph {et~al.}(1996)\citenamefont
  {Bennett}, \citenamefont {Bernstein}, \citenamefont {Popescu},\ and\
  \citenamefont {Schumacher}}]{Bennett1996}%
  \BibitemOpen
  \bibfield  {author} {\bibinfo {author} {\bibfnamefont {C.~H.}\ \bibnamefont
  {Bennett}}, \bibinfo {author} {\bibfnamefont {H.~J.}\ \bibnamefont
  {Bernstein}}, \bibinfo {author} {\bibfnamefont {S.}~\bibnamefont {Popescu}},
  \ and\ \bibinfo {author} {\bibfnamefont {B.}~\bibnamefont {Schumacher}},\
  }\href {\doibase 10.1103/PhysRevA.53.2046} {\bibfield  {journal} {\bibinfo
  {journal} {Phys. Rev. A}\ }\textbf {\bibinfo {volume} {53}},\ \bibinfo
  {pages} {2046} (\bibinfo {year} {1996})}\BibitemShut {NoStop}%
\bibitem [{\citenamefont {Vidal}\ \emph {et~al.}(2003)\citenamefont {Vidal},
  \citenamefont {Latorre}, \citenamefont {Rico},\ and\ \citenamefont
  {Kitaev}}]{Vidal2003}%
  \BibitemOpen
  \bibfield  {author} {\bibinfo {author} {\bibfnamefont {G.}~\bibnamefont
  {Vidal}}, \bibinfo {author} {\bibfnamefont {J.~I.}\ \bibnamefont {Latorre}},
  \bibinfo {author} {\bibfnamefont {E.}~\bibnamefont {Rico}}, \ and\ \bibinfo
  {author} {\bibfnamefont {A.}~\bibnamefont {Kitaev}},\ }\href {\doibase
  10.1103/PhysRevLett.90.227902} {\bibfield  {journal} {\bibinfo  {journal}
  {Phys. Rev. Lett.}\ }\textbf {\bibinfo {volume} {90}},\ \bibinfo {pages}
  {227902} (\bibinfo {year} {2003})}\BibitemShut {NoStop}%
\bibitem [{\citenamefont {Korepin}(2004)}]{Korepin2004}%
  \BibitemOpen
  \bibfield  {author} {\bibinfo {author} {\bibfnamefont {V.~E.}\ \bibnamefont
  {Korepin}},\ }\href {\doibase 10.1103/PhysRevLett.92.096402} {\bibfield
  {journal} {\bibinfo  {journal} {Phys. Rev. Lett.}\ }\textbf {\bibinfo
  {volume} {92}},\ \bibinfo {pages} {096402} (\bibinfo {year}
  {2004})}\BibitemShut {NoStop}%
\bibitem [{\citenamefont {Wolf}(2006)}]{Wolf2006}%
  \BibitemOpen
  \bibfield  {author} {\bibinfo {author} {\bibfnamefont {M.~M.}\ \bibnamefont
  {Wolf}},\ }\href {\doibase 10.1103/PhysRevLett.96.010404} {\bibfield
  {journal} {\bibinfo  {journal} {Phys. Rev. Lett.}\ }\textbf {\bibinfo
  {volume} {96}},\ \bibinfo {pages} {010404} (\bibinfo {year}
  {2006})}\BibitemShut {NoStop}%
\bibitem [{\citenamefont {Gioev}\ and\ \citenamefont
  {Klich}(2006)}]{Gioev2006}%
  \BibitemOpen
  \bibfield  {author} {\bibinfo {author} {\bibfnamefont {D.}~\bibnamefont
  {Gioev}}\ and\ \bibinfo {author} {\bibfnamefont {I.}~\bibnamefont {Klich}},\
  }\href {\doibase 10.1103/PhysRevLett.96.100503} {\bibfield  {journal}
  {\bibinfo  {journal} {Phys. Rev. Lett.}\ }\textbf {\bibinfo {volume} {96}},\
  \bibinfo {pages} {100503} (\bibinfo {year} {2006})}\BibitemShut {NoStop}%
\bibitem [{\citenamefont {Lattore}\ \emph {et~al.}(2004)\citenamefont
  {Lattore}, \citenamefont {Rico},\ and\ \citenamefont {Vidal}}]{Lat04}%
  \BibitemOpen
  \bibfield  {author} {\bibinfo {author} {\bibfnamefont {J.~I.}\ \bibnamefont
  {Lattore}}, \bibinfo {author} {\bibfnamefont {E.}~\bibnamefont {Rico}}, \
  and\ \bibinfo {author} {\bibfnamefont {G.}~\bibnamefont {Vidal}},\
  }\href@noop {} {\bibfield  {journal} {\bibinfo  {journal} {Quantum
  Information \& Computation}\ }\textbf {\bibinfo {volume} {4}},\ \bibinfo
  {pages} {48} (\bibinfo {year} {2004})}\BibitemShut {NoStop}%
\bibitem [{\citenamefont {Cincio}\ \emph {et~al.}(2008)\citenamefont {Cincio},
  \citenamefont {Dziarmaga},\ and\ \citenamefont {Rams}}]{Cin08}%
  \BibitemOpen
  \bibfield  {author} {\bibinfo {author} {\bibfnamefont {L.}~\bibnamefont
  {Cincio}}, \bibinfo {author} {\bibfnamefont {J.}~\bibnamefont {Dziarmaga}}, \
  and\ \bibinfo {author} {\bibfnamefont {M.~M.}\ \bibnamefont {Rams}},\ }\href
  {\doibase 10.1103/PhysRevLett.100.240603} {\bibfield  {journal} {\bibinfo
  {journal} {Phys. Rev. Lett.}\ }\textbf {\bibinfo {volume} {100}},\ \bibinfo
  {pages} {240603} (\bibinfo {year} {2008})}\BibitemShut {NoStop}%
\bibitem [{\citenamefont {Thomale}\ \emph {et~al.}(2010)\citenamefont
  {Thomale}, \citenamefont {Arovas},\ and\ \citenamefont
  {Bernevig}}]{Thomale2010}%
  \BibitemOpen
  \bibfield  {author} {\bibinfo {author} {\bibfnamefont {R.}~\bibnamefont
  {Thomale}}, \bibinfo {author} {\bibfnamefont {D.~P.}\ \bibnamefont {Arovas}},
  \ and\ \bibinfo {author} {\bibfnamefont {B.~A.}\ \bibnamefont {Bernevig}},\
  }\href {\doibase 10.1103/PhysRevLett.105.116805} {\bibfield  {journal}
  {\bibinfo  {journal} {Phys. Rev. Lett.}\ }\textbf {\bibinfo {volume} {105}},\
  \bibinfo {pages} {116805} (\bibinfo {year} {2010})}\BibitemShut {NoStop}%
\bibitem [{\citenamefont {Khaliullin}\ and\ \citenamefont
  {Maekawa}(2000)}]{Khaliullin2000}%
  \BibitemOpen
  \bibfield  {author} {\bibinfo {author} {\bibfnamefont {G.}~\bibnamefont
  {Khaliullin}}\ and\ \bibinfo {author} {\bibfnamefont {S.}~\bibnamefont
  {Maekawa}},\ }\href {\doibase 10.1103/PhysRevLett.85.3950} {\bibfield
  {journal} {\bibinfo  {journal} {Phys. Rev. Lett.}\ }\textbf {\bibinfo
  {volume} {85}},\ \bibinfo {pages} {3950} (\bibinfo {year}
  {2000})}\BibitemShut {NoStop}%
\bibitem [{\citenamefont {Khaliullin}\ \emph {et~al.}(2001)\citenamefont
  {Khaliullin}, \citenamefont {Horsch},\ and\ \citenamefont
  {Ole\ifmmode~\acute{s}\else \'{s}\fi{}}}]{Khaliullin2001}%
  \BibitemOpen
  \bibfield  {author} {\bibinfo {author} {\bibfnamefont {G.}~\bibnamefont
  {Khaliullin}}, \bibinfo {author} {\bibfnamefont {P.}~\bibnamefont {Horsch}},
  \ and\ \bibinfo {author} {\bibfnamefont {A.~M.}\ \bibnamefont
  {Ole\ifmmode~\acute{s}\else \'{s}\fi{}}},\ }\href {\doibase
  10.1103/PhysRevLett.86.3879} {\bibfield  {journal} {\bibinfo  {journal}
  {Phys. Rev. Lett.}\ }\textbf {\bibinfo {volume} {86}},\ \bibinfo {pages}
  {3879} (\bibinfo {year} {2001})}\BibitemShut {NoStop}%
\bibitem [{\citenamefont {Khaliullin}(2005)}]{Kha05}%
  \BibitemOpen
  \bibfield  {author} {\bibinfo {author} {\bibfnamefont {G.}~\bibnamefont
  {Khaliullin}},\ }\href@noop {} {\bibfield  {journal} {\bibinfo  {journal}
  {Prog. Theor. Phys. Suppl.}\ }\textbf {\bibinfo {volume} {160}},\ \bibinfo
  {pages} {155} (\bibinfo {year} {2005})}\BibitemShut {NoStop}%
\bibitem [{\citenamefont {Nakatsuji}\ \emph {et~al.}(2012)\citenamefont
  {Nakatsuji}, \citenamefont {Kuga}, \citenamefont {Kimura}, \citenamefont
  {Satake}, \citenamefont {Katayama}, \citenamefont {Nishibori}, \citenamefont
  {Sawa}, \citenamefont {Ishii}, \citenamefont {Hagiwara}, \citenamefont
  {Bridges}, \citenamefont {Ito}, \citenamefont {Higemoto}, \citenamefont
  {Karaki}, \citenamefont {Halim}, \citenamefont {Nugroho}, \citenamefont
  {Rodriguez-Rivera}, \citenamefont {Green},\ and\ \citenamefont
  {Broholm}}]{Nakatsuji2012}%
  \BibitemOpen
  \bibfield  {author} {\bibinfo {author} {\bibfnamefont {S.}~\bibnamefont
  {Nakatsuji}}, \bibinfo {author} {\bibfnamefont {K.}~\bibnamefont {Kuga}},
  \bibinfo {author} {\bibfnamefont {K.}~\bibnamefont {Kimura}}, \bibinfo
  {author} {\bibfnamefont {R.}~\bibnamefont {Satake}}, \bibinfo {author}
  {\bibfnamefont {N.}~\bibnamefont {Katayama}}, \bibinfo {author}
  {\bibfnamefont {E.}~\bibnamefont {Nishibori}}, \bibinfo {author}
  {\bibfnamefont {H.}~\bibnamefont {Sawa}}, \bibinfo {author} {\bibfnamefont
  {R.}~\bibnamefont {Ishii}}, \bibinfo {author} {\bibfnamefont
  {M.}~\bibnamefont {Hagiwara}}, \bibinfo {author} {\bibfnamefont
  {F.}~\bibnamefont {Bridges}}, \bibinfo {author} {\bibfnamefont {T.~U.}\
  \bibnamefont {Ito}}, \bibinfo {author} {\bibfnamefont {W.}~\bibnamefont
  {Higemoto}}, \bibinfo {author} {\bibfnamefont {Y.}~\bibnamefont {Karaki}},
  \bibinfo {author} {\bibfnamefont {M.}~\bibnamefont {Halim}}, \bibinfo
  {author} {\bibfnamefont {A.~A.}\ \bibnamefont {Nugroho}}, \bibinfo {author}
  {\bibfnamefont {J.~A.}\ \bibnamefont {Rodriguez-Rivera}}, \bibinfo {author}
  {\bibfnamefont {M.~A.}\ \bibnamefont {Green}}, \ and\ \bibinfo {author}
  {\bibfnamefont {C.}~\bibnamefont {Broholm}},\ }\href {\doibase
  10.1126/science.1212154} {\bibfield  {journal} {\bibinfo  {journal}
  {Science}\ }\textbf {\bibinfo {volume} {336}},\ \bibinfo {pages} {559}
  (\bibinfo {year} {2012})}\BibitemShut {NoStop}%
\bibitem [{\citenamefont {Ishiguro}\ \emph {et~al.}(2013)\citenamefont
  {Ishiguro}, \citenamefont {Kimura}, \citenamefont {Nakatsuji}, \citenamefont
  {Tsutsui}, \citenamefont {Baron}, \citenamefont {Kimura},\ and\ \citenamefont
  {Wakabayashi}}]{Ishiguro2013}%
  \BibitemOpen
  \bibfield  {author} {\bibinfo {author} {\bibfnamefont {Y.}~\bibnamefont
  {Ishiguro}}, \bibinfo {author} {\bibfnamefont {K.}~\bibnamefont {Kimura}},
  \bibinfo {author} {\bibfnamefont {S.}~\bibnamefont {Nakatsuji}}, \bibinfo
  {author} {\bibfnamefont {S.}~\bibnamefont {Tsutsui}}, \bibinfo {author}
  {\bibfnamefont {A.~Q.~R.}\ \bibnamefont {Baron}}, \bibinfo {author}
  {\bibfnamefont {T.}~\bibnamefont {Kimura}}, \ and\ \bibinfo {author}
  {\bibfnamefont {Y.}~\bibnamefont {Wakabayashi}},\ }\href
  {https://doi.org/10.1038/ncomms3022} {\bibfield  {journal} {\bibinfo
  {journal} {Nature Communications}\ }\textbf {\bibinfo {volume} {4}},\
  \bibinfo {pages} {2022} (\bibinfo {year} {2013})}\BibitemShut {NoStop}%
\bibitem [{\citenamefont {Man}\ \emph {et~al.}(2018)\citenamefont {Man},
  \citenamefont {Halim}, \citenamefont {Sawa}, \citenamefont {Hagiwara},
  \citenamefont {Wakabayashi},\ and\ \citenamefont {Nakatsuji}}]{Man18}%
  \BibitemOpen
  \bibfield  {author} {\bibinfo {author} {\bibfnamefont {H.}~\bibnamefont
  {Man}}, \bibinfo {author} {\bibfnamefont {M.}~\bibnamefont {Halim}}, \bibinfo
  {author} {\bibfnamefont {H.}~\bibnamefont {Sawa}}, \bibinfo {author}
  {\bibfnamefont {M.}~\bibnamefont {Hagiwara}}, \bibinfo {author}
  {\bibfnamefont {Y.}~\bibnamefont {Wakabayashi}}, \ and\ \bibinfo {author}
  {\bibfnamefont {S.}~\bibnamefont {Nakatsuji}},\ }\href@noop {} {\bibfield
  {journal} {\bibinfo  {journal} {J. Phys.: Condens. Matter}\ }\textbf
  {\bibinfo {volume} {30}},\ \bibinfo {pages} {443002} (\bibinfo {year}
  {2018})}\BibitemShut {NoStop}%
\bibitem [{\citenamefont {Pan}\ \emph {et~al.}(2019)\citenamefont {Pan},
  \citenamefont {Jang}, \citenamefont {Lee}, \citenamefont {Sutarto},
  \citenamefont {He}, \citenamefont {Zeng}, \citenamefont {Liu}, \citenamefont
  {Zhang}, \citenamefont {Feng}, \citenamefont {Hao}, \citenamefont {Zhao},
  \citenamefont {Xu}, \citenamefont {Chen}, \citenamefont {Hu},\ and\
  \citenamefont {Feng}}]{Pan19}%
  \BibitemOpen
  \bibfield  {author} {\bibinfo {author} {\bibfnamefont {B.~Y.}\ \bibnamefont
  {Pan}}, \bibinfo {author} {\bibfnamefont {H.}~\bibnamefont {Jang}}, \bibinfo
  {author} {\bibfnamefont {J.-S.}\ \bibnamefont {Lee}}, \bibinfo {author}
  {\bibfnamefont {R.}~\bibnamefont {Sutarto}}, \bibinfo {author} {\bibfnamefont
  {F.}~\bibnamefont {He}}, \bibinfo {author} {\bibfnamefont {J.~F.}\
  \bibnamefont {Zeng}}, \bibinfo {author} {\bibfnamefont {Y.}~\bibnamefont
  {Liu}}, \bibinfo {author} {\bibfnamefont {X.~W.}\ \bibnamefont {Zhang}},
  \bibinfo {author} {\bibfnamefont {Y.}~\bibnamefont {Feng}}, \bibinfo {author}
  {\bibfnamefont {Y.~Q.}\ \bibnamefont {Hao}}, \bibinfo {author} {\bibfnamefont
  {J.}~\bibnamefont {Zhao}}, \bibinfo {author} {\bibfnamefont {H.~C.}\
  \bibnamefont {Xu}}, \bibinfo {author} {\bibfnamefont {Z.~H.}\ \bibnamefont
  {Chen}}, \bibinfo {author} {\bibfnamefont {J.~P.}\ \bibnamefont {Hu}}, \ and\
  \bibinfo {author} {\bibfnamefont {D.~L.}\ \bibnamefont {Feng}},\ }\href
  {\doibase 10.1103/PhysRevX.9.021055} {\bibfield  {journal} {\bibinfo
  {journal} {Phys. Rev. X}\ }\textbf {\bibinfo {volume} {9}},\ \bibinfo {pages}
  {021055} (\bibinfo {year} {2019})}\BibitemShut {NoStop}%
\bibitem [{\citenamefont {Normand}\ and\ \citenamefont
  {Ole\ifmmode~\acute{s}\else \'{s}\fi{}}(2008)}]{Nor08}%
  \BibitemOpen
  \bibfield  {author} {\bibinfo {author} {\bibfnamefont {B.}~\bibnamefont
  {Normand}}\ and\ \bibinfo {author} {\bibfnamefont {A.~M.}\ \bibnamefont
  {Ole\ifmmode~\acute{s}\else \'{s}\fi{}}},\ }\href {\doibase
  10.1103/PhysRevB.78.094427} {\bibfield  {journal} {\bibinfo  {journal} {Phys.
  Rev. B}\ }\textbf {\bibinfo {volume} {78}},\ \bibinfo {pages} {094427}
  (\bibinfo {year} {2008})}\BibitemShut {NoStop}%
\bibitem [{\citenamefont {Normand}(2011)}]{Nor11}%
  \BibitemOpen
  \bibfield  {author} {\bibinfo {author} {\bibfnamefont {B.}~\bibnamefont
  {Normand}},\ }\href {\doibase 10.1103/PhysRevB.83.064413} {\bibfield
  {journal} {\bibinfo  {journal} {Phys. Rev. B}\ }\textbf {\bibinfo {volume}
  {83}},\ \bibinfo {pages} {064413} (\bibinfo {year} {2011})}\BibitemShut
  {NoStop}%
\bibitem [{\citenamefont {Chaloupka}\ and\ \citenamefont
  {Ole\ifmmode~\acute{s}\else \'{s}\fi{}}(2011)}]{Cha11}%
  \BibitemOpen
  \bibfield  {author} {\bibinfo {author} {\bibfnamefont {J.}~\bibnamefont
  {Chaloupka}}\ and\ \bibinfo {author} {\bibfnamefont {A.~M.}\ \bibnamefont
  {Ole\ifmmode~\acute{s}\else \'{s}\fi{}}},\ }\href {\doibase
  10.1103/PhysRevB.83.094406} {\bibfield  {journal} {\bibinfo  {journal} {Phys.
  Rev. B}\ }\textbf {\bibinfo {volume} {83}},\ \bibinfo {pages} {094406}
  (\bibinfo {year} {2011})}\BibitemShut {NoStop}%
\bibitem [{\citenamefont {Corboz}\ \emph {et~al.}(2012)\citenamefont {Corboz},
  \citenamefont {Lajk\'o}, \citenamefont {L\"auchli}, \citenamefont {Penc},\
  and\ \citenamefont {Mila}}]{Corboz2012}%
  \BibitemOpen
  \bibfield  {author} {\bibinfo {author} {\bibfnamefont {P.}~\bibnamefont
  {Corboz}}, \bibinfo {author} {\bibfnamefont {M.}~\bibnamefont {Lajk\'o}},
  \bibinfo {author} {\bibfnamefont {A.~M.}\ \bibnamefont {L\"auchli}}, \bibinfo
  {author} {\bibfnamefont {K.}~\bibnamefont {Penc}}, \ and\ \bibinfo {author}
  {\bibfnamefont {F.}~\bibnamefont {Mila}},\ }\href {\doibase
  10.1103/PhysRevX.2.041013} {\bibfield  {journal} {\bibinfo  {journal} {Phys.
  Rev. X}\ }\textbf {\bibinfo {volume} {2}},\ \bibinfo {pages} {041013}
  (\bibinfo {year} {2012})}\BibitemShut {NoStop}%
\bibitem [{\citenamefont {Brzezicki}\ \emph {et~al.}(2012)\citenamefont
  {Brzezicki}, \citenamefont {Dziarmaga},\ and\ \citenamefont
  {Ole\ifmmode~\acute{s}\else \'{s}\fi{}}}]{Brz12}%
  \BibitemOpen
  \bibfield  {author} {\bibinfo {author} {\bibfnamefont {W.}~\bibnamefont
  {Brzezicki}}, \bibinfo {author} {\bibfnamefont {J.}~\bibnamefont
  {Dziarmaga}}, \ and\ \bibinfo {author} {\bibfnamefont {A.~M.}\ \bibnamefont
  {Ole\ifmmode~\acute{s}\else \'{s}\fi{}}},\ }\href {\doibase
  10.1103/PhysRevLett.109.237201} {\bibfield  {journal} {\bibinfo  {journal}
  {Phys. Rev. Lett.}\ }\textbf {\bibinfo {volume} {109}},\ \bibinfo {pages}
  {237201} (\bibinfo {year} {2012})}\BibitemShut {NoStop}%
\bibitem [{\citenamefont {Ole\ifmmode~\acute{s}\else \'{s}\fi{}}\ \emph
  {et~al.}(2006)\citenamefont {Ole\ifmmode~\acute{s}\else \'{s}\fi{}},
  \citenamefont {Horsch}, \citenamefont {Feiner},\ and\ \citenamefont
  {Khaliullin}}]{Ole06}%
  \BibitemOpen
  \bibfield  {author} {\bibinfo {author} {\bibfnamefont {A.~M.}\ \bibnamefont
  {Ole\ifmmode~\acute{s}\else \'{s}\fi{}}}, \bibinfo {author} {\bibfnamefont
  {P.}~\bibnamefont {Horsch}}, \bibinfo {author} {\bibfnamefont {L.~F.}\
  \bibnamefont {Feiner}}, \ and\ \bibinfo {author} {\bibfnamefont
  {G.}~\bibnamefont {Khaliullin}},\ }\href {\doibase
  10.1103/PhysRevLett.96.147205} {\bibfield  {journal} {\bibinfo  {journal}
  {Phys. Rev. Lett.}\ }\textbf {\bibinfo {volume} {96}},\ \bibinfo {pages}
  {147205} (\bibinfo {year} {2006})}\BibitemShut {NoStop}%
\bibitem [{\citenamefont {Ole\'{s}}(2012)}]{Ole12}%
  \BibitemOpen
  \bibfield  {author} {\bibinfo {author} {\bibfnamefont {A.~M.}\ \bibnamefont
  {Ole\'{s}}},\ }\href@noop {} {\bibfield  {journal} {\bibinfo  {journal} {J.
  Phys.: Condens. Matter}\ }\textbf {\bibinfo {volume} {24}},\ \bibinfo {pages}
  {313201} (\bibinfo {year} {2012})}\BibitemShut {NoStop}%
\bibitem [{\citenamefont {Chen}\ \emph {et~al.}(2007)\citenamefont {Chen},
  \citenamefont {Wang}, \citenamefont {Li},\ and\ \citenamefont
  {Zhang}}]{Chen2007}%
  \BibitemOpen
  \bibfield  {author} {\bibinfo {author} {\bibfnamefont {Y.}~\bibnamefont
  {Chen}}, \bibinfo {author} {\bibfnamefont {Z.~D.}\ \bibnamefont {Wang}},
  \bibinfo {author} {\bibfnamefont {Y.~Q.}\ \bibnamefont {Li}}, \ and\ \bibinfo
  {author} {\bibfnamefont {F.~C.}\ \bibnamefont {Zhang}},\ }\href {\doibase
  10.1103/PhysRevB.75.195113} {\bibfield  {journal} {\bibinfo  {journal} {Phys.
  Rev. B}\ }\textbf {\bibinfo {volume} {75}},\ \bibinfo {pages} {195113}
  (\bibinfo {year} {2007})}\BibitemShut {NoStop}%
\bibitem [{\citenamefont {Goodenough}(1963)}]{Goodenough1963}%
  \BibitemOpen
  \bibfield  {author} {\bibinfo {author} {\bibfnamefont {J.~B.}\ \bibnamefont
  {Goodenough}},\ }\href@noop {} {\emph {\bibinfo {title} {Magnetism and the
  Chemical Bond}}}\ (\bibinfo  {publisher} {Interscience},\ \bibinfo {address}
  {New York},\ \bibinfo {year} {1963})\BibitemShut {NoStop}%
\bibitem [{\citenamefont {Kanamori}(1959)}]{Kanamori1959}%
  \BibitemOpen
  \bibfield  {author} {\bibinfo {author} {\bibfnamefont {J.}~\bibnamefont
  {Kanamori}},\ }\href {\doibase
  http://dx.doi.org/10.1016/0022-3697(59)90061-7} {\bibfield  {journal}
  {\bibinfo  {journal} {Journal of Physics and Chemistry of Solids}\ }\textbf
  {\bibinfo {volume} {10}},\ \bibinfo {pages} {87 } (\bibinfo {year}
  {1959})}\BibitemShut {NoStop}%
\bibitem [{\citenamefont {Miyasaka}\ \emph {et~al.}(2002)\citenamefont
  {Miyasaka}, \citenamefont {Okimoto},\ and\ \citenamefont {Tokura}}]{Miy02}%
  \BibitemOpen
  \bibfield  {author} {\bibinfo {author} {\bibfnamefont {S.}~\bibnamefont
  {Miyasaka}}, \bibinfo {author} {\bibfnamefont {Y.}~\bibnamefont {Okimoto}}, \
  and\ \bibinfo {author} {\bibfnamefont {Y.}~\bibnamefont {Tokura}},\ }\href
  {\doibase 10.1143/JPSJ.71.2086} {\bibfield  {journal} {\bibinfo  {journal}
  {J. Phys. Soc. Jpn.}\ }\textbf {\bibinfo {volume} {71}},\ \bibinfo {pages}
  {2082} (\bibinfo {year} {2002})}\BibitemShut {NoStop}%
\bibitem [{\citenamefont {Khaliullin}\ \emph {et~al.}(2004)\citenamefont
  {Khaliullin}, \citenamefont {Horsch},\ and\ \citenamefont
  {Ole\ifmmode~\acute{s}\else \'{s}\fi{}}}]{Kha04}%
  \BibitemOpen
  \bibfield  {author} {\bibinfo {author} {\bibfnamefont {G.}~\bibnamefont
  {Khaliullin}}, \bibinfo {author} {\bibfnamefont {P.}~\bibnamefont {Horsch}},
  \ and\ \bibinfo {author} {\bibfnamefont {A.~M.}\ \bibnamefont
  {Ole\ifmmode~\acute{s}\else \'{s}\fi{}}},\ }\href {\doibase
  10.1103/PhysRevB.70.195103} {\bibfield  {journal} {\bibinfo  {journal} {Phys.
  Rev. B}\ }\textbf {\bibinfo {volume} {70}},\ \bibinfo {pages} {195103}
  (\bibinfo {year} {2004})}\BibitemShut {NoStop}%
\bibitem [{\citenamefont {Snamina}\ and\ \citenamefont
  {Ole{\'{s}}}(2019)}]{Sna19}%
  \BibitemOpen
  \bibfield  {author} {\bibinfo {author} {\bibfnamefont {M.}~\bibnamefont
  {Snamina}}\ and\ \bibinfo {author} {\bibfnamefont {A.~M.}\ \bibnamefont
  {Ole{\'{s}}}},\ }\href {\doibase 10.1088/1367-2630/aaf0d5} {\bibfield
  {journal} {\bibinfo  {journal} {New J. Phys.}\ }\textbf {\bibinfo {volume}
  {21}},\ \bibinfo {pages} {023018} (\bibinfo {year} {2019})}\BibitemShut
  {NoStop}%
\bibitem [{\citenamefont {Schlappa}\ \emph {et~al.}(2012)\citenamefont
  {Schlappa}, \citenamefont {Wohlfeld}, \citenamefont {Zhou}, \citenamefont
  {Mourigal}, \citenamefont {Haverkort}, \citenamefont {Strocov}, \citenamefont
  {Hozoi}, \citenamefont {Monney}, \citenamefont {Nishimoto}, \citenamefont
  {Singh}, \citenamefont {Revcolevschi}, \citenamefont {Caux}, \citenamefont
  {Patthey}, \citenamefont {Rønnow}, \citenamefont {van~den Brink},\ and\
  \citenamefont {Schmitt}}]{Schlappa2012}%
  \BibitemOpen
  \bibfield  {author} {\bibinfo {author} {\bibfnamefont {J.}~\bibnamefont
  {Schlappa}}, \bibinfo {author} {\bibfnamefont {K.}~\bibnamefont {Wohlfeld}},
  \bibinfo {author} {\bibfnamefont {K.~J.}\ \bibnamefont {Zhou}}, \bibinfo
  {author} {\bibfnamefont {M.}~\bibnamefont {Mourigal}}, \bibinfo {author}
  {\bibfnamefont {M.~W.}\ \bibnamefont {Haverkort}}, \bibinfo {author}
  {\bibfnamefont {V.~N.}\ \bibnamefont {Strocov}}, \bibinfo {author}
  {\bibfnamefont {L.}~\bibnamefont {Hozoi}}, \bibinfo {author} {\bibfnamefont
  {C.}~\bibnamefont {Monney}}, \bibinfo {author} {\bibfnamefont
  {S.}~\bibnamefont {Nishimoto}}, \bibinfo {author} {\bibfnamefont
  {S.}~\bibnamefont {Singh}}, \bibinfo {author} {\bibfnamefont
  {A.}~\bibnamefont {Revcolevschi}}, \bibinfo {author} {\bibfnamefont
  {J.}~\bibnamefont {Caux}}, \bibinfo {author} {\bibfnamefont {L.}~\bibnamefont
  {Patthey}}, \bibinfo {author} {\bibfnamefont {H.~M.}\ \bibnamefont
  {Rønnow}}, \bibinfo {author} {\bibfnamefont {J.}~\bibnamefont {van~den
  Brink}}, \ and\ \bibinfo {author} {\bibfnamefont {T.}~\bibnamefont
  {Schmitt}},\ }\href {\doibase 10.1038/nature10974} {\bibfield  {journal}
  {\bibinfo  {journal} {Nature}\ }\textbf {\bibinfo {volume} {485}},\ \bibinfo
  {pages} {82} (\bibinfo {year} {2012})}\BibitemShut {NoStop}%
\bibitem [{\citenamefont {Bisogni}\ \emph {et~al.}(2015)\citenamefont
  {Bisogni}, \citenamefont {Wohlfeld}, \citenamefont {Nishimoto}, \citenamefont
  {Monney}, \citenamefont {Trinckauf}, \citenamefont {Zhou}, \citenamefont
  {Kraus}, \citenamefont {Koepernik}, \citenamefont {Sekar}, \citenamefont
  {Strocov}, \citenamefont {B\"uchner}, \citenamefont {Schmitt}, \citenamefont
  {van~den Brink},\ and\ \citenamefont {Geck}}]{Bisogni2015}%
  \BibitemOpen
  \bibfield  {author} {\bibinfo {author} {\bibfnamefont {V.}~\bibnamefont
  {Bisogni}}, \bibinfo {author} {\bibfnamefont {K.}~\bibnamefont {Wohlfeld}},
  \bibinfo {author} {\bibfnamefont {S.}~\bibnamefont {Nishimoto}}, \bibinfo
  {author} {\bibfnamefont {C.}~\bibnamefont {Monney}}, \bibinfo {author}
  {\bibfnamefont {J.}~\bibnamefont {Trinckauf}}, \bibinfo {author}
  {\bibfnamefont {K.}~\bibnamefont {Zhou}}, \bibinfo {author} {\bibfnamefont
  {R.}~\bibnamefont {Kraus}}, \bibinfo {author} {\bibfnamefont
  {K.}~\bibnamefont {Koepernik}}, \bibinfo {author} {\bibfnamefont
  {C.}~\bibnamefont {Sekar}}, \bibinfo {author} {\bibfnamefont
  {V.}~\bibnamefont {Strocov}}, \bibinfo {author} {\bibfnamefont
  {B.}~\bibnamefont {B\"uchner}}, \bibinfo {author} {\bibfnamefont
  {T.}~\bibnamefont {Schmitt}}, \bibinfo {author} {\bibfnamefont
  {J.}~\bibnamefont {van~den Brink}}, \ and\ \bibinfo {author} {\bibfnamefont
  {J.}~\bibnamefont {Geck}},\ }\href {\doibase 10.1103/PhysRevLett.114.096402}
  {\bibfield  {journal} {\bibinfo  {journal} {Phys. Rev. Lett.}\ }\textbf
  {\bibinfo {volume} {114}},\ \bibinfo {pages} {096402} (\bibinfo {year}
  {2015})}\BibitemShut {NoStop}%
\bibitem [{\citenamefont {Wohlfeld}\ \emph {et~al.}(2011)\citenamefont
  {Wohlfeld}, \citenamefont {Daghofer}, \citenamefont {Nishimoto},
  \citenamefont {Khaliullin},\ and\ \citenamefont {van~den
  Brink}}]{Wohlfeld2011}%
  \BibitemOpen
  \bibfield  {author} {\bibinfo {author} {\bibfnamefont {K.}~\bibnamefont
  {Wohlfeld}}, \bibinfo {author} {\bibfnamefont {M.}~\bibnamefont {Daghofer}},
  \bibinfo {author} {\bibfnamefont {S.}~\bibnamefont {Nishimoto}}, \bibinfo
  {author} {\bibfnamefont {G.}~\bibnamefont {Khaliullin}}, \ and\ \bibinfo
  {author} {\bibfnamefont {J.}~\bibnamefont {van~den Brink}},\ }\href {\doibase
  10.1103/PhysRevLett.107.147201} {\bibfield  {journal} {\bibinfo  {journal}
  {Phys. Rev. Lett.}\ }\textbf {\bibinfo {volume} {107}},\ \bibinfo {pages}
  {147201} (\bibinfo {year} {2011})}\BibitemShut {NoStop}%
\bibitem [{\citenamefont {Wohlfeld}\ \emph {et~al.}(2013)\citenamefont
  {Wohlfeld}, \citenamefont {Nishimoto}, \citenamefont {Haverkort},\ and\
  \citenamefont {van~den Brink}}]{Wohlfeld2013}%
  \BibitemOpen
  \bibfield  {author} {\bibinfo {author} {\bibfnamefont {K.}~\bibnamefont
  {Wohlfeld}}, \bibinfo {author} {\bibfnamefont {S.}~\bibnamefont {Nishimoto}},
  \bibinfo {author} {\bibfnamefont {M.~W.}\ \bibnamefont {Haverkort}}, \ and\
  \bibinfo {author} {\bibfnamefont {J.}~\bibnamefont {van~den Brink}},\ }\href
  {\doibase 10.1103/PhysRevB.88.195138} {\bibfield  {journal} {\bibinfo
  {journal} {Phys. Rev. B}\ }\textbf {\bibinfo {volume} {88}},\ \bibinfo
  {pages} {195138} (\bibinfo {year} {2013})}\BibitemShut {NoStop}%
\bibitem [{\citenamefont {Chen}\ \emph {et~al.}(2015)\citenamefont {Chen},
  \citenamefont {van Veenendaal}, \citenamefont {Devereaux},\ and\
  \citenamefont {Wohlfeld}}]{Chen2015}%
  \BibitemOpen
  \bibfield  {author} {\bibinfo {author} {\bibfnamefont {C.-C.}\ \bibnamefont
  {Chen}}, \bibinfo {author} {\bibfnamefont {M.}~\bibnamefont {van
  Veenendaal}}, \bibinfo {author} {\bibfnamefont {T.~P.}\ \bibnamefont
  {Devereaux}}, \ and\ \bibinfo {author} {\bibfnamefont {K.}~\bibnamefont
  {Wohlfeld}},\ }\href {\doibase 10.1103/PhysRevB.91.165102} {\bibfield
  {journal} {\bibinfo  {journal} {Phys. Rev. B}\ }\textbf {\bibinfo {volume}
  {91}},\ \bibinfo {pages} {165102} (\bibinfo {year} {2015})}\BibitemShut
  {NoStop}%
\bibitem [{\citenamefont {{Witczak-Krempa}}\ \emph {et~al.}(2014)\citenamefont
  {{Witczak-Krempa}}, \citenamefont {{Chen}}, \citenamefont {{Kim}},\ and\
  \citenamefont {{Balents}}}]{WitczakKrempa2014}%
  \BibitemOpen
  \bibfield  {author} {\bibinfo {author} {\bibfnamefont {W.}~\bibnamefont
  {{Witczak-Krempa}}}, \bibinfo {author} {\bibfnamefont {G.}~\bibnamefont
  {{Chen}}}, \bibinfo {author} {\bibfnamefont {Y.~B.}\ \bibnamefont {{Kim}}}, \
  and\ \bibinfo {author} {\bibfnamefont {L.}~\bibnamefont {{Balents}}},\
  }\href@noop {} {\bibfield  {journal} {\bibinfo  {journal} {Annual Review of
  Condensed Matter Physics}\ }\textbf {\bibinfo {volume} {5}},\ \bibinfo
  {pages} {57} (\bibinfo {year} {2014})}\BibitemShut {NoStop}%
\bibitem [{\citenamefont {Bertinshaw}\ \emph {et~al.}(2019)\citenamefont
  {Bertinshaw}, \citenamefont {Kim}, \citenamefont {Khaliullin},\ and\
  \citenamefont {Kim}}]{Ber19}%
  \BibitemOpen
  \bibfield  {author} {\bibinfo {author} {\bibfnamefont {J.}~\bibnamefont
  {Bertinshaw}}, \bibinfo {author} {\bibfnamefont {Y.~K.}\ \bibnamefont {Kim}},
  \bibinfo {author} {\bibfnamefont {G.}~\bibnamefont {Khaliullin}}, \ and\
  \bibinfo {author} {\bibfnamefont {B.~J.}\ \bibnamefont {Kim}},\ }\href
  {\doibase 10.1146/annurev-conmatphys-031218-013113} {\bibfield  {journal}
  {\bibinfo  {journal} {Annu. Rev. Condens. Matter Physics}\ }\textbf {\bibinfo
  {volume} {10}},\ \bibinfo {pages} {315} (\bibinfo {year} {2019})}\BibitemShut
  {NoStop}%
\bibitem [{\citenamefont {Ishikawa}\ \emph {et~al.}(2019)\citenamefont
  {Ishikawa}, \citenamefont {Takayama}, \citenamefont {Kremer}, \citenamefont
  {Nuss}, \citenamefont {Dinnebier}, \citenamefont {Kitagawa}, \citenamefont
  {Ishii},\ and\ \citenamefont {Takagi}}]{Ish19}%
  \BibitemOpen
  \bibfield  {author} {\bibinfo {author} {\bibfnamefont {H.}~\bibnamefont
  {Ishikawa}}, \bibinfo {author} {\bibfnamefont {T.}~\bibnamefont {Takayama}},
  \bibinfo {author} {\bibfnamefont {R.~K.}\ \bibnamefont {Kremer}}, \bibinfo
  {author} {\bibfnamefont {J.}~\bibnamefont {Nuss}}, \bibinfo {author}
  {\bibfnamefont {R.}~\bibnamefont {Dinnebier}}, \bibinfo {author}
  {\bibfnamefont {K.}~\bibnamefont {Kitagawa}}, \bibinfo {author}
  {\bibfnamefont {K.}~\bibnamefont {Ishii}}, \ and\ \bibinfo {author}
  {\bibfnamefont {H.}~\bibnamefont {Takagi}},\ }\href {\doibase
  10.1103/PhysRevB.100.045142} {\bibfield  {journal} {\bibinfo  {journal}
  {Phys. Rev. B}\ }\textbf {\bibinfo {volume} {100}},\ \bibinfo {pages}
  {045142} (\bibinfo {year} {2019})}\BibitemShut {NoStop}%
\bibitem [{\citenamefont {Lefran\ifmmode~\mbox{\c{c}}\else \c{c}\fi{}ois}\
  \emph {et~al.}(2016)\citenamefont {Lefran\ifmmode~\mbox{\c{c}}\else
  \c{c}\fi{}ois}, \citenamefont {Pradipto}, \citenamefont {Moretti~Sala},
  \citenamefont {Chapon}, \citenamefont {Simonet}, \citenamefont {Picozzi},
  \citenamefont {Lejay}, \citenamefont {Petit},\ and\ \citenamefont
  {Ballou}}]{Lef16}%
  \BibitemOpen
  \bibfield  {author} {\bibinfo {author} {\bibfnamefont {E.}~\bibnamefont
  {Lefran\ifmmode~\mbox{\c{c}}\else \c{c}\fi{}ois}}, \bibinfo {author}
  {\bibfnamefont {A.-M.}\ \bibnamefont {Pradipto}}, \bibinfo {author}
  {\bibfnamefont {M.}~\bibnamefont {Moretti~Sala}}, \bibinfo {author}
  {\bibfnamefont {L.~C.}\ \bibnamefont {Chapon}}, \bibinfo {author}
  {\bibfnamefont {V.}~\bibnamefont {Simonet}}, \bibinfo {author} {\bibfnamefont
  {S.}~\bibnamefont {Picozzi}}, \bibinfo {author} {\bibfnamefont
  {P.}~\bibnamefont {Lejay}}, \bibinfo {author} {\bibfnamefont
  {S.}~\bibnamefont {Petit}}, \ and\ \bibinfo {author} {\bibfnamefont
  {R.}~\bibnamefont {Ballou}},\ }\href {\doibase 10.1103/PhysRevB.93.224401}
  {\bibfield  {journal} {\bibinfo  {journal} {Phys. Rev. B}\ }\textbf {\bibinfo
  {volume} {93}},\ \bibinfo {pages} {224401} (\bibinfo {year}
  {2016})}\BibitemShut {NoStop}%
\bibitem [{\citenamefont {Kitagawa}\ \emph {et~al.}(2018)\citenamefont
  {Kitagawa}, \citenamefont {Takayama}, \citenamefont {Matsumoto},
  \citenamefont {Kato}, \citenamefont {Takano}, \citenamefont {Kishimoto},
  \citenamefont {Bette}, \citenamefont {Dinnebier}, \citenamefont {Jackeli},\
  and\ \citenamefont {Takagi}}]{Kit18}%
  \BibitemOpen
  \bibfield  {author} {\bibinfo {author} {\bibfnamefont {K.}~\bibnamefont
  {Kitagawa}}, \bibinfo {author} {\bibfnamefont {T.}~\bibnamefont {Takayama}},
  \bibinfo {author} {\bibfnamefont {Y.}~\bibnamefont {Matsumoto}}, \bibinfo
  {author} {\bibfnamefont {A.}~\bibnamefont {Kato}}, \bibinfo {author}
  {\bibfnamefont {R.}~\bibnamefont {Takano}}, \bibinfo {author} {\bibfnamefont
  {Y.}~\bibnamefont {Kishimoto}}, \bibinfo {author} {\bibfnamefont
  {S.}~\bibnamefont {Bette}}, \bibinfo {author} {\bibfnamefont
  {R.}~\bibnamefont {Dinnebier}}, \bibinfo {author} {\bibfnamefont
  {G.}~\bibnamefont {Jackeli}}, \ and\ \bibinfo {author} {\bibfnamefont
  {H.}~\bibnamefont {Takagi}},\ }\href {https://doi.org/10.1038/nature25482}
  {\bibfield  {journal} {\bibinfo  {journal} {Nature}\ }\textbf {\bibinfo
  {volume} {554}},\ \bibinfo {pages} {341} (\bibinfo {year}
  {2018})}\BibitemShut {NoStop}%
\bibitem [{\citenamefont {Veenstra}\ \emph {et~al.}(2014)\citenamefont
  {Veenstra}, \citenamefont {Zhu}, \citenamefont {Raichle}, \citenamefont
  {Ludbrook}, \citenamefont {Nicolaou}, \citenamefont {Slomski}, \citenamefont
  {Landolt}, \citenamefont {Kittaka}, \citenamefont {Maeno}, \citenamefont
  {Dil}, \citenamefont {Elfimov}, \citenamefont {Haverkort},\ and\
  \citenamefont {Damascelli}}]{Vee14}%
  \BibitemOpen
  \bibfield  {author} {\bibinfo {author} {\bibfnamefont {C.~N.}\ \bibnamefont
  {Veenstra}}, \bibinfo {author} {\bibfnamefont {Z.-H.}\ \bibnamefont {Zhu}},
  \bibinfo {author} {\bibfnamefont {M.}~\bibnamefont {Raichle}}, \bibinfo
  {author} {\bibfnamefont {B.~M.}\ \bibnamefont {Ludbrook}}, \bibinfo {author}
  {\bibfnamefont {A.}~\bibnamefont {Nicolaou}}, \bibinfo {author}
  {\bibfnamefont {B.}~\bibnamefont {Slomski}}, \bibinfo {author} {\bibfnamefont
  {G.}~\bibnamefont {Landolt}}, \bibinfo {author} {\bibfnamefont
  {S.}~\bibnamefont {Kittaka}}, \bibinfo {author} {\bibfnamefont
  {Y.}~\bibnamefont {Maeno}}, \bibinfo {author} {\bibfnamefont {J.~H.}\
  \bibnamefont {Dil}}, \bibinfo {author} {\bibfnamefont {I.~S.}\ \bibnamefont
  {Elfimov}}, \bibinfo {author} {\bibfnamefont {M.~W.}\ \bibnamefont
  {Haverkort}}, \ and\ \bibinfo {author} {\bibfnamefont {A.}~\bibnamefont
  {Damascelli}},\ }\href {\doibase 10.1103/PhysRevLett.112.127002} {\bibfield
  {journal} {\bibinfo  {journal} {Phys. Rev. Lett.}\ }\textbf {\bibinfo
  {volume} {112}},\ \bibinfo {pages} {127002} (\bibinfo {year}
  {2014})}\BibitemShut {NoStop}%
\bibitem [{\citenamefont {Zhang}\ \emph {et~al.}(2016)\citenamefont {Zhang},
  \citenamefont {Gorelov}, \citenamefont {Sarvestani},\ and\ \citenamefont
  {Pavarini}}]{Zha16}%
  \BibitemOpen
  \bibfield  {author} {\bibinfo {author} {\bibfnamefont {G.}~\bibnamefont
  {Zhang}}, \bibinfo {author} {\bibfnamefont {E.}~\bibnamefont {Gorelov}},
  \bibinfo {author} {\bibfnamefont {E.}~\bibnamefont {Sarvestani}}, \ and\
  \bibinfo {author} {\bibfnamefont {E.}~\bibnamefont {Pavarini}},\ }\href
  {\doibase 10.1103/PhysRevLett.116.106402} {\bibfield  {journal} {\bibinfo
  {journal} {Phys. Rev. Lett.}\ }\textbf {\bibinfo {volume} {116}},\ \bibinfo
  {pages} {106402} (\bibinfo {year} {2016})}\BibitemShut {NoStop}%
\bibitem [{\citenamefont {Khaliullin}(2013)}]{Kha13}%
  \BibitemOpen
  \bibfield  {author} {\bibinfo {author} {\bibfnamefont {G.}~\bibnamefont
  {Khaliullin}},\ }\href {\doibase 10.1103/PhysRevLett.111.197201} {\bibfield
  {journal} {\bibinfo  {journal} {Phys. Rev. Lett.}\ }\textbf {\bibinfo
  {volume} {111}},\ \bibinfo {pages} {197201} (\bibinfo {year}
  {2013})}\BibitemShut {NoStop}%
\bibitem [{\citenamefont {Jain}\ \emph {et~al.}(2017)\citenamefont {Jain},
  \citenamefont {Krautloher}, \citenamefont {Porras}, \citenamefont {Ryu},
  \citenamefont {Chen}, \citenamefont {Abernathy}, \citenamefont {Park},
  \citenamefont {Ivanov}, \citenamefont {Chaloupka}, \citenamefont
  {Khaliullin}, \citenamefont {Keimer},\ and\ \citenamefont {Kim}}]{Jain2017}%
  \BibitemOpen
  \bibfield  {author} {\bibinfo {author} {\bibfnamefont {A.}~\bibnamefont
  {Jain}}, \bibinfo {author} {\bibfnamefont {M.}~\bibnamefont {Krautloher}},
  \bibinfo {author} {\bibfnamefont {J.}~\bibnamefont {Porras}}, \bibinfo
  {author} {\bibfnamefont {G.~H.}\ \bibnamefont {Ryu}}, \bibinfo {author}
  {\bibfnamefont {D.~P.}\ \bibnamefont {Chen}}, \bibinfo {author}
  {\bibfnamefont {D.~L.}\ \bibnamefont {Abernathy}}, \bibinfo {author}
  {\bibfnamefont {J.~T.}\ \bibnamefont {Park}}, \bibinfo {author}
  {\bibfnamefont {A.}~\bibnamefont {Ivanov}}, \bibinfo {author} {\bibfnamefont
  {J.}~\bibnamefont {Chaloupka}}, \bibinfo {author} {\bibfnamefont
  {G.}~\bibnamefont {Khaliullin}}, \bibinfo {author} {\bibfnamefont
  {B.}~\bibnamefont {Keimer}}, \ and\ \bibinfo {author} {\bibfnamefont {B.~J.}\
  \bibnamefont {Kim}},\ }\href {https://doi.org/10.1038/nphys4077} {\bibfield
  {journal} {\bibinfo  {journal} {Nature Physics}\ }\textbf {\bibinfo {volume}
  {13}},\ \bibinfo {pages} {633} (\bibinfo {year} {2017})}\BibitemShut
  {NoStop}%
\bibitem [{\citenamefont {Chaloupka}\ \emph {et~al.}(2010)\citenamefont
  {Chaloupka}, \citenamefont {Jackeli},\ and\ \citenamefont
  {Khaliullin}}]{Cha10}%
  \BibitemOpen
  \bibfield  {author} {\bibinfo {author} {\bibfnamefont {J.}~\bibnamefont
  {Chaloupka}}, \bibinfo {author} {\bibfnamefont {G.}~\bibnamefont {Jackeli}},
  \ and\ \bibinfo {author} {\bibfnamefont {G.}~\bibnamefont {Khaliullin}},\
  }\href {\doibase 10.1103/PhysRevLett.105.027204} {\bibfield  {journal}
  {\bibinfo  {journal} {Phys. Rev. Lett.}\ }\textbf {\bibinfo {volume} {105}},\
  \bibinfo {pages} {027204} (\bibinfo {year} {2010})}\BibitemShut {NoStop}%
\bibitem [{\citenamefont {Winter}\ \emph {et~al.}(2017)\citenamefont {Winter},
  \citenamefont {Tsirlin}, \citenamefont {Daghofer}, \citenamefont {van~den
  Brink}, \citenamefont {Singh}, \citenamefont {Gegenwart},\ and\ \citenamefont
  {Valent{\'{\i}}}}]{Winter2017}%
  \BibitemOpen
  \bibfield  {author} {\bibinfo {author} {\bibfnamefont {S.~M.}\ \bibnamefont
  {Winter}}, \bibinfo {author} {\bibfnamefont {A.~A.}\ \bibnamefont {Tsirlin}},
  \bibinfo {author} {\bibfnamefont {M.}~\bibnamefont {Daghofer}}, \bibinfo
  {author} {\bibfnamefont {J.}~\bibnamefont {van~den Brink}}, \bibinfo {author}
  {\bibfnamefont {Y.}~\bibnamefont {Singh}}, \bibinfo {author} {\bibfnamefont
  {P.}~\bibnamefont {Gegenwart}}, \ and\ \bibinfo {author} {\bibfnamefont
  {R.}~\bibnamefont {Valent{\'{\i}}}},\ }\href {\doibase
  10.1088/1361-648x/aa8cf5} {\bibfield  {journal} {\bibinfo  {journal} {Journal
  of Physics: Condensed Matter}\ }\textbf {\bibinfo {volume} {29}},\ \bibinfo
  {pages} {493002} (\bibinfo {year} {2017})}\BibitemShut {NoStop}%
\bibitem [{\citenamefont {Kitaev}(2006)}]{Kitaev2006}%
  \BibitemOpen
  \bibfield  {author} {\bibinfo {author} {\bibfnamefont {A.}~\bibnamefont
  {Kitaev}},\ }\href {\doibase http://dx.doi.org/10.1016/j.aop.2005.10.005}
  {\bibfield  {journal} {\bibinfo  {journal} {Annals of Physics}\ }\textbf
  {\bibinfo {volume} {321}},\ \bibinfo {pages} {2 } (\bibinfo {year}
  {2006})}\BibitemShut {NoStop}%
\bibitem [{\citenamefont {Lundgren}\ \emph {et~al.}(2012)\citenamefont
  {Lundgren}, \citenamefont {Chua},\ and\ \citenamefont {Fiete}}]{Lun12}%
  \BibitemOpen
  \bibfield  {author} {\bibinfo {author} {\bibfnamefont {R.}~\bibnamefont
  {Lundgren}}, \bibinfo {author} {\bibfnamefont {V.}~\bibnamefont {Chua}}, \
  and\ \bibinfo {author} {\bibfnamefont {G.~A.}\ \bibnamefont {Fiete}},\ }\href
  {\doibase 10.1103/PhysRevB.86.224422} {\bibfield  {journal} {\bibinfo
  {journal} {Phys. Rev. B}\ }\textbf {\bibinfo {volume} {86}},\ \bibinfo
  {pages} {224422} (\bibinfo {year} {2012})}\BibitemShut {NoStop}%
\bibitem [{\citenamefont {You}\ \emph {et~al.}(2012)\citenamefont {You},
  \citenamefont {Ole\ifmmode~\acute{s}\else \'{s}\fi{}},\ and\ \citenamefont
  {Horsch}}]{You12}%
  \BibitemOpen
  \bibfield  {author} {\bibinfo {author} {\bibfnamefont {W.-L.}\ \bibnamefont
  {You}}, \bibinfo {author} {\bibfnamefont {A.~M.}\ \bibnamefont
  {Ole\ifmmode~\acute{s}\else \'{s}\fi{}}}, \ and\ \bibinfo {author}
  {\bibfnamefont {P.}~\bibnamefont {Horsch}},\ }\href {\doibase
  10.1103/PhysRevB.86.094412} {\bibfield  {journal} {\bibinfo  {journal} {Phys.
  Rev. B}\ }\textbf {\bibinfo {volume} {86}},\ \bibinfo {pages} {094412}
  (\bibinfo {year} {2012})}\BibitemShut {NoStop}%
\bibitem [{\citenamefont {L\"auchli}\ and\ \citenamefont
  {Schliemann}(2012)}]{Lauchli2012}%
  \BibitemOpen
  \bibfield  {author} {\bibinfo {author} {\bibfnamefont {A.~M.}\ \bibnamefont
  {L\"auchli}}\ and\ \bibinfo {author} {\bibfnamefont {J.}~\bibnamefont
  {Schliemann}},\ }\href {\doibase 10.1103/PhysRevB.85.054403} {\bibfield
  {journal} {\bibinfo  {journal} {Phys. Rev. B}\ }\textbf {\bibinfo {volume}
  {85}},\ \bibinfo {pages} {054403} (\bibinfo {year} {2012})}\BibitemShut
  {NoStop}%
\bibitem [{\citenamefont {You}\ \emph {et~al.}(2015{\natexlab{a}})\citenamefont
  {You}, \citenamefont {Horsch},\ and\ \citenamefont
  {Ole\ifmmode~\acute{s}\else \'{s}\fi{}}}]{You2015}%
  \BibitemOpen
  \bibfield  {author} {\bibinfo {author} {\bibfnamefont {W.-L.}\ \bibnamefont
  {You}}, \bibinfo {author} {\bibfnamefont {P.}~\bibnamefont {Horsch}}, \ and\
  \bibinfo {author} {\bibfnamefont {A.~M.}\ \bibnamefont
  {Ole\ifmmode~\acute{s}\else \'{s}\fi{}}},\ }\href {\doibase
  10.1103/PhysRevB.92.054423} {\bibfield  {journal} {\bibinfo  {journal} {Phys.
  Rev. B}\ }\textbf {\bibinfo {volume} {92}},\ \bibinfo {pages} {054423}
  (\bibinfo {year} {2015}{\natexlab{a}})}\BibitemShut {NoStop}%
\bibitem [{\citenamefont {You}\ \emph {et~al.}(2015{\natexlab{b}})\citenamefont
  {You}, \citenamefont {Ole{\'{s}}},\ and\ \citenamefont {Horsch}}]{You2015b}%
  \BibitemOpen
  \bibfield  {author} {\bibinfo {author} {\bibfnamefont {W.-L.}\ \bibnamefont
  {You}}, \bibinfo {author} {\bibfnamefont {A.~M.}\ \bibnamefont {Ole{\'{s}}}},
  \ and\ \bibinfo {author} {\bibfnamefont {P.}~\bibnamefont {Horsch}},\ }\href
  {\doibase 10.1088/1367-2630/17/8/083009} {\bibfield  {journal} {\bibinfo
  {journal} {New J. Phys.}\ }\textbf {\bibinfo {volume} {17}},\ \bibinfo
  {pages} {083009} (\bibinfo {year} {2015}{\natexlab{b}})}\BibitemShut
  {NoStop}%
\bibitem [{\citenamefont {{Valiulin}}\ \emph {et~al.}(2019)\citenamefont
  {{Valiulin}}, \citenamefont {{Mikheyenkov}}, \citenamefont {{Kugel}},\ and\
  \citenamefont {{Barabanov}}}]{Val19}%
  \BibitemOpen
  \bibfield  {author} {\bibinfo {author} {\bibfnamefont {V.~E.}\ \bibnamefont
  {{Valiulin}}}, \bibinfo {author} {\bibfnamefont {A.~V.}\ \bibnamefont
  {{Mikheyenkov}}}, \bibinfo {author} {\bibfnamefont {K.~I.}\ \bibnamefont
  {{Kugel}}}, \ and\ \bibinfo {author} {\bibfnamefont {A.~F.}\ \bibnamefont
  {{Barabanov}}},\ }\href@noop {} {\bibfield  {journal} {\bibinfo  {journal}
  {JETP Letters}\ }\textbf {\bibinfo {volume} {109}},\ \bibinfo {pages} {546}
  (\bibinfo {year} {2019})}\BibitemShut {NoStop}%
\bibitem [{\citenamefont {Kugel}\ and\ \citenamefont
  {Khomskii}(1982)}]{Kugel1982}%
  \BibitemOpen
  \bibfield  {author} {\bibinfo {author} {\bibfnamefont {K.~I.}\ \bibnamefont
  {Kugel}}\ and\ \bibinfo {author} {\bibfnamefont {D.~I.}\ \bibnamefont
  {Khomskii}},\ }\href {\doibase 10.1070/PU1982v025n04ABEH004537} {\bibfield
  {journal} {\bibinfo  {journal} {Sov. Phys. Usp.}\ }\textbf {\bibinfo {volume}
  {25}},\ \bibinfo {pages} {231} (\bibinfo {year} {1982})}\BibitemShut
  {NoStop}%
\bibitem [{Note1()}]{Note1}%
  \BibitemOpen
  \bibinfo {note} {Note that in Ref.~\cite {Lun12} the impact of relatively
  small spin-orbit coupling on the entanglement spectra was
  discussed.}\BibitemShut {Stop}%
\bibitem [{\citenamefont {P\"arschke}\ and\ \citenamefont
  {Ray}(2018)}]{PaerschkeRay2018}%
  \BibitemOpen
  \bibfield  {author} {\bibinfo {author} {\bibfnamefont {E.~M.}\ \bibnamefont
  {P\"arschke}}\ and\ \bibinfo {author} {\bibfnamefont {R.}~\bibnamefont
  {Ray}},\ }\href {\doibase 10.1103/PhysRevB.98.064422} {\bibfield  {journal}
  {\bibinfo  {journal} {Phys. Rev. B}\ }\textbf {\bibinfo {volume} {98}},\
  \bibinfo {pages} {064422} (\bibinfo {year} {2018})}\BibitemShut {NoStop}%
\bibitem [{\citenamefont {Jackeli}\ and\ \citenamefont
  {Khaliullin}(2009{\natexlab{a}})}]{Jackeli2009}%
  \BibitemOpen
  \bibfield  {author} {\bibinfo {author} {\bibfnamefont {G.}~\bibnamefont
  {Jackeli}}\ and\ \bibinfo {author} {\bibfnamefont {G.}~\bibnamefont
  {Khaliullin}},\ }\href {\doibase 10.1103/PhysRevLett.102.017205} {\bibfield
  {journal} {\bibinfo  {journal} {Phys. Rev. Lett.}\ }\textbf {\bibinfo
  {volume} {102}},\ \bibinfo {pages} {017205} (\bibinfo {year}
  {2009}{\natexlab{a}})}\BibitemShut {NoStop}%
\bibitem [{\citenamefont {Eremin}\ \emph {et~al.}(2011)\citenamefont {Eremin},
  \citenamefont {Deisenhofer}, \citenamefont {Eremina}, \citenamefont
  {Teyssier}, \citenamefont {van~der Marel},\ and\ \citenamefont
  {Loidl}}]{Ere11}%
  \BibitemOpen
  \bibfield  {author} {\bibinfo {author} {\bibfnamefont {M.~V.}\ \bibnamefont
  {Eremin}}, \bibinfo {author} {\bibfnamefont {J.}~\bibnamefont {Deisenhofer}},
  \bibinfo {author} {\bibfnamefont {R.~M.}\ \bibnamefont {Eremina}}, \bibinfo
  {author} {\bibfnamefont {J.}~\bibnamefont {Teyssier}}, \bibinfo {author}
  {\bibfnamefont {D.}~\bibnamefont {van~der Marel}}, \ and\ \bibinfo {author}
  {\bibfnamefont {A.}~\bibnamefont {Loidl}},\ }\href {\doibase
  10.1103/PhysRevB.84.212407} {\bibfield  {journal} {\bibinfo  {journal} {Phys.
  Rev. B}\ }\textbf {\bibinfo {volume} {84}},\ \bibinfo {pages} {212407}
  (\bibinfo {year} {2011})}\BibitemShut {NoStop}%
\bibitem [{\citenamefont {Choukroun}(2011)}]{Cho11}%
  \BibitemOpen
  \bibfield  {author} {\bibinfo {author} {\bibfnamefont {J.}~\bibnamefont
  {Choukroun}},\ }\href {\doibase 10.1103/PhysRevB.84.014415} {\bibfield
  {journal} {\bibinfo  {journal} {Phys. Rev. B}\ }\textbf {\bibinfo {volume}
  {84}},\ \bibinfo {pages} {014415} (\bibinfo {year} {2011})}\BibitemShut
  {NoStop}%
\bibitem [{\citenamefont {Svoboda}\ \emph {et~al.}(2017)\citenamefont
  {Svoboda}, \citenamefont {Randeria},\ and\ \citenamefont {Trivedi}}]{Svo17}%
  \BibitemOpen
  \bibfield  {author} {\bibinfo {author} {\bibfnamefont {C.}~\bibnamefont
  {Svoboda}}, \bibinfo {author} {\bibfnamefont {M.}~\bibnamefont {Randeria}}, \
  and\ \bibinfo {author} {\bibfnamefont {N.}~\bibnamefont {Trivedi}},\ }\href
  {\doibase 10.1103/PhysRevB.95.014409} {\bibfield  {journal} {\bibinfo
  {journal} {Phys. Rev. B}\ }\textbf {\bibinfo {volume} {95}},\ \bibinfo
  {pages} {014409} (\bibinfo {year} {2017})}\BibitemShut {NoStop}%
\bibitem [{\citenamefont {Koga}\ \emph {et~al.}(2018)\citenamefont {Koga},
  \citenamefont {Nakauchi},\ and\ \citenamefont {Nasu}}]{Koga2018}%
  \BibitemOpen
  \bibfield  {author} {\bibinfo {author} {\bibfnamefont {A.}~\bibnamefont
  {Koga}}, \bibinfo {author} {\bibfnamefont {S.}~\bibnamefont {Nakauchi}}, \
  and\ \bibinfo {author} {\bibfnamefont {J.}~\bibnamefont {Nasu}},\ }\href
  {\doibase 10.1103/PhysRevB.97.094427} {\bibfield  {journal} {\bibinfo
  {journal} {Phys. Rev. B}\ }\textbf {\bibinfo {volume} {97}},\ \bibinfo
  {pages} {094427} (\bibinfo {year} {2018})}\BibitemShut {NoStop}%
\bibitem [{\citenamefont {Sutherland}(1975)}]{Sutherland1975}%
  \BibitemOpen
  \bibfield  {author} {\bibinfo {author} {\bibfnamefont {B.}~\bibnamefont
  {Sutherland}},\ }\href {\doibase 10.1103/PhysRevB.12.3795} {\bibfield
  {journal} {\bibinfo  {journal} {Phys. Rev. B}\ }\textbf {\bibinfo {volume}
  {12}},\ \bibinfo {pages} {3795} (\bibinfo {year} {1975})}\BibitemShut
  {NoStop}%
\bibitem [{\citenamefont {Yamashita}\ \emph {et~al.}(1998)\citenamefont
  {Yamashita}, \citenamefont {Shibata},\ and\ \citenamefont
  {Ueda}}]{Yamashita1998}%
  \BibitemOpen
  \bibfield  {author} {\bibinfo {author} {\bibfnamefont {Y.}~\bibnamefont
  {Yamashita}}, \bibinfo {author} {\bibfnamefont {N.}~\bibnamefont {Shibata}},
  \ and\ \bibinfo {author} {\bibfnamefont {K.}~\bibnamefont {Ueda}},\ }\href
  {\doibase 10.1103/PhysRevB.58.9114} {\bibfield  {journal} {\bibinfo
  {journal} {Phys. Rev. B}\ }\textbf {\bibinfo {volume} {58}},\ \bibinfo
  {pages} {9114} (\bibinfo {year} {1998})}\BibitemShut {NoStop}%
\bibitem [{\citenamefont {Li}\ \emph {et~al.}(1999)\citenamefont {Li},
  \citenamefont {Ma}, \citenamefont {Shi},\ and\ \citenamefont
  {Zhang}}]{Li1999}%
  \BibitemOpen
  \bibfield  {author} {\bibinfo {author} {\bibfnamefont {Y.-Q.}\ \bibnamefont
  {Li}}, \bibinfo {author} {\bibfnamefont {M.}~\bibnamefont {Ma}}, \bibinfo
  {author} {\bibfnamefont {D.-N.}\ \bibnamefont {Shi}}, \ and\ \bibinfo
  {author} {\bibfnamefont {F.-C.}\ \bibnamefont {Zhang}},\ }\href {\doibase
  10.1103/PhysRevB.60.12781} {\bibfield  {journal} {\bibinfo  {journal} {Phys.
  Rev. B}\ }\textbf {\bibinfo {volume} {60}},\ \bibinfo {pages} {12781}
  (\bibinfo {year} {1999})}\BibitemShut {NoStop}%
\bibitem [{\citenamefont {Kolezhuk}\ and\ \citenamefont
  {Mikeska}(1998)}]{Kolezhuk1998}%
  \BibitemOpen
  \bibfield  {author} {\bibinfo {author} {\bibfnamefont {A.~K.}\ \bibnamefont
  {Kolezhuk}}\ and\ \bibinfo {author} {\bibfnamefont {H.-J.}\ \bibnamefont
  {Mikeska}},\ }\href {\doibase 10.1103/PhysRevLett.80.2709} {\bibfield
  {journal} {\bibinfo  {journal} {Phys. Rev. Lett.}\ }\textbf {\bibinfo
  {volume} {80}},\ \bibinfo {pages} {2709} (\bibinfo {year}
  {1998})}\BibitemShut {NoStop}%
\bibitem [{\citenamefont {Kolezhuk}\ \emph {et~al.}(2001)\citenamefont
  {Kolezhuk}, \citenamefont {Mikeska},\ and\ \citenamefont
  {Schollw\"ock}}]{Kolezhuk2000}%
  \BibitemOpen
  \bibfield  {author} {\bibinfo {author} {\bibfnamefont {A.~K.}\ \bibnamefont
  {Kolezhuk}}, \bibinfo {author} {\bibfnamefont {H.-J.}\ \bibnamefont
  {Mikeska}}, \ and\ \bibinfo {author} {\bibfnamefont {U.}~\bibnamefont
  {Schollw\"ock}},\ }\href {\doibase 10.1103/PhysRevB.63.064418} {\bibfield
  {journal} {\bibinfo  {journal} {Phys. Rev. B}\ }\textbf {\bibinfo {volume}
  {63}},\ \bibinfo {pages} {064418} (\bibinfo {year} {2001})}\BibitemShut
  {NoStop}%
\bibitem [{\citenamefont {Arovas}\ and\ \citenamefont
  {Auerbach}(1995)}]{Arovas1995}%
  \BibitemOpen
  \bibfield  {author} {\bibinfo {author} {\bibfnamefont {D.~P.}\ \bibnamefont
  {Arovas}}\ and\ \bibinfo {author} {\bibfnamefont {A.}~\bibnamefont
  {Auerbach}},\ }\href {\doibase 10.1103/PhysRevB.52.10114} {\bibfield
  {journal} {\bibinfo  {journal} {Phys. Rev. B}\ }\textbf {\bibinfo {volume}
  {52}},\ \bibinfo {pages} {10114} (\bibinfo {year} {1995})}\BibitemShut
  {NoStop}%
\bibitem [{\citenamefont {Pati}\ \emph {et~al.}(1998)\citenamefont {Pati},
  \citenamefont {Singh},\ and\ \citenamefont {Khomskii}}]{Pati1998}%
  \BibitemOpen
  \bibfield  {author} {\bibinfo {author} {\bibfnamefont {S.~K.}\ \bibnamefont
  {Pati}}, \bibinfo {author} {\bibfnamefont {R.~R.~P.}\ \bibnamefont {Singh}},
  \ and\ \bibinfo {author} {\bibfnamefont {D.~I.}\ \bibnamefont {Khomskii}},\
  }\href {\doibase 10.1103/PhysRevLett.81.5406} {\bibfield  {journal} {\bibinfo
   {journal} {Phys. Rev. Lett.}\ }\textbf {\bibinfo {volume} {81}},\ \bibinfo
  {pages} {5406} (\bibinfo {year} {1998})}\BibitemShut {NoStop}%
\bibitem [{\citenamefont {Itoi}\ \emph {et~al.}(2000)\citenamefont {Itoi},
  \citenamefont {Qin},\ and\ \citenamefont {Affleck}}]{Itoi2000}%
  \BibitemOpen
  \bibfield  {author} {\bibinfo {author} {\bibfnamefont {C.}~\bibnamefont
  {Itoi}}, \bibinfo {author} {\bibfnamefont {S.}~\bibnamefont {Qin}}, \ and\
  \bibinfo {author} {\bibfnamefont {I.}~\bibnamefont {Affleck}},\ }\href
  {\doibase 10.1103/PhysRevB.61.6747} {\bibfield  {journal} {\bibinfo
  {journal} {Phys. Rev. B}\ }\textbf {\bibinfo {volume} {61}},\ \bibinfo
  {pages} {6747} (\bibinfo {year} {2000})}\BibitemShut {NoStop}%
\bibitem [{\citenamefont {Zheng}\ and\ \citenamefont
  {Oitmaa}(2001)}]{Zheng2001}%
  \BibitemOpen
  \bibfield  {author} {\bibinfo {author} {\bibfnamefont {W.}~\bibnamefont
  {Zheng}}\ and\ \bibinfo {author} {\bibfnamefont {J.}~\bibnamefont {Oitmaa}},\
  }\href {\doibase 10.1103/PhysRevB.64.014410} {\bibfield  {journal} {\bibinfo
  {journal} {Phys. Rev. B}\ }\textbf {\bibinfo {volume} {64}},\ \bibinfo
  {pages} {014410} (\bibinfo {year} {2001})}\BibitemShut {NoStop}%
\bibitem [{\citenamefont {Li}\ and\ \citenamefont {Shen}(2005)}]{Li2005}%
  \BibitemOpen
  \bibfield  {author} {\bibinfo {author} {\bibfnamefont {P.}~\bibnamefont
  {Li}}\ and\ \bibinfo {author} {\bibfnamefont {S.-Q.}\ \bibnamefont {Shen}},\
  }\href {\doibase 10.1103/PhysRevB.72.214439} {\bibfield  {journal} {\bibinfo
  {journal} {Phys. Rev. B}\ }\textbf {\bibinfo {volume} {72}},\ \bibinfo
  {pages} {214439} (\bibinfo {year} {2005})}\BibitemShut {NoStop}%
\bibitem [{\citenamefont {Kugel}\ \emph {et~al.}(2015)\citenamefont {Kugel},
  \citenamefont {Khomskii}, \citenamefont {Sboychakov},\ and\ \citenamefont
  {Streltsov}}]{Kugel2015}%
  \BibitemOpen
  \bibfield  {author} {\bibinfo {author} {\bibfnamefont {K.~I.}\ \bibnamefont
  {Kugel}}, \bibinfo {author} {\bibfnamefont {D.~I.}\ \bibnamefont {Khomskii}},
  \bibinfo {author} {\bibfnamefont {A.~O.}\ \bibnamefont {Sboychakov}}, \ and\
  \bibinfo {author} {\bibfnamefont {S.~V.}\ \bibnamefont {Streltsov}},\ }\href
  {\doibase 10.1103/PhysRevB.91.155125} {\bibfield  {journal} {\bibinfo
  {journal} {Phys. Rev. B}\ }\textbf {\bibinfo {volume} {91}},\ \bibinfo
  {pages} {155125} (\bibinfo {year} {2015})}\BibitemShut {NoStop}%
\bibitem [{\citenamefont {Gorshkov}\ \emph {et~al.}(2010)\citenamefont
  {Gorshkov}, \citenamefont {Hermele}, \citenamefont {Gurarie}, \citenamefont
  {Xu}, \citenamefont {Julienne}, \citenamefont {Ye}, \citenamefont {Zoller},
  \citenamefont {Demler}, \citenamefont {Lukin},\ and\ \citenamefont
  {Rey}}]{Gorshkov2010}%
  \BibitemOpen
  \bibfield  {author} {\bibinfo {author} {\bibfnamefont {A.~V.}\ \bibnamefont
  {Gorshkov}}, \bibinfo {author} {\bibfnamefont {M.}~\bibnamefont {Hermele}},
  \bibinfo {author} {\bibfnamefont {V.}~\bibnamefont {Gurarie}}, \bibinfo
  {author} {\bibfnamefont {C.}~\bibnamefont {Xu}}, \bibinfo {author}
  {\bibfnamefont {P.~S.}\ \bibnamefont {Julienne}}, \bibinfo {author}
  {\bibfnamefont {J.}~\bibnamefont {Ye}}, \bibinfo {author} {\bibfnamefont
  {P.}~\bibnamefont {Zoller}}, \bibinfo {author} {\bibfnamefont
  {E.}~\bibnamefont {Demler}}, \bibinfo {author} {\bibfnamefont {M.~D.}\
  \bibnamefont {Lukin}}, \ and\ \bibinfo {author} {\bibfnamefont {A.~M.}\
  \bibnamefont {Rey}},\ }\href {https://doi.org/10.1038/nphys1535} {\bibfield
  {journal} {\bibinfo  {journal} {Nature Physics}\ }\textbf {\bibinfo {volume}
  {6}},\ \bibinfo {pages} {289} (\bibinfo {year} {2010})}\BibitemShut {NoStop}%
\bibitem [{\citenamefont {Zhang}\ \emph {et~al.}(2014)\citenamefont {Zhang},
  \citenamefont {Bishof}, \citenamefont {Bromley}, \citenamefont {Kraus},
  \citenamefont {Safronova}, \citenamefont {Zoller}, \citenamefont {Rey},\ and\
  \citenamefont {Ye}}]{Zhang2014}%
  \BibitemOpen
  \bibfield  {author} {\bibinfo {author} {\bibfnamefont {X.}~\bibnamefont
  {Zhang}}, \bibinfo {author} {\bibfnamefont {M.}~\bibnamefont {Bishof}},
  \bibinfo {author} {\bibfnamefont {S.~L.}\ \bibnamefont {Bromley}}, \bibinfo
  {author} {\bibfnamefont {C.~V.}\ \bibnamefont {Kraus}}, \bibinfo {author}
  {\bibfnamefont {M.~S.}\ \bibnamefont {Safronova}}, \bibinfo {author}
  {\bibfnamefont {P.}~\bibnamefont {Zoller}}, \bibinfo {author} {\bibfnamefont
  {A.~M.}\ \bibnamefont {Rey}}, \ and\ \bibinfo {author} {\bibfnamefont
  {J.}~\bibnamefont {Ye}},\ }\href {\doibase 10.1126/science.1254978}
  {\bibfield  {journal} {\bibinfo  {journal} {Science}\ }\textbf {\bibinfo
  {volume} {345}},\ \bibinfo {pages} {1467} (\bibinfo {year}
  {2014})}\BibitemShut {NoStop}%
\bibitem [{\citenamefont {Dou}\ \emph {et~al.}(2016)\citenamefont {Dou},
  \citenamefont {Kotov},\ and\ \citenamefont {Uchoa}}]{Dou2016}%
  \BibitemOpen
  \bibfield  {author} {\bibinfo {author} {\bibfnamefont {X.}~\bibnamefont
  {Dou}}, \bibinfo {author} {\bibfnamefont {V.~N.}\ \bibnamefont {Kotov}}, \
  and\ \bibinfo {author} {\bibfnamefont {B.}~\bibnamefont {Uchoa}},\ }\href
  {https://doi.org/10.1038/srep31737} {\bibfield  {journal} {\bibinfo
  {journal} {Scientific Reports}\ }\textbf {\bibinfo {volume} {6}},\ \bibinfo
  {pages} {31737} (\bibinfo {year} {2016})}\BibitemShut {NoStop}%
\bibitem [{\citenamefont {Horsch}\ \emph {et~al.}(2003)\citenamefont {Horsch},
  \citenamefont {Khaliullin},\ and\ \citenamefont {Ole\ifmmode~\acute{s}\else
  \'{s}\fi{}}}]{Horsch2003}%
  \BibitemOpen
  \bibfield  {author} {\bibinfo {author} {\bibfnamefont {P.}~\bibnamefont
  {Horsch}}, \bibinfo {author} {\bibfnamefont {G.}~\bibnamefont {Khaliullin}},
  \ and\ \bibinfo {author} {\bibfnamefont {A.~M.}\ \bibnamefont
  {Ole\ifmmode~\acute{s}\else \'{s}\fi{}}},\ }\href {\doibase
  10.1103/PhysRevLett.91.257203} {\bibfield  {journal} {\bibinfo  {journal}
  {Phys. Rev. Lett.}\ }\textbf {\bibinfo {volume} {91}},\ \bibinfo {pages}
  {257203} (\bibinfo {year} {2003})}\BibitemShut {NoStop}%
\bibitem [{\citenamefont {Jackeli}\ and\ \citenamefont
  {Khaliullin}(2009{\natexlab{b}})}]{Jackeli2009b}%
  \BibitemOpen
  \bibfield  {author} {\bibinfo {author} {\bibfnamefont {G.}~\bibnamefont
  {Jackeli}}\ and\ \bibinfo {author} {\bibfnamefont {G.}~\bibnamefont
  {Khaliullin}},\ }\href {\doibase 10.1103/PhysRevLett.103.067205} {\bibfield
  {journal} {\bibinfo  {journal} {Phys. Rev. Lett.}\ }\textbf {\bibinfo
  {volume} {103}},\ \bibinfo {pages} {067205} (\bibinfo {year}
  {2009}{\natexlab{b}})}\BibitemShut {NoStop}%
\bibitem [{\citenamefont {Solovyev}(2008)}]{Solovyev2008}%
  \BibitemOpen
  \bibfield  {author} {\bibinfo {author} {\bibfnamefont {I.~V.}\ \bibnamefont
  {Solovyev}},\ }\href {\doibase 10.1088/1367-2630/10/1/013035} {\bibfield
  {journal} {\bibinfo  {journal} {New J. Phys.}\ }\textbf {\bibinfo {volume}
  {10}},\ \bibinfo {pages} {013035} (\bibinfo {year} {2008})}\BibitemShut
  {NoStop}%
\bibitem [{\citenamefont {Koch}(2011)}]{Koch}%
  \BibitemOpen
  \bibfield  {author} {\bibinfo {author} {\bibfnamefont {E.}~\bibnamefont
  {Koch}},\ }\href@noop {} {\emph {\bibinfo {title} {The Lanczos Method, in:
  The LDA+DMFT approach to strongly correlated materials, \textrm{edited by E.
  Pavarini, E. Koch, D. Vollhardt, and A. Lichtenstein}}}}\ (\bibinfo
  {publisher} {Forschungszentrum J\"ulich},\ \bibinfo {address} {J\"ulich},\
  \bibinfo {year} {2011})\BibitemShut {NoStop}%
\bibitem [{\citenamefont {Kitaev}\ and\ \citenamefont {Preskill}(2006)}]{vNS}%
  \BibitemOpen
  \bibfield  {author} {\bibinfo {author} {\bibfnamefont {A.}~\bibnamefont
  {Kitaev}}\ and\ \bibinfo {author} {\bibfnamefont {J.}~\bibnamefont
  {Preskill}},\ }\href {\doibase 10.1103/PhysRevLett.96.110404} {\bibfield
  {journal} {\bibinfo  {journal} {Phys. Rev. Lett.}\ }\textbf {\bibinfo
  {volume} {96}},\ \bibinfo {pages} {110404} (\bibinfo {year}
  {2006})}\BibitemShut {NoStop}%
\bibitem [{\citenamefont {Li}\ \emph {et~al.}(1998)\citenamefont {Li},
  \citenamefont {Ma}, \citenamefont {Shi},\ and\ \citenamefont {Zhang}}]{Li98}%
  \BibitemOpen
  \bibfield  {author} {\bibinfo {author} {\bibfnamefont {Y.~Q.}\ \bibnamefont
  {Li}}, \bibinfo {author} {\bibfnamefont {M.}~\bibnamefont {Ma}}, \bibinfo
  {author} {\bibfnamefont {D.~N.}\ \bibnamefont {Shi}}, \ and\ \bibinfo
  {author} {\bibfnamefont {F.~C.}\ \bibnamefont {Zhang}},\ }\href {\doibase
  10.1103/PhysRevLett.81.3527} {\bibfield  {journal} {\bibinfo  {journal}
  {Phys. Rev. Lett.}\ }\textbf {\bibinfo {volume} {81}},\ \bibinfo {pages}
  {3527} (\bibinfo {year} {1998})}\BibitemShut {NoStop}%
\bibitem [{\citenamefont {Ole\ifmmode~\acute{s}\else \'{s}\fi{}}\ \emph
  {et~al.}(1985)\citenamefont {Ole\ifmmode~\acute{s}\else \'{s}\fi{}},
  \citenamefont {Tr\'eglia}, \citenamefont {Spanjaard},\ and\ \citenamefont
  {Jullien}}]{Ole85}%
  \BibitemOpen
  \bibfield  {author} {\bibinfo {author} {\bibfnamefont {A.~M.}\ \bibnamefont
  {Ole\ifmmode~\acute{s}\else \'{s}\fi{}}}, \bibinfo {author} {\bibfnamefont
  {G.}~\bibnamefont {Tr\'eglia}}, \bibinfo {author} {\bibfnamefont
  {D.}~\bibnamefont {Spanjaard}}, \ and\ \bibinfo {author} {\bibfnamefont
  {R.}~\bibnamefont {Jullien}},\ }\href {\doibase 10.1103/PhysRevB.32.2167}
  {\bibfield  {journal} {\bibinfo  {journal} {Phys. Rev. B}\ }\textbf {\bibinfo
  {volume} {32}},\ \bibinfo {pages} {2167} (\bibinfo {year}
  {1985})}\BibitemShut {NoStop}%
\bibitem [{\citenamefont {Osterloh}\ \emph {et~al.}(2002)\citenamefont
  {Osterloh}, \citenamefont {Amico}, \citenamefont {Falci},\ and\ \citenamefont
  {Fazio}}]{Osterloh2002}%
  \BibitemOpen
  \bibfield  {author} {\bibinfo {author} {\bibfnamefont {A.}~\bibnamefont
  {Osterloh}}, \bibinfo {author} {\bibfnamefont {L.}~\bibnamefont {Amico}},
  \bibinfo {author} {\bibfnamefont {G.}~\bibnamefont {Falci}}, \ and\ \bibinfo
  {author} {\bibfnamefont {R.}~\bibnamefont {Fazio}},\ }\href {\doibase
  10.1038/416608a} {\bibfield  {journal} {\bibinfo  {journal} {Nature}\
  }\textbf {\bibinfo {volume} {416}},\ \bibinfo {pages} {608} (\bibinfo {year}
  {2002})}\BibitemShut {NoStop}%
\bibitem [{\citenamefont {Jia}\ \emph {et~al.}(2008)\citenamefont {Jia},
  \citenamefont {Subramaniam}, \citenamefont {Gruzberg},\ and\ \citenamefont
  {Chakravarty}}]{Xun2008}%
  \BibitemOpen
  \bibfield  {author} {\bibinfo {author} {\bibfnamefont {X.}~\bibnamefont
  {Jia}}, \bibinfo {author} {\bibfnamefont {A.~R.}\ \bibnamefont
  {Subramaniam}}, \bibinfo {author} {\bibfnamefont {I.~A.}\ \bibnamefont
  {Gruzberg}}, \ and\ \bibinfo {author} {\bibfnamefont {S.}~\bibnamefont
  {Chakravarty}},\ }\href {\doibase 10.1103/PhysRevB.77.014208} {\bibfield
  {journal} {\bibinfo  {journal} {Phys. Rev. B}\ }\textbf {\bibinfo {volume}
  {77}},\ \bibinfo {pages} {014208} (\bibinfo {year} {2008})}\BibitemShut
  {NoStop}%
\bibitem [{\citenamefont {Hijii}\ \emph {et~al.}(2005)\citenamefont {Hijii},
  \citenamefont {Kitazawa},\ and\ \citenamefont {Nomura}}]{Hijii2005}%
  \BibitemOpen
  \bibfield  {author} {\bibinfo {author} {\bibfnamefont {K.}~\bibnamefont
  {Hijii}}, \bibinfo {author} {\bibfnamefont {A.}~\bibnamefont {Kitazawa}}, \
  and\ \bibinfo {author} {\bibfnamefont {K.}~\bibnamefont {Nomura}},\ }\href
  {\doibase 10.1103/PhysRevB.72.014449} {\bibfield  {journal} {\bibinfo
  {journal} {Phys. Rev. B}\ }\textbf {\bibinfo {volume} {72}},\ \bibinfo
  {pages} {014449} (\bibinfo {year} {2005})}\BibitemShut {NoStop}%
\bibitem [{Note2()}]{Note2}%
  \BibitemOpen
  \bibinfo {note} {$ \lambda _{\protect \rm {CRIT}}$ depends on the particular
  values of the model parameters, see Sec. \ref {sec:resultsB}.}\BibitemShut
  {Stop}%
\end{thebibliography}
%

\end{document}